\RequirePackage{fix-cm}
\documentclass[natbib,smallextended]{svjour3}       % onecolumn (second format)
\smartqed  % flush right qed marks, e.g. at end of proof
\usepackage{graphicx}
\usepackage{color,soul}
\usepackage[colorlinks=true,linkcolor=blue,citecolor=blue,urlcolor=blue]{hyperref}
\usepackage{amsmath}
\usepackage{amssymb}
\usepackage{aas_macros}
\newcommand{\gaia}{GAIA}
\newcommand{\MV}{MultiView}
\newcommand{\SFPR}{SFPR}
\newcommand{\MFPR}{MFPR}
\newcommand{\uas}{$\mu$as}
\newcommand{\uasyr}{$\mu$as\,yr$^{-1}$}
\newcommand{\amin}{$^\prime$}
\newcommand{\arcmin}{$^\prime$}

\newcommand{\Msun}{\mbox{$M_{\odot}$}}

\sf %tipo de letra utilizado!!!!!!!!

\newcommand{\kms}{\mbox{km~s$^{\sf -1}$}}

\def\mnras{MNRAS}
\def\apj{ApJ}
\def\apjl{ApJL}
\def\aj{AJ}
\def\aap{A\&A}
\def\pasj{PASJ}

\def\wat{H$_{2}$O}

\def\nh3{NH$_{3}$}
\def\kms{km~s$^{-1}$}

\newcommand{\degr}{\ensuremath{^\circ}}

\def\arcsec{\hbox{$^{\prime\prime}$}}

\journalname{The Astronomy and Astrophysics Review}
\begin{document}

\title{Precise radio astrometry and new developments for the next generation of instruments}

\titlerunning{New developments in astrometry}

\author{Mar\'{\i}a J. Rioja \and Richard Dodson 
}

\institute{M.\ J. Rioja \at
  %International Centre for Radio Astronomy Research, M468, The University of Western Australia, 35 Stirling Hwy, Crawley, Western Australia, 6009
 CSIRO Astronomy and Space Science, 26 Dick Perry Avenue, Kensington WA 6151, Australia and \\
 International Centre for Radio Astronomy Research, The University of Western Australia, 35 Stirling Hwy, Western Australia and \\
 Observatorio Astron\'omico Nacional (IGN), Alfonso XII, 3 y 5, 28014 Madrid, Spain \\
 \email{maria.rioja@uwa.edu.au}
 \and
 R. Dodson \at
 International Centre for Radio Astronomy Research, The University of Western Australia, 35 Stirling Hwy, Western Australia
}

\date{Received: date / Accepted: date}
% The correct dates will be entered by the editor

\maketitle

\begin{abstract}
%look at: 
% check TLA if Sec 2 moved.
% Terms for SFPR/MFPR,
%all bold, coloured and struck out text,\\
%section headings
%VGOS RadSci paper -reference. Importance?

%Submission.tex 

We present a technique-led review of the progression of precise radio astrometry, from the first demonstrations, half a century ago, until to date and into the future. We cover the developments that have been fundamental to allow high accuracy and precision astrometry to be regularly achieved. 
We review the opportunities provided by the next-generation of instruments coming online, which are primarily: SKA, ngVLA and pathfinders, along with EHT and other (sub)mm-wavelength arrays, Space-VLBI, Geodetic arrays and optical astrometry from \gaia.

From the historical development we predict the future potential astrometric performance, and therefore the instrumental requirements that must be provided to deliver these.
The next-generation of methods will allow ultra-precise astrometry to be performed at a much wider range of frequencies (hundreds of MHz to hundreds of GHz). 
One of the key potentials is that astrometry will become generally applicable, and therefore unbiased large surveys can be performed. The next-generation methods are fundamental in allowing this.
We review the small but growing number of major astrometric surveys in the radio, to highlight the scientific impact that such projects can provide. 

Based on these perspectives, the future of radio astrometry is bright. 
We foresee a revolution coming from: ultra-high precision radio astrometry, large surveys of many objects, improved sky coverage and at new frequency bands other than those available today. These will enable the addressing of a host of innovative open scientific questions in astrophysics.

\end{abstract}

\setcounter{tocdepth}{3} % TOC subsubsections
\tableofcontents

\section{Introduction}
\typeout{The future of radio astrometry is bright!}
The field of astrometry, through the measurement of the precise positions, distances and motions of astronomical objects, 
is a fundamental tool for astrophysics, and has been revolutionised in recent years.
The level of accuracy and precision achieved, along with its applicability, determines its potential. 
Currently radio astrometry with Very Long Baseline Interferometry (VLBI) provides the highest accuracy and precision in astronomy.
The relentless increase of the precision in these measurements over time has extended the application into many new areas of fundamental astronomy, astrophysics and cosmology. 
As a result astrometry is assisting in the revision of our knowledge of the physical parameters of individual sources, such as 
size, luminosity, mass and age, and the understanding of stellar birth and stellar evolution, 
using direct measurements of the trigonometric parallax distances to objects across the Galaxy, with unprecedented $\pm10\%$ accuracy. 
Furthermore, combining with the proper motion measurements these reveal the 3-D spiral structure and the values of the fundamental dynamical parameters of our Galaxy, using the six-dimensional phase-space distributions of Galactic sources. 
In many cases the ``gold standard'' VLBI distances measured 
by trigonometric parallax is significantly different from the
kinematic distances and also occasionally to Hipparcos measurements, hence the fundamental role of VLBI to serve as a cross-check for its successor, \gaia.
For a recent review of the astrophysical applications of these measurements in radio astrometry see \citet{reid_micro}, and for optical astrometry see \citet{optical_astro}.

\subsection{VLBI astrometry}
\begin{figure}[htb]
    \centering
    \includegraphics[width=\textwidth]{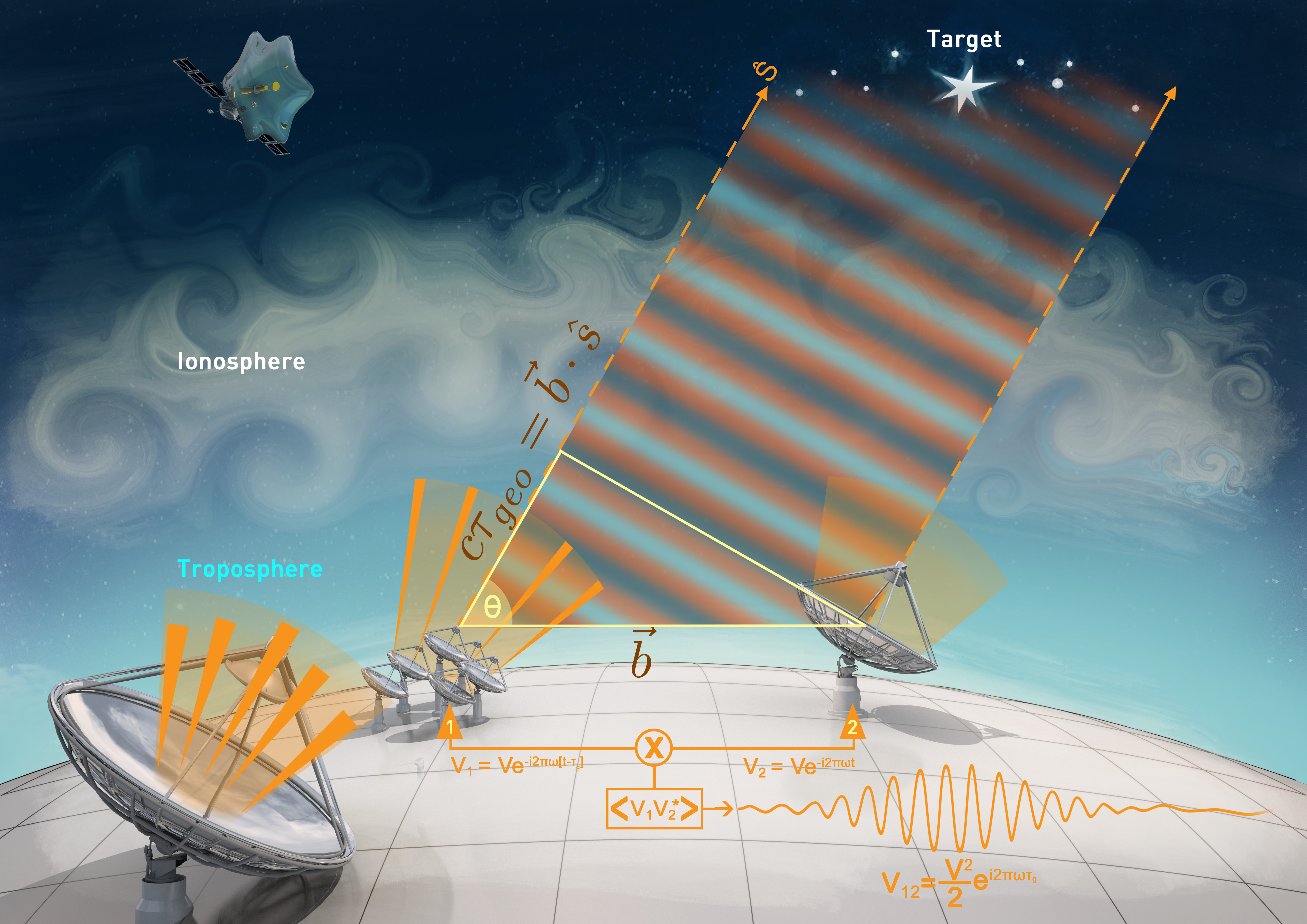}
    \caption{A schematic of a VLBI interferometric array, comprising diverse elements on the ground and in space.
    VLBI measures the total delay in the arrival time of the signal from a target source, along the direction $\hat{s}$, to pairs of elements that can be separated by thousands of kilometres and form the baseline $\protect\overrightarrow{b}$.
    The dot product of the two vectors corresponds to the geometric delay $\tau_{\rm geo}$. %, and thus the path-length difference. 
    The propagation through the turbulent tropospheric and ionospheric components of the atmosphere introduces additional contaminating delays, as does the instrumentation, which need to be removed in the analysis (Sects.~\ref{sec:methods_old} and \ref{sec:methods_new}).
    The long baseline allows an extremely accurate determination of $\hat{s}$. 
    Each pair of elements output two voltages, V$_1$ and V$_2$, which are multiplied and averaged in the correlator, creating a time-dependent interferometric fringe pattern. The phase and delay from this complex correlated data product are the fundamental VLBI observables in astrometry. %
    %For each pair of elements the baseline vector is . 
    %Also shown is a large multiple-pixel telescope and a space-VLBI antenna.
    { 
    The figure shows a variety in the nature and size of the elements in a VLBI-array in the era of the next-generation of instruments (Sect.~\ref{sec:next}). 
    Alongside the moderate-sized single dish telescope there are the large collecting areas provided by a massive single dish or many smaller dishes from a connected array phased-up, and also the extremely long baselines provided by a radio telescope on a space platform.  
    %To provide a homogeneous FoV.
    %To match the FoV represented by the light orange cone, 
    New technologies can provide matching FoVs (light orange cones) from all of these elements and the capability for multi-frequency observations (Sect.~\ref{sec:tech}).
    Multiple-pixel capabilities are illustrated by the multiple tied-array beams from the connected array and the multiple beams from a phased array feed on the large telescope, both of which match the wider FoV of the moderate-sized single pixel telescope. 
    The capability for simultaneous multi-frequency observations is indicated by the two-tone incoming wavefront. Image credit: M.J. Rioja and R. Dodson}}
    \label{fig:radio_int}
\end{figure}

Very Long Baseline Interferometry  is a geometric technique that provides the highest angular resolutions in astronomy, as sketched in Fig.~\ref{fig:radio_int}. 
{
% From mm-VLBI
% A radio interferometer is an instrument which enables to combine the radio waves coming from an astronomical object to form interference fringes. By correlating the signals collected simultaneously by each telescope forming the array, radio interferometers measure the complex visibility function V(u, v), which is the (noise-corrupted) Fourier transform of the brightness distribution of the sky. The (u, v) coordinates define, in units of wavelength, the East–West and the North–South component of each baseline projected in the sky, as seen from the source. Thus, the (u, v)-plane contains information about the existence or absence of a visibility measurement in a certain point. The filling of the (u, v)-plane, i.e., the (u, v)-coverage, is by definition incomplete for any interferometer, and can be improved by adding more telescopes to the array and by increasing the on-source time up to 12 h, so that the Earth rotation enables a single baseline to sample a full track (an ellipse) in the (u, v)-plane. Each baseline of projected length b is characterized by its own fringe pattern, and is only sensitive to source structures on scales comparable to the fringe spacing ��/��. Therefore, the better the sampling of the (u, v)-plane, the more reliable will be the reconstruction of the sky brightness distribution. The smallest angular scale an interferometer can probe, its resolution, coincides, in general, with the diffraction limit ∼��/��max, where ��max is the maximum baseline length.
% From mm-VLBI
Interferometry consists of the measurements of the coherence function, at a spatial separation. In astronomy these correspond to the correlation of two signals, taken by observatories separated by a `baseline' vector and, as the signal is travelling from astronomical distances, represents the Fourier term of the transform of the brightness distribution of the observable sky, at the spatial frequency given by the baseline \citep[e.g., Van Citter 1934, discussed in][]{prin_optics}. % \citep{vancitter}. %\citep{TMSv3}[Chpt. 15]. % Or
Larger separations (i.e. longer baselines) represent higher `spatial frequencies' and provide higher image angular resolutions.
The more complete the sampling the more reliable will be the reconstruction of the sky brightness distribution. 
Radio interferometry is blessed in that the phase of the correlated signal can be measured, so the coherence function can be recovered. This complex quantity is called the `visibility function' and depends only on the antenna pair separation (the baseline), measured in uv-space: $(u,v,w)$, and the brightness distribution. Thus the reconstruction of the observed sky can be performed with a simple Fourier transform of these complex quantities, sampled at the uv-points.
In comparison, for optical interferometry or X-ray crystallography the phase terms are lost, introducing significant complexities in the reconstruction \citep[e.g.,][]{perutz_lec}.
The phase of a complex number has the greatest impact on image reconstruction,
%(see the Fourier Duck, or Some Chapter in some white book).
thus in radio the prime observable quantity is considered to be the phase and its derivatives in frequency (delay) and time (rate), particularly in astrometry where peak position is more important than image fidelity.
}
VLBI uses simultaneous observations between an array of widely separated telescopes, 
each equipped  with an extremely precise atomic clock,
to measure the delay or difference in the arrival time of a radio wavefront at pairs of antennae when combined at the time of the correlation; the output of the correlator becomes what we call the VLBI observables. 
{ In the ideal case without errors, as shown Fig.~\ref{fig:radio_int}, this measures the geometric delay $\tau_{\rm geo}$. That is $\tau_{\rm geo} = \overrightarrow{b}.\hat{s}/c$, where $\hat{s}$ is the unit vector in the direction of the source, $\overrightarrow{b}$ is the baseline vector and $c$ is the speed of light.
Astrometry is concerned with the accurate measurement of this term, but in reality the direct measurement is contaminated with delay contributions arising from instrumental and atmospheric propagation effects. 
%errors from the instrument and the atmosphere. 
Moreover, the wide separations between the antennas makes the stabilisation of phases difficult because the lines of sight from each antenna pass through totally uncorrelated atmospheres (ionosphere and troposphere). 
The art of astrometry is the careful identification and correction for these non-astronomical contributions, through special observational and analytical techniques, and by using accurate geometric and atmospheric a-priori models; that is `calibration'. 
%The VLBI observables 
Because these measurements are precise to a few picoseconds VLBI has the potential to determine
the celestial source positions to micro-arcsecond (\uas) level when using very long baselines, providing there is sufficient signal-to-noise ratio (SNR). % (i.e., sensitivity is important for astrometry). 
For a full description of the fundamentals of radio interferometry we recommend \citet{TMSv3}. 
The individual elements of the array can be very different in character, as indicated in Fig. \ref{fig:radio_int}. 
With the new technical capabilties now available the elements can be comprised of: connected-arrays forming multiple tied-array beams, large telescopes with multi-beam feeds and/or smaller telescopes with a single pixel feed. 
Therefore, despite their difference in character, they can have a matching, larger, Field of View (FoV) (see the light orange cone, Fig. \ref{fig:radio_int}).
}
%
%\sout{Many of the difficulties with phases can be overcome by careful observing techniques and by using accurate geometric and atmospheric models.}
%Because these measurements are precise to a few picoseconds VLBI has the potential to determine  the relative position of the antennas to a few millimeters and the celestial source positions to a fraction of a milli-arcsecond when using very long baselines 

The majority of VLBI post-processing analysis uses self-calibration techniques, 
which are very powerful and widely used to generate the highest spatial resolution images in astronomy. Self-calibration relies on phase-closure constraints to separate 
the contributions from the (baseline-based) intrinsic source structure from all  other (antenna-based) {contaminating contributions that are cancelled out in the summation around the three baselines}. The latter includes the source position errors, which is therefore lost in the image product.
Special post-processing analysis techniques are therefore required to preserve the astrometric information, whilst removing the bulk of errors from the contaminating contributions.
The conditions for astrometry with VLBI are more stringent than { those} for imaging, which results in astrometric surveys being constrained by selection effects and biases, and limited to a restricted region of the spectrum. 
This paper focuses on the exciting opportunities (and challenges) for ultra-high precision ($\sim$\uas) radio astrometric surveys with the arrival of the next-generation instruments, planned and under construction, from the combined power of highest sensitivity and long baselines, along with a historic perspective of the evolution of the precision astrometric measurements in the past 50 years. 

The key lies both in the ongoing advances in the accuracy and precision of astrometric measurements and its extended applicability to many objects.
Nowadays, 10\,\uas{} high precision astrometry is  achievable, for a limited range of frequencies;  
this level of accuracy is approaching the potential (i.e., the thermal noise level) of the observations with current instruments. 
The power and potential of astrometric surveys have been demonstrated, for example with the 3-D Galactic mapping projects carried by BeSSeL (Bar and Spiral Structure Legacy) and VERA (VLBI Exploration of Radio Astrometry).
Great benefits will be obtained from comprehensive surveys of high-accuracy radio astrometric measurements, 
as demonstrated by the optical \gaia\ astrometry mission. This of course requires that the astrometric methods for radio are robust and widely applicable for more targets and at a wider range of frequencies; that is, that they are not limited to carefully selected cases. 
%{\bf something about Gaia here?}

The arrival of the next-generation radio observatories, { such as the Square Kilometre Array \citep[SKA;][]{ska_aas} and the next-generation Very Large Array \citep[ngVLA;][]{ngvla},} will provide a huge increase in sensitivity, by up to several orders of magnitude, 
spanning the whole radio frequency band; these observatories will participate in VLBI observations,
as powerful phased-up telescopes, in conjunction with other existing telescopes, resulting in a dramatic increase in the sensitivity of the VLBI array.
The combination of enhanced sensitivity and long baselines offers great benefits for astrometry, and will result in a ``breakthrough'' with respect to what we can achieve today.
On one hand we will have the ability to significantly increase the astrometric measurement accuracy, 
on the other, the capability to study in exquisite detail larger samples  through the inclusion of weaker objects. 
That is, the realistic possibility to carry out large radio $\mu$as astrometric surveys of complete samples of objects, to advance the understanding of astrophysical phenomena and classes.

The next-generation instruments bring opportunities and challenges for astrometry.
Increased sensitivity (alone) does not result in increased astrometric accuracy and precision in the presence of systematic errors, which often set the ultimate limit. 
In order to realise the full potential from the enhanced sensitivity with the next-generation instruments, calibration algorithms for systematic errors must achieve the same level of refinement; 
that is, to mitigate the systematic residual errors to the reduced level of the thermal noise limits of the observations. 
The on-going development of new calibration methods and new technologies are breaking new ground and providing solutions to overcome the current limitations, in particular the frequency dependent systematic errors arising from the propagation of a wave-front through the Earth's atmosphere, which
constitutes the dominant source of astrometric errors that limit the applicability. 
The distinct nature of these errors across the spectrum requires different calibration strategies depending on the observing frequency, and this drives the layout followed in this review. 
Additionally, one will have to deal with the new sources of errors that will become significant at these reduced thermal noise levels.   
The goal is that a combination of all these new developments will lead to the improvement of astrometric performance by an order of magnitude,
for a wide range of radio frequencies and for many objects with the prospect to expand its applications into new areas of research.

\textit{Contents of this paper:} The sections in this paper are  organised to provide:
a common framework and description for astrometric methods;
an overview of the achievements so far, with highlights from surveys with current instruments;
and a projection into the future of radio astrometry.
The imminent arrival of new instruments and new technological and analytical developments will enable new astrophysical applications, examples of which we describe.

% Sect.~\ref{sec:next} is an overview of the next-generation instruments planned or under construction across the radio frequency band, from the perspective of the new astrometric opportunities offered by their extremely high sensitivity and the new science enabled.

% Sect.~\ref{sec:formula} is concerned with the fundamental constraints in astrometry. Sections \ref{sec:methods_old} and \ref{sec:methods_new} review the progress in calibration methods, comprising established methods and recent innovative solutions of interest for next-generation instruments, respectively.
% Sect.~\ref{sec:past} looks back to the historical results and compares them to the expected systematic errors, to identify the important advances in the past. 
% Sect.~\ref{sec:tech} describes key technological developments and studies relevant to the next-generation methods described above, to match the potential of the next-generation instruments. 

% Sect.~\ref{sec:applications} is a review of a subset of astrometric surveys with existing instruments, 
% selected to highlight the benefits of high-precision astrometry.
% They are limited to publications since the review of \citet{reid_micro}. 
% Sect.~\ref{sec:future} explores the future possibilities opened by the new techniques and available or planned infrastructure. Sect.~\ref{sec:summary} is the summary of the review.

% Alternative Order 
{ Section~\ref{sec:formula} is concerned with the fundamental constraints in astrometry and provides the framework for the characterisation and comparison of the methods. Section~\ref{sec:methods_old} reviews the past progress in calibration, with established methods.
Section~\ref{sec:next} is an overview of the next-generation instruments planned or under construction across the radio frequency band, from the perspective of the new astrometric opportunities offered by their extremely high sensitivity and the new science enabled.
Section~\ref{sec:methods_new} introduces recent innovations in methods, and enabling technologies, of interest for these next-generation instruments.
Section~\ref{sec:past} looks back at the evolution of the published astrometric results and compares them to the expected systematic errors, to identify the important advances so far. 
Section~\ref{sec:tech} describes key technological developments and recent atmospheric propagation studies relevant to the next-generation methods described above, to match the potential of the next-generation instruments. 
Section~\ref{sec:applications} is a review of a subset of astrometric surveys with existing instruments, 
selected to highlight the benefits of high-precision astrometry.
They are limited to publications since the review of \citet{reid_micro}. 
Section~\ref{sec:future} explores the future scientific possibilities opened by the new methods and available or planned infrastructure. Section~\ref{sec:summary} is the summary of the review.}

\section{Fundamental constraints in astrometry} % (Requirements for Astrometry)}
\label{sec:formula}% Formulae layout the inputs to the errors
% ionospehere/troposphere
% dynamic/static

\typeout{This section Lists the Systematic errors that limit the goals in the introduction}

We aim at reviewing the astrometric constraints, along with a focus on methods and technological developments that have contributed to significant advances in the field, as a launchpad for inferring the direction of new developments into the future. 
Benefits have come from increasingly stable~instruments, receiver and backend developments that resulted in increased sensitivity through low system temperatures, wider bandwidths, frequency agility, improved telescope networks to optimise uv-coverage and higher slew speeds of radio telescopes.

These combined with advances in calibration methods, the increased number and frequency coverage of suitable~calibrator sources, improved a-priori knowledge of the contributions to the observables, among others, 
have paved the way towards the huge expansion of the field of precise astrometry with VLBI.

Next-generation instruments provide unprecedented sensitivity and have the potential for an order of magnitude improvement, delivering \uas{} astrometric surveys, provided the requirements for optimum astrometric calibration methods and usability are not designed out.
%This is Important to ensure the new astrometry requirements are not `designed’ out.

\subsection{Micro-arcsecond astrometry basics}\label{sec:basics}
{ 
The primary VLBI observables for high precision astrometry are: the interferometric phase ${\phi}$, the phase difference between the signals received at a pair of antenna, or equivalently the phase delay (the phase in radians divided by the frequency in Hz, in seconds; $\tau_{\phi}$  =${1 \over 2\pi} {\phi \over \nu}$), the frequency derivative or group delay (the rate of change of phase with frequency, in seconds; $\tau_g$ =${1 \over 2\pi} {\delta\phi \over \delta\nu}$) and the phase time derivative (usually as delay rate in seconds/second; ${1 \over 2\pi \nu} {\delta\phi \over \delta t}$). }
%, albeit with some differences. 
%{\bf insert here reduced version - complement with error phase expressions } \\
%
Following the standard nomenclature \citep[e.g.,][]{TMSv3} the residual phase  values or errors, after subtracting the {\it a-priori} model 
%(values (for the various contributing terms) 
contributions from the measured values at the time of correlation or post-processing, for observations of the target source ($A$) for a given baseline as a function of time and frequency, are shown as a sum of contributions:
%(For each interferometer, or baseline, of a VLBI array the interferometer phase, $\phi$(t), can be written as follows:)

%\eqnum{1}
\begin{equation}\label{eq:basis}
%$
  \phi_{A} (t,\nu) = \phi_{A, \rm pos} + \phi_{A, \rm str} + \phi_{A, \rm geo} +  \phi_{A, \rm tro} + \phi_{A, \rm ion}
  +\phi_{A, \rm inst} + \phi_{A, \rm thermal} + 
  2\pi n \hspace*{1cm}
  n \in {\rm integer}
 %$ 
\end{equation}
where the most relevant in astrometry is $\phi_{A, \rm pos} = 2\pi \overrightarrow{b} . (\hat{s}_{\rm true}-\hat{s}_{\rm model})\,/\,\lambda$ arises from inadequacies in the a-priori knowledge of the source position, with $\hat{s}_{\rm true}$ the unit vector in the actual direction of the source  and $\hat{s}_{\rm model}$ is the a-priori position used in the correlation,
$\overrightarrow{b}$ is the baseline vector and $\lambda$ is the wavelength of the observing frequency $\nu$.
$\phi_{A, \rm str}$ stands for the contribution from the structure in extended sources. 
%The difference between the true and model unit vectors $\hat{s}_{\rm true}-\hat{s}_{\rm model}$ is $\delta\theta$.
These terms represent the intrinsic properties of the source of interest to astrometry and are
mixed with other extrinsic or contaminating contributions such as: $\phi_{A, \rm geo}$ arising from model errors in the geometry of the array including the orientation of the Earth among others, $\phi_{A, \rm tro}$ and $\phi_{A, \rm ion}$ arise from mismodelling of propagation medium effects as the cosmic signal transverses the Earth's atmosphere, due to unpredictable refractivity variations in the troposphere and ionosphere along the line of sight, respectively. 
$\phi_{A, \rm inst}$ is due to the unaccounted instrumental instabilities.
The purpose of astrometric calibration is the identification and elimination of these extrinsic terms.
$\phi_{A, \rm thermal}$ stands for the thermal noise error or measurement error and is related to the sensitivity of the instrument, setting the ultimate potential astrometric limit, with a standard deviation given by $\sigma\phi_{A, \rm  thermal} \sim{1 \over SNR}$ in radians, %{1 \over 2 \pi \nu} 
%($\sigma (\phi) = 1 / SNR$; $\sigma _{\tau_{phi}} = (1/ 2 \pi \nu) x (1/ SNR)$) 
where SNR is the signal-to-noise ratio of the fringe detection. 
Finally, $2\pi n$ stands for an unknown integer number $n$ of $2 \pi$ phase cycles that represents the inherent ambiguous nature of the measured phases; $n$ can vary between observations of the same source at different times, and between different sources.
It sets the greatest challenge for astrometry using the phase observable.

The group delay observable is determined from the phase slope over the discrete frequencies in the spanned observed bandwidth.
A similar qualitative description as shown in Eq.~\eqref{eq:basis}  applies to the residual group delay values, but without the $2\pi$ ambiguity term, which makes it more usable for absolute position measurements. 
The group delay is the observable used for VLBI geodesy and absolute astrometry.
The phase delay observable is significantly more precise than the group delay, 
by a factor of ${\nu \over \Delta \nu_{\rm eff}}$, where $\nu$ stands for the observing frequency, $\Delta \nu_{\rm eff}$ is related to the { effective spanned (synthesised) bandwidth
\citep[$\Delta\nu_{rms}$ in][Eq. A12.32]{TMSv3}, and is the order of
30 for $\nu=8$\,GHz and $\Delta \nu_{\rm eff}$ = 0.3 GHz.} 

Using the extremely precise phase observable has the potential to reach \uas\ level astrometry;
%providing the other contaminating contributions are accounted for, to the same level, (either by monitoring them separately or by estimating them as variables in the astrometric analysis); 
%
on the other hand the inherent ambiguous nature of the measured phases (i.e., only known to modulo 2$\pi$) 
%results in an unknown integer number of cycles in the measured phases (where n can vary for observations at different times for the same source, and for different sources, 
%which
complicates its direct interpretation. To resolve this issue requires a sufficiently accurate a-priori knowledge of the value of all the contributions in Eq.~\eqref{eq:basis} to a small fraction of the phase ambiguity spacings, which corresponds to the length of the observing wavelength.
%with uncertainties much smaller than $2\pi$ to the observed phases to a level of a fraction of the observing wavelength, e.g., 5 mm or 20 ps at 8 GHz (change) 
This is the so called ``phase connection'' process, which is complicated for the long baselines of VLBI. 
%and  this applies only to special cases (see below). 

An approach to overcome this issue is to use a differential or relative analysis.
%to carry out
For example, the differential observables of near-contemporaneous observations of two nearby sources on the sky, a target (T) and a reference (R), reduces the sensitivity to the mismodelling of common contributions.
Such an observing schedule consists in general of interleaving scans between a pair of sources for which the angular separation or switching angle and the source switching time are significantly smaller than the isoplanatic patch size and the atmospheric coherence time, respectively, at the observing frequency. The coherence time is conventionally defined as the time for the phase change to be a radian and can be very short at high frequencies.

Following Eq.~\eqref{eq:basis} the residual differential phase observable is given by:
%
%$ \phi_T - \mathcal{R} \times \phi_R= \phi_{1} - \mathcal{R} \times \phi_{2} = \phi_{T,str} + (\phi_{T,pos} - \mathcal{R} \times \phi_{R,pos}) + \sigma_{\phi} + \sigma_{\rm thermal}  + 2\pi n^\prime $
\begin{equation*}
  \begin{split}
    \phi_T (t, \nu) - \phi_R (t', \nu) = (\phi_{T, \rm pos} - \phi_{R, \rm pos}) +  (\phi_{T, \rm str} - \phi_{R, \rm str})\\
    + (\phi_{T, \rm geo} - \phi_{R,\rm geo} + \phi_{T, \rm tro} - \phi_{R, \rm tro} + \phi_{T, \rm ion} - \phi_{R,\rm ion} + \phi_{T, \rm inst} - \phi_{R, \rm inst})\\
        \pm \sigma\phi_{\rm thermal} +  2\pi (n_T - n_R) \hspace*{1cm} n_T,n_R \in {\rm integer}
\label{eq:terms_diff_conv}
  \end{split}
\end{equation*} 
where $\phi_{T, \rm geo} - \phi_{R, \rm geo}\sim0; \phi_{T, \rm tro} - \phi_{R, \rm tro}\sim0; \phi_{T, \rm ion} - \phi_{R, \rm ion}\sim0; \phi_{T, \rm inst} - \phi_{R, \rm inst}\sim0$, when the conditions described above are fulfilled.
%{ as long as the cycle time  ($t-t'$) is sufficiently short, as described above. } %smaller than the atmospheric (and instrumental) variation time and the ionospheric isoplanatic patch.} 
In this case the differential observable is largely free from the contaminating error contributions
%$\phi_{\rm geo}, \phi_{\rm tro}, \phi_{\rm ion}, \phi_{\rm inst} \sim0$ 
and therefore yields an accurate estimate of the relative or differential astrometry between T and R, which is derived from the term $\phi_{T,pos} - \phi_{R,pos}$. 
The source structure terms and the noise terms do not cancel; the former can be estimated using self-calibration techniques and hence are not an astrometric challenge, the latter, $\sigma\phi_{\rm thermal}$, is the combined thermal error term.
To a first order approximation, the magnitude of the residual differential errors for the contaminating contributions are reduced by a so called ``dilution factor'' given by the pair angular separation, or switching angle, expressed in radians; additionally the switching time is relevant.
%; their deviation from zero amount for/results in calibration errors. 
Similar difference equations can be written for both the other observables the delay and the rate, but the phase observable has the potential to provide the most accurate astrometry and it is the focus of this paper. 
Implementations of this concept are the so called differential phase delay and phase reference imaging (or simply phase referencing) astrometric techniques.

The main challenge of differential astrometry is to connect the phases and resolve the ambiguity issue %, that is, to determine the value of $\Delta n$ { (i.e., $n_T - n_R$)} integer, 
for each source, baseline and time. This has been successfully achieved using two approaches:  phase-delay fitting  
\citep[e.g.,][]{wittels_75_phd,shapiro_79,marcaide_84,bartel_86} and phase reference mapping \citep[e.g.,][]{alef_88,lestrade_90}. %{\bf CLARIFY REFERENCES: include here Alef 1989, Lestrate 1991 reference, as appears in Beasley\&Conway 1995}. 
The former uses an iterative process to determine the value of $\Delta n$ between consecutive measurements of the phase for both sources using the individual and differential observables; %; that is ``phase-connection''.
this is followed by a least-squares fit of the unambiguous differential phase delays to 
measure the relative source separation \citep{robertson_75}. % (Robertson 1975).
The latter does not explicitly determine the $\Delta n$ integer for the target source, but instead uses a Fourier transform of the differential observable to form an image, 
where the shift in the position of the peak intensity of the source with respect to the centre of the image directly conveys the relative astrometric information in $\phi_{T,pos} - \phi_{R,pos}$ or $\delta\theta$.
{ 
In general, the residual rates are used for the phase connection between consecutive  measurements of the reference source and to interpolate the value to the time of the target source.}  
%When the two sources are not observed simultaneously the residual rates are used for the phase connection between consecutive  measurements of the reference source and to interpolate the value to the time of the target source. 
%
% the phase connection between consecutive measurements of the reference source using the rates are interpolated to the time of the target source observations. This ensures that the differential observable does not have jumps in the phase.
%
The approaches of phase-delay fitting or phase-referenced mapping are equivalent, albeit with some differences as described in the next section; 
they have been evolving since first proposed but without changing their essence. 
In this paper we refer to both methods as conventional phase referencing or simply PR.

In the last 45+ years there have been steady incremental improvements of phase-based relative astrometry.
Important steps forward have come from: i) improved instrumental stability and sensitivity,  ii) longer baselines, iii) improved a-priori models applied to both target and reference data, vi) advanced calibration strategies to mitigate the propagation of ``contaminating'' residual errors into astrometric errors at different frequency regimes, and vii) improved astrometric VLBI calibrator catalogues. 

The low frequency regime has lagged behind, but have had a renaissance with the arrival of the next-generation instruments such as the SKA.
We foresee a new era for astrometry with the combined power from the next-generation instruments and breakthrough technologies planned and under construction, and new calibration algorithms. 

We highlight that innovative recent calibration developments have come from extending the concept of a `reference' beyond observations of another source, as in established methods. %; this applies to the differential phase observable too.
Therefore we use a generalised expression for the differential phase observable, as a linear combination of the target and reference observables, where the latter can comprise observations of the same target source or another, or multiple other sources, at the same frequency as the target ($\nu_T$) or at a different frequency ($\nu_R$), or a combination of these strategies:
\begin{equation}
%$
%  \phi_T - \mathcal{R} \times \phi_R= \phi_{1} - \mathcal{R} \times \phi_{2} = \phi_{T,str} + (\phi_{T, \rm pos} - \mathcal{R} \times \phi_{R, \rm pos}) + \sigma_{\phi} + \sigma_{\rm thermal}  + 2\pi n^\prime 
    \phi_T (t, \nu_T) - \mathcal{R} \times \phi_R (t^\prime, \nu_R) = (\phi_{T, \rm pos} - \mathcal{R} \times \phi_{R, \rm pos}) \pm \sigma\phi^{\rm cal} \pm \sigma\phi_{\rm thermal}  + 2\pi (n_T - \mathcal{R} n_R) %,  {\sc with n_T, n_R integer, R real} %{\sc with R real,  n^\prime integer} n does not have to be interger if R is not \phi_{T,str} + 
\label{eq:terms_diff}\end{equation} 
where  $(\phi_{T, \rm pos} - \mathcal{R} \times \phi_{R, \rm pos})$ retains the relative astrometry information of interest,
%(between the reference and the target), 
with $\mathcal{R}$ a real linear scale factor for the reference phases. { $\mathcal{R}$ is method dependent, for example it is equal to 1 for PR, or to the frequency ratio for frequency phase transfer.}
%or ${\nu^{high} \over \nu^{\rm low}}$ in source frequency phase referencing, 
$\sigma\phi^{\rm cal}$ stands for the combined residual differential phase errors from mismodelling of the contaminating contributions (\textit{hereafter} calibration errors) for a given calibration method. % and $n_T$ and $n_R$ are integer.
%, hence called calibration errors; 
The other terms are as before.
%
%Residual systematic errors from atmospheric mismodeling are the dominant source of calibration errors and limit the accuracy of the astrometric measurements (i.e., one is in the systematic error limited regime);
Residual contributions from atmospheric mismodelling are the dominant source of calibration errors and limit the accuracy of the astrometric measurements (i.e., in general one is in the systematic error limited regime);
%
%Inadequacies in the modelling of the propagation medium effects are usually a dominant source of calibration errors and limit the achieved accuracy of the astrometric measurements (i.e., one is in the systematic errors limited regime); 
their magnitude is quantified next.
Errors in the a-priori geometric parameters, such as in telescope and reference source coordinates, Earth orientation parameters etc.,  usually have a lesser effect, thanks to improvements from VLBI geodetic observations;
all of these are usually larger than the instrumental contribution, due to modern stable electronic systems and maser clocks. 
In an ideal calibration scenario $\sigma\phi^{cal} \sim 0$ and $\sigma\phi_{\rm thermal}$
sets the ultimate limit in the astrometric accuracy (i.e., one is in the thermal noise limited regime). 
The structure terms $\phi_{T, \rm str}$ and  $\phi_{R, \rm str}$ can be estimated from the images and are excluded from Eq.~\eqref{eq:terms_diff} to simplify the presentation.
Nevertheless, see Sect.~\ref{sec:ref_pts} on the stability of reference points for details on some considerations.

%{{\bf Formulae to quantify PM residual relative calibration phase errors:} \\}

%The residual propagation effects from the radio signals through the Earth's atmosphere are in general a dominant contribution to the calibration errors.
\citet{a07} presented semi-analytic expressions to estimate the magnitude of residual propagation effects for conventional PR from unaccounted refractivity changes over time and space (dynamic and static, respectively) in the troposphere and the ionosphere; we find the latter the most complete description of the various similar investigations \citep[e.g.,][]{pradel_06,honma_08_trop}.
These are the dominant contributions to $\sigma\phi^{cal}$.
%Eq.~\eqref{eq:a07_terms} is a modified version, adjusted to 8GHz and for ground baselines.
We will use a modified version of those to estimate the calibration errors using different astrometric techniques presented in the next sections.
The errors comprise of dynamic  (or rapidly varying)  components (hereafter $\sigma\phi^{\rm cal}_{\rm dyn,tro}$ and $\sigma\phi^{\rm cal}_{\rm dyn,ion}$) arising from short term fluctuations in the distribution of water vapour (the ``wet'' component) in the troposphere and of the total electron content (TEC, measured in TEC Units where 1 TECU = $10^{16} {\rm el /m^2}$) in the ionosphere, along the line of sight of the observations, respectively. These introduce random fluctuations with zero mean. 
The static (or slowly varying) components  (hereafter $\sigma\phi^{\rm cal}_{\rm sta, tro}$  and $\sigma\phi^{\rm cal}_{\rm sta,ion}$) arise from relatively stable contributions predominately from the hydrostatic (``dry'') components in the troposphere, and the TEC in the ionosphere, respectively. % \sout{and the ``wet''}
Eq.~\eqref{eq:a07_terms} is a modified version of those in \citet{a07}, adjusted to 8\,GHz and for { ground to ground baselines  (i.e., scaling by 5.4 and multiplying by $\sqrt{2}$, respectively)}, in units of degrees:

\begin{equation}\label{eq:a07_terms}
  \begin{split}
\sigma\phi^{\rm cal}_{\rm sta, tro}(\nu) &= 20.0 (\nu/8{\rm GHz}) (\Delta \ell_z/3{\rm cm}) {(\Delta\theta_{\rm tro} / 2\degr)}\\ 
%20.0 {(\nu_R / 8{\rm GHz})} {(\Delta\ell / 3{\rm cm})} \times \\
& \times {(\sec(Z_g) / \sec(45\degr))} {(\tan(Z_g) / \tan(45\degr))} \\
\sigma\phi^{\rm cal}_{\rm sta, ion}(\nu) &= 20.5 (\nu/8{\rm GHz})^{-1} (\Delta{\rm I}/6{\rm TECU}) (\Delta\theta_{\rm ion}/2\degr)\\ 
& \times (\sec(Z_F)/\sec(41\degr))(\tan(Z_F)/\tan(41\degr))\\
% \sigma\phi^{\nu}_{\rm geo}{\rm [deg]} &=  4.8 {(\nu_R / 8{\rm GHz})} {(\Delta P / 1{\rm cm})} 
\sigma\phi^{\rm cal}_{\rm dyn, tro}(\nu) &= 7.1 \, C_w (\nu/8{\rm GHz})(\sec(Z_g)/\sec(45\degr))^{1/2}\\ 
& \times [T_{\rm swt,tro}/60s+0.16(\sec(Z_g)/\sec(45\degr))(\Delta\theta_{\rm tro}/2\degr)]^{5/6}\\
\sigma\phi^{\rm cal}_{\rm dyn, ion}(\nu) &= 3.5 (\nu/8 {\rm GHz})^{-1} (\sec(Z_i)/\sec(43\degr))^{1/2}\\ 
& \times [0.21(T_{\rm swt,ion}/60s)+(\sec(Z_i)/\sec(43\degr))(\Delta\theta_{\rm ion}/2\degr)]^{5/6}
  \end{split}
\end{equation}
%
%{\it \color{purple}
where $\nu$ is the observing frequency in GHz,
%$\nu_T, \nu_R$ are the target and the reference observing frequencies in GHz. 
$Z_g$ is the mean zenith angle for the two sources in degrees at the ground level for a terrestrial telescope,
%$Z_i$ and $Z_F$ are the zenith angles at the ionospheric pierce point for at the altitude of the ionospheric turbulent screen (bottom of the F-layer, $\sim$300\,km) and from the ionospheric pierce point at the altitude of the electron density peak (peak of the F-layer, $\sim$450\,km), 
$Z_i$ and $Z_F$ are the zenith angles at the ionosphere piercing points, for the bottom ($\sim$300\,km) and the peak ($\sim$450\,km) of the F-layer, respectively. 
$\Delta\ell_z$ is the residual tropospheric zenith excess path length in cm and $\Delta{\rm I}$ is the residual total electron content (TEC) in TECU in the zenith direction at the ionosphere piercing point.
$\Delta{\rm I}$ can be expressed as a frequency dependent excess path length, in cm, as $40.3\Delta{\rm I}\,\nu^{-2}$ with $\Delta{\rm I}$ in TECU and $\nu$ in GHz.
$C_w$ is a unitless factor to characterise tropospheric conditions (with values of 1, 2 or 4 for good, normal and poor).
 %, ZF is the zenith angle at the altitude of the electron density peak
Some parameters are specific to the calibration strategy: 
%\sout{$C_1, C_2$ are multiplicative factors. }
$\Delta\theta_{tro}$ and $\Delta\theta_{\rm ion}$ are the angular separations, between T and R, relevant for the mitigation of the spatial structure of the tropospheric and 
ionospheric disturbances, respectively, in degrees; 
$T_{\rm swt,tro}$ and $T_{\rm swt,ion}$ are the equivalent in the temporal domain, corresponding to the switching time between interleaving observations of T and R, respectively, in seconds. 

For example, typical values for $\Delta\ell_z$ and $\Delta{\rm I}$, after calibration solely using ground based measurements for water vapour and GPS-derived TEC corrections, are 3--5\,cm and 6--10\,TECU, respectively. 

Inspection of Eq.~\eqref{eq:a07_terms} leads to several immediate conclusions for the contributions to calibration errors in conventional PR: 
\begin{itemize}
    \item That the tropospheric residual errors scale with frequency and the ionospheric contributions scale with wavelength.
    \item That the balance between tropospheric and ionospheric calibration errors falls around 8\,GHz
    \item That the smaller the angular separation $(\Delta \theta$) between  T and R the smaller are all the phase errors. 
    \item That the faster the switching time ($T_{\rm swt}$) between  T and R the smaller the dynamic errors.
    \item That the static terms, all other contributions being nominal, are the most significant for astrometry.
    \item That the static tropospheric phase errors scale linearly with the residual path length ($\Delta\ell_z$).
    \item That the static ionospheric phase errors scale linearly with the residual TEC ($\Delta {\rm I}$).
    \item That the  errors grow very large at low elevations, particularly for the troposphere.\footnote{Compare Figs.~13.6 and 14.3 in \citet{TMSv3} for the difference between the dependence of the troposphere and the ionosphere.}
    \item That, for the tropospheric dynamical terms, the errors related to switching time dominate those related to the angular separation.
    \item That, for the ionospheric dynamical terms, the errors related to the angular separation dominate those related to the switching time.
\end{itemize}

%As we are focusing on the dominant contributions to calibration errors, e
Expressions for the calibration errors arising from the geometrical model are presented in \citet{a07, shapiro_79, reid_micro} and other publications and are functionally similar to the static troposphere, but of a smaller scale.
The consequences of errors in the reference source position are discussed, for phase reference mapping, in \citet[Sec. 5.4]{reid_micro}; to avoid the second-order effects the absolute position of the reference source should be known to better than 10\,mas.

%{{\bf Quantify relative astrometric errors:} \\}

Finally we discuss the impact of the calibration errors onto the accuracy of the astrometric measurements or astrometric errors. %, $\sigma\Delta\theta_{\rm AB}$.
It is intuitively easy to understand that the propagation onto astrometric errors ($\sigma\Delta\theta^{\rm cal}$) will vary depending on the nature of the calibration errors, as well as the array geometry. 
For example, the dynamic components introduce random fluctuations with zero mean and average away rapidly. 
Therefore, if they are not large enough to prevent phase connection, these do not introduce systematic bias in the astrometric measurements. 
The static components, however, introduce long-term phase gradients above a telescope,
which will not average out over the experiment and would propagate into a position error in the image (without loss in the peak flux, if coherent across the array); 
therefore they are the most important astrometric error contribution. 
Both components would lead to degraded image quality, with the fractional flux ratio (FFR) being $e^{-\sigma\Phi^2/2}$ \citep{TMSv3}, where $\sigma\Phi$ is the standard deviation of the phase errors.

Here, we follow a simple approach to estimate the systematic astrometric error for a single epoch of observations:
$\sigma\Delta\theta^{\rm cal} \approx {{\sigma\phi^{\rm cal}_{\rm sta}[\rm deg]} \over {{360\degr}}} {\theta_{\rm beam}}$,
or equivalently $\approx {\Delta \ell \over |b|} \sec{Z}\tan{Z} \Delta \theta$,
where $|b|$ is the maximum baseline length, %$\Delta \theta$ is the angular separation between sources, 
$\theta_{\rm beam}$ is the synthesised beam size, $Z$ is the mean Zenith angle, and $\sigma\phi^{\rm cal}_{\rm sta}$ and $\Delta \ell$ correspond  to  
adding in quadrature the residual error terms arising from static tropospheric and ionospheric disturbances for the phase and excess pathlength, respectively. This simple expression assumes that $\Delta\theta$ is common for all contributions;
in PR this corresponds to the angular separation between the target and the calibrator sources. 
Smaller contributions from geometric errors and thermal noise ($\sigma\Delta\theta_{\rm thermal}$; see next section) will also contribute.
The simple approximation used here is acceptable for observations with moderate zenith angles with most arrays. 
A complete characterisation of the astrometric error propagation requires detailed analysis using simulation studies, such as carried out in \citet{pradel_06,a07} or \citet{honma_08_trop}, to correctly account for the impact of the array configuration and source declination, among other considerations. 

In Sects.~\ref{sec:methods_old} and \ref{sec:methods_new}  we review the differential astrometric methods, with emphasis on the relevant aspects to reduce the dominant errors and scope of application in each case.
We use Eq.~\eqref{eq:a07_terms} to estimate the phase errors and the astrometric errors as described above to characterise and compare 
the expected performance of the different methods.

\subsection{Definition and stability of fiducial reference points}\label{sec:ref_pts}

Astrometry involves the identification of reference points within the source images to which the measurements are referred.
Obviously a fixed compact calibrator source 
%{ recognizable and stable over time reference point in the calibrator source,} 
provides a solid fiducial point for tracking changes of the target position and facilitates the interpretation in multi-epoch observations;  the radio core components of 
AGNs (QSOs) at high red shifts are good candidates to provide such points in the sky, as suggested by the standard theory of extragalactic radio sources \citep{bk_79} %(Blandford \& Konigl, ??)
and supported by measurements of the stringent upper limits on the stability of AGN cores of a few \uasyr. 
%\ \citep{bartel_86}, and of 10\uasyr\ \citep{rioja_00}. % (between center of mass of two quasars 1038+528AB). 

The temporal stability and uncertainties in the definition of the reference  points (within the target \emph{and} calibrator source images)
are expected to become the next dominant source of astrometric errors, once tropospheric and ionospheric effects are precisely calibrated out, particularly in observations at low frequencies.
%
%Their magnitude is difficult to estimate, but one can intuitively guess that it depends on the properties of the unresolved emission near the reference point, and the so called thermal noise error, respectively.
The position measurement uncertainty in an image due to thermal noise is given by
%The latter represents the instrumental limit based solely on the finite SNR in the maps of both sources; the main contribution is therefore from the weaker source of the pair.
$\sigma\Delta\theta_{\rm thermal}\approx\theta_{\rm beam}$/(1.2 DR) where DR is the Dynamic Range: the ratio of the flux density at the position of the reference point to the Root Mean Squared (RMS) noise in the maps \citep{condon_98}. %; see also \citet{reid_88}
% Reid 88 is \theta_{\rm beam}$/(2 DR) 
It represents the lower bound of the astrometric error; higher resolution (from observations at higher frequencies and with longer baselines) and higher DR (from higher source flux and higher array sensitivity) result in smaller thermal errors. 
Larger errors are expected to be related to the effects from underlying source structure and variability, which can introduce temporal changes in the physical conditions at the base of the jet 
resulting in intrinsic changes in the position of the core, defined as the region where the optical depth is equal to unity; this effect is known as core-jitter.
Moreover, even in the case of a true stationary core, apparent position shifts can arise, for example, from structural changes on scales below the interferometer resolution near the reference point caused, for example, by the ejection of new components at the base of the jet; these are called structure-blending effects. 
The magnitude of the apparent shifts can depend on the resolution and is expected to change between multi-epoch observations, following the direction of the source axis. 
Related structure-blending effects must be considered in any comparison between observations at very different resolutions of sources with asymmetrical source structure, even in the absence of any structural changes. We call these extrinsic effects, to differentiate from intrinsic position changes.
Observations at higher frequencies and with higher angular resolutions are less vulnerable to this effect; also, these can indicate the onset of the occurrence of this at lower frequencies, where the effects will be larger. 
In all cases, using multiple background calibrator sources can help identify  unsuitable calibrators.

Examples of detailed studies on the stability of the reference points in AGNs
%level of stability of radio core positions with focus on the determination of the reference points
%limits imposed by AGN source structure and variability 
are presented in \citet{rioja_00}, with upper limits to secular trends (``core motion'') along the source axis for quasars 1038+528 A and B 
$\le$5\uas\, yr$^{-1}$ and deviations (i.e., core jitter) $<$20\uas, from 4 epochs of observations spanning 14 years, at 8 GHz. 
Similar limits were found by \citet{bartel_86} between 3C345 and NRAO0512 from 12 epochs over nine years and
\citet{fomalont_11} in the relative positions between multiple pairs of quasars in observations spanning one year, at 43 and 23 GHz. 

Image quality is important for astrometry and observations should be designed to optimise both source structure and position information.
While the improved quality in the measured brightness distributions of sources with next-generation instruments is expected to partially alleviate this effect, 
careful strategies to mitigate this source of systematic error are mandatory, particularly for high DR observations, %, at lower frequencies with ground arrays, 
i.e., VLBI with SKA;
for example, incorporating extra calibrators to allow for cross-checks and
%investigating weak sources as indicators of less structure effects, 
considerations of the uv-coverage to recover the source structure.
The requirements on the precision of the amplitude calibration required to achieve a DR of 1000:1 are under study, but will undoubtedly have a significant effect at these levels of sensitivity. 
%{\bf concluding sentence: no}

\subsection{Absolute radio astrometry catalogues and weak reference source searches}\label{sec:cals}
% - read Petrov 2019 for more context
For differential astrometry applications a calibrator list with accurate positions of extragalactic radio sources is fundamental and should comprise
of a dense grid of sources for all directions and at all frequencies. 
%One of the prime requirements for astrometry is an initial list of calibrator sources with accurate positions. These accurate source positions are vital, as the absolute astrometric error contributes to the total error, scaled by the dilution factor. 
%
% \subsection{Absolute Astrometry}\label{sec:applications_icrf}
% { \bf move to calibrators?\\}
The most comprehensive catalogue available to date is the third realisation of the International Celestial Reference Frame, ICRF3 \citep{charlot_icrf3}, adopted by the IAU in August 2018. 
%, and since then serving as the fundamental reference frame for all astrometric applications.
ICRF3 is the first multi-frequency reference frame ever realised, comprising positions for 4536 sources at S/X band (2.3/8.4 GHz), 824 sources at K band (24 GHz) and 678 sources at X/Ka band (8.4/32 GHz), where 600 sources have positions available at all three datasets.
%\sout{A major feature of ICRF3 aside from its multi-frequency nature is the consideration of Galactocentric acceleration in its realization, implying that source positions now depend on epoch, unlike in previous realizations. In this respect, all positions away from the adopted reference epoch of 2015.0 must be derived by propagating the ICRF3 positions, accounting for a Galactocentric acceleration of 5.8\uasyr.}
These positions have been estimated independently at each of the frequencies in order to preserve the underlying astrophysical content behind such positions.
The ICRF3 frame shows median positional errors of the order of 100\,\uas{} in right ascension and 200\,\uas{} in declination, with a noise floor of 30\,\uas{} in the individual source coordinates.
%To compensate for the the sparse distribution of ICRF sources on the sky, 
%Efforts to produce a more dense grid of calibrator sources have
%The VLBA has driven the effort to 
%For relative astrometry we require a greater source density than that of the ICRF. 
Since 1994 the VLBA has carried out  a sequence of calibrator surveys (VCS-1 to 10) that have provided a more dense grid of calibrator sources at declinations $>-45\degr$
%For this purpose since 1994 the VLBA (VCS-1 to 10) have performed calibrator surveys for declination $>$-45\degr, 
(e.g., \citealt{petrov_08} and references therein; \citealt{petrov_16}). 
More recently the LBA has performed calibrator surveys for the Southern hemisphere (LCS-1 and 2) \citep[and references therein]{petrov_19_lba}. % respectively for the latest published results. % 
%
%The work of L. Petrov and the Astrogeo Center\footnote{http://astrogeo.org} continues to improve the calibrator listings for cm-VLBI; for example the 9$^{\rm th}$ VLBA Calibrator Survey (VCS-9) \citep{petrov_16} and 2$^{\rm nd}$ LBA Calibrator Survey have been recently released \citep{petrov_19_lba}.
These surveys increase the number of known calibrators to 16,466, which can be compared against the number in the ICRF3 catalogue. Note that the ICRF sources have significantly more accurate positions, as they have been regularly monitored for many years; errors in the coordinates of reference sources propagate into astrometric errors. 
%We note the suggestions in \citet{petrov_19_lba} on building the initial source listings, to ensure a high detection success rate. Along with the traditional flat spectral index selection criteria they suggest the use of IR colours, which is particularly useful near the Galactic plane where cross identification in surveys at different frequencies  can be difficult.

At the higher frequencies, a Korean VLBI Network (KVN) Legacy program, the Multi-frequency AGN Survey on KVN (MASK), will greatly increase the number of known VLBI calibrators at the highest frequencies.
The only previous large high-frequency VLBI surveys were \citet{sslee_86} and \citet{nair_19}, which give a total of 162 detections on VLBI baselines at 86GHz.
%Neither are astrometric. 
MASK uses the calibration solutions at the lowest frequencies to extend the coherence-time at the highest frequencies. 
Data release of the MASK results are expected shortly, and over 600 sources are currently detected at 86\,GHz and about half of that number at 130\,GHz.

Having a very close calibrator has great benefits but the probability for finding a known reference less than a degree away from the target is still low. 
Calibrators that can be observed simultaneously with the target are known as `in-beam' calibrators and
the benefits from these can be sufficient that it is worthwhile searching for a previously unknown VLBI source. 
There are calibrator search strategies for when a suitable catalogued calibrator sufficiently close can not be found. 
%The default in-beam PR calibrator search strategy is simple.
An effective one is to observe in the direction of the target source but cross-correlate the data for the positions of all suitable compact, flat or steep spectrum sources that fall within the FoV of the telescope.  
In the past this required multiple passes through the correlator, changing the coordinates of the individual pointing centres, thus had a significant impact on the processing time and no more than a few pointings would ever be provided.
The breakthrough came with the software correlator DiFX-2 \citep{difx}, which could form multiple phase centres at the first averaging stage on the intermediate data products. This massively reduced the computational cost, and DiFX-2 has been used to correlate hundreds of individual phase centres from a single experiment 
(e.g., 556 phase centres in the Orion Star Forming Region, \citet[submitted]{forbrich_20}).
%,forbrich_20}. % orion had more
%\citep[e.g., 556 phase centres in the Orion Star Forming Region, Forbrich 2019 in prep][]{forbrich_19}. %, middelberg_11}. %,forbrich_16}. % orion had more
%
Recently LOw Frequency ARray (LOFAR) long-baseline studies have proposed strategies to find suitable compact calibrators for hither-to unexplored low frequency ranges \citep[see][]{moldon_15,lofar_lbs}. They find the source density to be about 1 suitable source per sq. deg. at 150\,MHz.

\section{Established precise astrometric calibration methods}
\label{sec:methods_old}

\begin{table}[h]
  \caption{Indicative temporal sequence of source/frequency scans in observing schedules, for implementation of some of the astrometric techniques discussed in Sects.~\ref{sec:methods_old} and \ref{sec:methods_new}. 
Conventional phase referencing (PR) with source switching, in-beam PR with simultaneous source observations, Advanced Tropospheric Calibration with Geodetic blocks (ATC), \MV{} (here with source switching but could also be simultaneous), SFPR with two sources (here with simultaneous multi-frequency observations but could also be with fast frequency switching) and MFPR, here with simultaneous multi-frequency observations and ICE-blocks, using only the target source.
`$P,\,{\color{red}R},\, {\color{green}T}$' are the scans on the primary calibrator, reference and target sources, respectively; `$G$' stands for the GeoBlocks. `--' represents  frequency switching, `$\cdots$' represents source switching and one over the other represents simultaneous source/frequency observations. Multiple calibrators combined in the analysis with different weights is represented by $\Sigma_i \alpha_i$, where $\alpha_i$ are real values. 
{ Different frequencies $\nu$ are indicated by the subscripts $R$ and $T$ for reference and target frequencies respectively, and $i,j$ and $k$ for the ICE-block calibration.}}
\label{tab:sched}
    \centering
\begin{tabular}{|l|c|}
\hline
% conventional%
PR & $P\cdots{\color{red}R}\cdots{\color{green}T}\cdots{\color{red}R}\cdots{\color{green}T}\cdots{\color{red}R}\cdots{\color{green}T}\cdots{\color{red}R}\cdots$\\
\hline
%in-beam
In-beam PR&$P\cdots{{\color{red}R}\over{\color{green}T}} \phantom{0000} 
{{\color{red}R}\over{\color{green}T}} \phantom{0000} 
{{\color{red}R}\over{\color{green}T}} \phantom{0000}
{{\color{red}R}\over{\color{green}T}} \phantom{0000}
{{\color{red}R}\over{\color{green}T}} \phantom{0000} {{\color{red}R}\over{\color{green}T}} \cdots$\\
\hline
% conventional%
ATC & $P\cdots{\color{red}R}\cdots{\color{green}T}\cdots{\color{red}R}\cdots{\color{black}G}\cdots{\color{red}R}\cdots{\color{green}T}\cdots{\color{red}R}\cdots$\\
\hline
% MV 
\MV &$P\cdots{\color{red}\Sigma_i \alpha_i R_i}\cdots{\color{green}T}\cdots{\color{red}\Sigma_i \alpha_i R_i}\cdots{\color{green}T}\cdots{\color{red}\Sigma_i \alpha_i R_i} \cdots$\\
\hline
% SFPR 
%SFPR&$P\cdots{T({\color{red}\nu_R}}\cdots{\color{green}\nu_T})\cdots{R({\color{red}\nu_R}}\cdots{\color{green}\nu_T})\cdots{T({\color{red}\nu_R}}\cdots{\color{green}\nu_T}) \cdots$\\
SFPR&$P\cdots{T({{\color{red}\nu_R}\over{\color{green}\nu_T}})} \cdots{R({{\color{red}\nu_R}\over{\color{green}\nu_T}})} \cdots{T({{\color{red}\nu_R}\over{\color{green}\nu_T}})} 
\cdots{R({{\color{red}\nu_R}\over{\color{green}\nu_T}})} \cdots$\\
\hline
MFPR&$P\cdots{T({{\color{red}\nu_R}\over{\color{green}\nu_T}})} -{T({\color{red}\nu_i}-{\color{red}\nu_j}-{\color{red}\nu_k})} -{T({{\color{red}\nu_R}\over{\color{green}\nu_T}})}\cdots$\\
\hline
\end{tabular}
\end{table}

\subsection{Conventional PR astrometry}
%/{\bf Conventional Relative Astrometry - 2 sources, T, R at same frequency}}
\label{sec:convPR}

The standard approach for phase-based astrometric measurements of a program source (A) 
%consists of using interleaving observations, with switching time $T_{\rm swt}$, of a nearby calibrator source (R), which is separated by  $\Delta\theta_{AB}$ of up to a few degrees, at the same observing frequency, $\nu$. 
consists of using interleaving observations of a nearby calibrator source (B), which is separated up to a few degrees, at the same observing frequency, $\nu$. 
%A differential analysis is used to retain the accurate astrometric signature, whilst removing the bulk of contaminating residual contributions. 
%
This approach has been used since the beginnings of phase-based differential or relative astrometry in the 1970s; 
here we refer to it as conventional relative astrometry (or conventional PR, or simply PR). 
See Table~\ref{tab:sched} for an indicative observing schedule.

Following Eq.~\eqref{eq:terms_diff} for the differential phase observables, with $\mathcal{R}$=1 and $\nu_R = \nu_T = \nu$, the PR-calibrated dataset is given by:
%formed by applying the calibration from the reference to the target source, 
\begin{equation*}
  \phi_T -  \mathcal{R} \phi_R  = \phi_{A}(t_1, \nu) -\phi_B (t_2, \nu) =  (\phi_{A, \rm pos} - \phi_{B,pos}) \pm \sigma\phi^{PR} \pm \sigma\phi_{\rm thermal} + 2\pi \Delta n \hspace{1cm} \Delta n \in {\rm integer}
\end{equation*}
where the indexes A and B refer to the two sources; $(\phi_{A, \rm pos} - \phi_{B,pos})$ is the astrometric term of interest, for a precise measurement 
%retains the astrometric information on 
of the angular separation between both sources at the observing frequency $\nu$ and $\sigma\phi^{PR}$ stands for the conventional PR calibration errors.
%using PR (residual relative phase error term).
%After which the data can be Fourier transformed and deconvolved to provide the image, in which case we measure $\delta\theta$ as difference in the astrometric position from the assumed separation. Normally one assumes that the position of the reference is known, so the measurement is just the astrometric offset for the target. 

Note that, in general, the observations of A and B are carried out at different times and along different lines of sight. 
The source switching interval ($T_{\rm swt}$) and pair switching angle ($\Delta \theta_{AB}$) are crucial parameters to determine 
the quality of the compensation of the temporal and spatial structure of the residual errors in the term $\sigma\phi^{PR}$, 
respectively. %(i.e., residual calibration errors) 
%(, and hence the astrometric error).
A careful planning of the observations is the first step towards improving the astrometric accuracy, by prioritising the compensation of the dominant errors at the observing frequency: fast tropospheric fluctuations in the high frequency regime (i.e., with short $T_{\rm swt}$) and ionospheric spatial gradients above the antennas in the low frequency regime (i.e., with small $\Delta \theta_{AB}$). In practise, the best outcome comes from 
%using short $T_{\rm swt}$ and small $\Delta \theta_{AB}$, where possible (OR 
taking both considerations into account.

The magnitude of the dominant contributions to $\sigma\phi^{PR}$ are given by Eq.~\eqref{eq:a07_terms} using
$T_{\rm swt,tro}=T_{\rm swt,ion}=T_{\rm swt}$ and $\Delta \theta_{tro}=\Delta \theta_{\rm ion}=\Delta \theta_{AB}$.
The resultant astrometric error  $\sigma\Delta\theta_{AB}$ for the high and low frequency regimes,  using the approximation as described in Sect.~\ref{sec:basics}, are given by:
\begin{equation*}\begin{split}
%$
&[\sigma\Delta \theta_{AB}]^2 \sim [{\Delta \ell_z \over |b|}\,\sec{Z_g
}\tan{Z_g} \Delta \theta_{AB}]^2 + [O_{\rm ion}({\Delta \ell_I \over {|b| \nu^2}},\Delta \theta_{AB})]^2 + [O_{\rm geo}(\Delta \theta_{AB})]^2 + [\sigma\Delta\theta_{\rm thermal}]^2;\hspace{.2cm} \nu > 8 {\rm GHz}\\
%{\rm with\ \Delta \ell_z \sim3\,cm}\\
%
&[\sigma\Delta \theta_{AB}]^2 \sim [{\Delta \ell_I \over {|b| \nu^2}}\,\,\sec{Z_F
}\tan{Z_F} \Delta \theta_{AB}]^2 + [O_{\rm tro}({{\Delta \ell_z}\over{|b|}},\Delta \theta_{AB})]^2 + [O_{\rm geo}(\Delta \theta_{AB})]^2 + [\sigma\Delta\theta_{\rm thermal}]^2;\hspace{.1cm}\nu < 8 {\rm GHz}\\
%{\rm with \Delta I\ \sim2.4m\ at\ 1GHz} 
\label{eq:res}
\end{split}\end{equation*}
where %$\Delta \ell_z$ is the tropospheric residual path length at the zenith and 
$\Delta \ell_I$ is the residual ionospheric excess path length at 1\,GHz equal to 40.3$\Delta I$\,cm \citep{sovers_98}, %\sout{Typical errors would be for example, 18\uas/\degr{} at 22GHz and 560\uas/\degr{} at 1.6GHz.}
% The residuals are dominated by the dispersive delay, which typically are the order of 6TECU, or 1\,m at 1.5GHz. 
which corresponds to 2.4\,m for the typical value of $\Delta{\rm I}$=6\,TECU.
Terms in $O()$ represent smaller contributions. 
Then, for example, for $|{b}|$=6,000 km, $Z\sim$40\degr, atmospheric errors of $\Delta\ell_z\sim$5\,cm % $$\Delta \theta_{AB} = 1\degr$,
and $\Delta I$=6\,TECU gives astrometric accuracy 
$\sigma\Delta\theta_{AB} \sim$30\uas{} per degree of pair angular separation (hereafter \uas/deg), for strong sources, at $\nu >$8GHz. 
Note that the (astrometric) accuracy is frequency independent 
in the high frequency regime, being dominated by non-dispersive tropospheric fluctuations.
Instead, in the low frequency regime it is frequency dependent as a consequence of the dominant dispersive ionospheric fluctuations. 
With the parameters as above, $\sigma\Delta\theta_{AB} \sim$63\uas/deg and $800$\uas/deg at 5GHz and 1.4 GHz, respectively. 
Smaller contributions from the thermal noise astrometric errors amount $\le$10\uas{} at $\nu \ge$ 10\,GHz and $\sim$20\uas{} and $\sim$100\uas{} at 5 and 1.4 GHz (L-band), respectively, for DR 100:1.

%{{\bf Scope of application and Estimates of achieved astrometry performance OR astrometric performance:} \\}

The current scope of application of PR is in the moderate frequency regime from $\sim$ 1.4 and up to 43 GHz \citep[with the single unique exception at 86 GHz,][]{porcas_02},
albeit with a rapidly deteriorating performance and usability
towards the edges, particularly in the low frequency regime.
The upper frequency limit is imposed by the increasingly fast tropospheric fluctuations and the 
mechanical limitations of the telescopes to perform source switching within the short coherence times, along with sensitivity considerations and scarcity of calibrators. 
The lower frequency limit arises from the increasingly large spatial ionospheric disturbances, which result in very large calibration errors.

%{{\bf  Solving Phase Ambiguities: PR vs DPDA - this maybe better in intro above? not sure...} \\}

Our definition of PR applies to both phase reference mapping and differential phase delay astrometry techniques. % \sout{in its basic form}, 
%which uses a phase-based differential analysis of observations of a pair of sources A and B to improve the accuracy of the (relative) coordinates of the source A, using either Fourier transform or phase-fitting, respectively.  
\citet{rioja_00} made a detailed comparison of these approaches and showed that the two methods produce equivalent results, albeit there are differences between them. Differential phase delay astrometry requires direct detection of both sources, which are then modelled using a least-squares-fitting procedure 
%\sout{(phase delay and group delay fitting) }
to determine the source coordinates, and possibly also other parameters used in the a-priori models, which in principle allows for larger angular separations. 
In phase-reference mapping it is sufficient to detect one of the two sources, therefore it is also applicable for imaging sources too weak for self-calibration.
The Fourier transform approach determines the source coordinates only, but has the advantage of being very easy to use and therefore has been very widely applied.
%(\sout{\bf say here something about implementation in general use package AIPS? Relevant to scope of application theme...}.
Both methods have evolved since their first implementations e.g., VLBI3 \citep{robertson_75} to UVPAP \citep{marti-vidal_08} for differential phase delay astrometry and SPRINT \citep{lestrade_90} to AIPS \citep{aips}, VEDA  \citep[VEra Data Analyzer][]{veda} or now VLBI-capable CASA \citep{casa_evn_18}, for phase reference mapping.

%{\it  On instrumental requirements:} \\

%{\it  Historic Evolution of PR performance/Scope of application:} \\

% In 1995 astrometry was limited to be between 22\,GHz and 1.6\,GHz. During the 2000's, explorations of 
% new technical solutions for 22\,GHz 
% provided a several-fold improvement in the tropospheric models. This resulted in a massive increase in applications,  because one could achieve tens of \uas{} astrometry using a calibrator several degrees away.
% On the other hand, the low frequencies have lagged behind, using the angular separation as the main approach to mitigate ionospheric model errors. 
% %
% This highlights the relevance of advances in calibration strategies, along side the instrumental developments.

Improved instrumental backends and clock stability, rapid switching and settling capabilities and
%sTable~instrumental performance, so that the clock, electronics and mechanical drifts are inconsequential.
massively increased receiver bandwidths and recording rates 
(i.e., sensitivity), improved the probability of finding a close calibrator and reducing $\Delta \theta_{AB}$ and $T_{\rm swt}$; the primary calibration mechanism in PR.
Furthermore the more complete astrometric catalogues, improved geometric parameters (station coordinates, reference source coordinates, EOP, UT1 and so on) and tropospheric and ionospheric modelling 
%plus inclusion of advanced calibration features (see Sect. \ref{sec:ATC} and \ref{sec:AIC})
have led to an overall improvement in the astrometric performance, but not 
any conceptual change from that in its inception \citep[e.g.,][]{shapiro_79, alef_88}. % (two sources and single frequency). % (correct??).

Further improvements have come from dedicated engineering solutions, such as the dual-beam system of VERA that allows for simultaneous observations of pairs of sources up to 2.2\degr\ apart, i.e., $T_{\rm swt}$=0. 
Looking ahead, the increased sensitivity with the next-generation instruments and arrival of 
innovative technologies to extend the FoV of large single telescopes and arrays and superior frequency agility
%multi-beam instruments/capabilities  for large single telescope and arrays 
will result in even better performance. % by reducing $\Delta \theta$ and $T_{\rm swt}$.
%
%{ END SENTENCE (?):} 
The overall astrometric performance can be improved using advanced strategies and instruments that enable them, as discussed in the following sections.  

\subsection{In-beam PR astrometry}
\label{sec:inbeam}

In-beam PR astrometric VLBI refers to a particularly favourable configuration of conventional PR where the target and the calibrator sources 
%when both sources are sufficiently nearby that they 
lie within the primary beam of the telescopes (hence the name ``in-beam'')  and can be observed simultaneously. 
See Table~\ref{tab:sched} for an indicative observing schedule.
This  configuration results in a superior error compensation, as a result of small $\Delta\theta_{AB}$. % and $T_{\rm swt}\sim$0). 
For extremely close sources, the differential analysis 
can reach the thermal noise limit, $\sigma\Delta\theta_{\rm thermal}$
and results in $\mu$as astrometry. %and $\Delta \theta_{AB} = thermal noise = FWHM / SNR$. 
%
%{\it  Scope of application} \\
The first demonstrations of %in-beam PR $\mu$as systematic error-level astrometry  
in-beam PR astrometry with $\mu$as-level systematic errors
at $\sim$8\,GHz, and few tens of $\mu$as at $\sim$2\,GHz, date from the 1980's using 1058+328A/B, a pair of sources 33\arcsec{} apart \citep{marcaide_84,rioja_phd,rioja_inbeam}.

The considerations for in-beam PR astrometry are largely unchanged since then, but the prospects for in-beam PR, particularly at low frequencies, have improved
because the probabilities for finding suitable in-beam calibrators in the direction of a target of interest have increased significantly.
This is a result of observations with higher sensitivity, advances in (weak) calibrator search methods (see Sect.~\ref{sec:cals})
%\sout{\bf xx but this is here? the only justification of an inbeam section is the search - otherwise it belongs in conventional)} 
and multi-stage analysis using weak nearby calibrators \citep{fomalont_99}.
Currently it is regularly possible to find a suitable calibrator source within the FoV of the 25-m VLBA dishes at 1.4\,GHz (i.e., $\sim$30\amin) in any arbitrary direction { (e.g., in \citet{deller_18} the minimum calibrator flux density was 5\,mJy.)}
This has had a significant impact, particularly in the low frequency regime, where PR astrometry has very large errors ($\sim$1\,mas/deg at L-band). % (see Sect.~\ref{sec:convPR}). 
At L-band a nominal goal of 100\uas\ per epoch (or equivalent differential delay errors $\sim$3\,mm) can be achieved with a reference source $\le$10\arcmin\ away from the target, following Eq.~\eqref{eq:a07_terms} for the systematic residual ionospheric errors. This would match the typical thermal astrometric limits with current instruments assuming a DR of 100:1.
%Under typical conditions, for a baseline length of 6,000\,km, a nominal goal of 100\uas\ per epoch (or equivalent differential delay errors $\sim 3$mm) 
%can be achieved with a reference source $\le$10\arcmin\ away from the target, following Eq.~\eqref{eq:a07_terms} for the systematic residual ionospheric errors at L-band.
% The residuals are dominated by the dispersive delay, which typically are the order of 6TECU, or 1\,m at 1.5GHz. 

However the larger uncertainties in the absolute coordinates of weak reference sources
%larger uncertainties in the weak source coordinates 
is a consideration for the astrometric accuracy 
%of the angular pair separation astrometric measurements 
and should be taken into account.
In science cases that rely on the changes in the sky position of the target across multi-epoch observations (i.e., parallax, proper motion) these uncertainties introduce a constant offset and so are of little relevance. %, as the 
%as long as the weak source provides a fixed fiducial point in the sky (i.e., that its position does not change, even that the absolute coordinates are not accurately known). 
In all cases, for the astrometric error budget one has to take into account the increased thermal noise errors arising from weak sources, along with the systematic residual errors.

% {{\bf Sec 6?}
% For state of the art in-beam PR observations with a calibrator $\sim$10\arcmin{} away with current instruments the level of thermal noise and the systematic error regimes are similar, at frequencies $\nu\le$3\,GHz, and agree with estimates above.
% {\bf say here?} We note that it requires to push through both thermal and systematic limits in order to further improve the astrometric accuracy, that is, would require increased signal-to-noise ratio and closer sources; therefore, that limits the prospects for improved accuracy with in-beam PR even with next-generation instruments ({\bf is that correct?})
% {\bf And use later for the argument of in-beam MV much superior to in-beam PR with next gen instruments. I say loose whole paragraph}}

%At higher frequencies the use of in-beam PR continues to be very/extremely limited.
In-beam PR astrometry is suitable at any frequency, 
although the probability of finding a calibrator within the smaller FoV of the telescope at higher frequencies, and shorter coherence times, becomes very low. Table~\ref{tab:inbeam_mv} lists the number of sources one would expect to find within the FoV of a 20m radio telescope, with the sensitivity levels of current instruments (Col. 6) and SKA (Col. 8), for a range of frequencies. Therefore at high frequencies $>$8\,GHz even with SKA Phase-2 or ngVLA, 
the usability of in-beam PR continues to be extremely limited and unless one is lucky, one would have to slew between the source pair, even with next-generation instruments.

% {{\bf summarized above - delete from here? } \\
% Considerations for the astrometric errors arising from uncertainties in the calibrator/weak source coordinates ({\bf is this correct? calculate} and thermal noise are likely to be a significant contribution, along with the ionospheric errors.
% In science cases that rely on the changes in the position of the target in the sky (i.e., parallax, proper motion)  across multi-epoch observations, the requirement for a suitable reference source is that it provides a fiducial point in the sky (i.e., that its position does not change, even that the absolute coordinates are not accurately known); in these cases the thermal noise is the relevant contribution. 
% In cases where the pair relative separation is the measurement of interest (such as in the definition of celestial reference frame),
% {\sout{then the coordinates of the reference/calibrator source must be accurately known and in-beam calibration is less suitable {\bf OR}}}
% then both contributions are relevant and accurate calibrator coordinates are important. In all science cases, the systematic ionospheric astrometric errors, of the order of $\sim 100 \mu$as even for a 10\arcmin{} pair separation at L-band, pose an ultimate limit.
% This limit can be overcome using recently developed advanced calibration strategies, such as MultiView (section 5.1) {{\bf why have you commented that out?}}.
% %
% {{\bf summarized above - delete to here? } \\}}

If the target and calibrator sources can be co-observed there is no source switching and one would assume that $T_{\rm swt}$ would be zero, nevertheless we note that this term also encompasses the solution interval on the calibrator. In-beam calibrators tend to be weaker, the thermal limits dominate the error budget and $T_{\rm swt}$ becomes the calibration solution interval. %{\bf better??}

\begin{table}
    \caption{
    %table for \MV{} and in-beam 
    Table to characterise the performance of \MV{} and its feasibility for current and next-generation instruments, across the spectrum. 
    Col. 1 is the observing frequency, Col. 2 is the spatial resolution for a 6,000km baseline.
    Col. 3 is the estimated systematic astrometric error using \MV, as discussed in Sect.~\ref{sec:study_mv}.
    Col. 4 is the required DR for a matching thermal-noise astrometric error ($\sigma\Delta\theta_{\rm thermal}$).
    Col. 5 lists the source fluxes that are 100 times the 1 hour image sensitivity with current VLBI arrays (derived from EVNCalc with bandwidths of 16MHz at $\nu<$ 1\,GHz, 128MHz at 1.6\,GHz, otherwise 256MHz). We use DR 100:1 as it is a typical value with current instruments.
    % The observing frequency (Col. 1) and spatial resolution (Col. 2) for a 6,000km baseline, allows us to calculate the estimated systematic astrometric error using \MV, as discussed in Sect.~\ref{sec:study_mv} (Col 3). The ratio of Col. 2 and 3 gives the equivalent matching dynamic range (Col. 4). Col. 5 gives the 1 hour sensitivity limits at current VLBI sensitivities (derived from EVNCalc with bandwidths of 16MHz at $\nu<$ 1\,GHz, 128MHz at 1.6\,GHz, otherwise 256MHz) at a DR of 100 { (i.e., 100$\times {\rm SEFD}^{\rm current}/\sqrt{\Delta\nu\tau}$)}. 
    Col. 6 is the number of sources, with fluxes larger than Col. 5, expected within the { FoV} of a single pixel 20m antenna if FoV $\le$1\degr{}, otherwise 1\degr{} (marked with $^{\dagger}$), using the parameterisation from \citet{TREC}.
    Col. 7 lists the source flux that is the matching DR value (Col. 4) times the 1 hour image sensitivity with SKA-VLBI Phase-1 (taken from \citet{jj_wp10} { for the full array combined, 256MHz for SKA1-Low, 2GHz for SKA1-Mid}). 
    Col. 8 is the number of sources, with fluxes larger than Col. 7, expected within the { FoV} of a single pixel 20m antenna if FoV $\le$1\degr{}, otherwise 1\degr  (marked with $^{\dagger}$). 
%    Col. 7 is the 1 hour sensitivity limits for SKA-VLBI Phase-1 (with sensitivities taken from \citet{jj_wp10} { for the full array combined, 256MHz for SKA1-Low, 2GHz for SKA1-Mid}) and the listed matching DR, { i.e., Col.4 $\times {\rm SEFD}^{\rm SKA}/\sqrt{\Delta\nu\tau}$)} and the corresponding number of in-beam sources expected (Col. 8). 
    For the higher frequencies we also included in brackets the number of in-beam sources that could be expected for SKA-Phase 2.
    The number of in-beam sources for ngVLA observations would fall between these two values.
    Based on Col. 6 and 8, in-beam \MV\ would be feasible at frequencies $<$1.4, $<$2 and $<$6.7\,GHz, with the sensitivities of current VLBI, SKA-VLBI Phase 1 and Phase 2, respectively, and with switched \MV\ at higher frequencies. Therefore the astrometric performance can be significantly improved with the next-generation instruments and \MV, reaching the \MV{} astrometric precision in Col. 3.
    %and lower frequencies, currently. With SKA-VLBI Phase-1 sensitivities sufficient sources for in-beam \MV{} would be available at frequencies at 2\,GHz and below and for Phase-2 at 6.7\,GHz and below.
    %We note that in-beam conventional PR would be possible at 2GHz and below (but without achieving $\sigma\Delta\theta^{Mv}$).
    %skamidref(n=freq,d=20.,m=1e-3) %% Dish d;
    \label{tab:inbeam_mv}}
    \centering
    \begin{tabular}{|c|ccc||cl|cl|}
    \hline
    Frequency & Resolution & MV error & Matching  &  $\Delta I_{\rm m}^{\rm current, 1h}$ & No. of  & $\Delta I_{\rm m}^{\rm SKA, 1h}$  & No. of \\%& within 1$^o$\\
    $\nu$ & $\theta_{\rm beam}$ & $\sigma\Delta\theta^{MV}$ & DR &  $\times$100 & in-beam  & $\times$ DR  & in-beam \\%& within 1$^o$\\
     (GHz) & (mas) & ($\mu$as) &  & (mJy/beam) & sources & (mJy/beam) & sources \\
        \hline 
        0.3 & 34 & 150    & 230    & 120 &1.2$^{\dagger}$ &5.1 & 14$^{\dagger}$ \\%& 123\\ % 16MHz VLBA 1hr
        0.9 & 11 & 17     & 674    & 20  &3.5 &3.1 & 15  \\%& 16 \\ % 16MHz EVN-
        1.6 & 6.4 & 6     & $>$1000& 4.9 &2.9 &2.1 & 5.5  \\%& 16 \\ % 128MHz VLBA
        5.0 & 2.1 &$\sim$1& $>$1000& 2.3 &0.4 &2.4 & 0.4 (6) \\%& 13 \\ % 256MHz VLBA
        8.0 & 1.3 &$\sim$1& $>$1000& 3.6 &0.1 &2.6 & 0.1 (2) \\%& 12 \\ % 256MHz VLBA
        15.0& 0.7 &$\sim$1& 687    & 6.0 &0.0 &3.0 & 0.0 (0.4) \\%& 13 \\ % 256MHz VLBA
        %22.0& 0.5 & $\sim$1 & 468 & 7.1 &0.0 &2.4 0& 0.0 \\%& 10\\ % 256MHz VLBA
        \hline
    \end{tabular}
\end{table}

The ideal instrumental requirements to benefit from in-beam PR are high sensitivity, such as provided by large powerful telescopes, and wide FoV, such as provided by small telescopes. %, which is almost a contradiction. 
%(However this requirement can be addressed by, on a connected array, the ability to form multiple tied-array beams on the individual sources and by, on single large dishes, the ability to form multiple beams. These are discussed later in Sect.~\ref{sec:tech}.) {\bf OR} 
This contradiction can be addressed with new technology developments to enlarge the FoV, e.g.
using receivers with multiple-pixel capabilities (such as PAFs or multi-beam feeds) on large single telescopes, 
%for large single telescopes, with multi-pixel (or multi-beam) receivers (i.e., Phased Array Feeds, PAFs), 
and the capabilities for multiple tied-array beams in the directions of nearby sources with antenna arrays, as shown in Fig. \ref{fig:radio_int}. These are discussed later in Sect.~\ref{sec:tech}.

\subsubsection{Stacking to increase SNR} % - ``stacking'' signals from multiple sources to increase SNR}
\label{sec:hdm}
%{\bf For Elsewhere .. Tech?}\\

Hybrid Double Mapping (HDM)  \citep{rioja_00} is an alternative astrometric path to in-beam PR, where the signals from both sources are 
combined or stacked into a single hybrid dataset to increase the SNR; hence this is an advanced strategy of interest for weak sources.
%The signature of the relative astrometric separation between both sources is preserved in the hybrid dataset 
In HDM the hybrid dataset preserves the signature of the relative (astrometric) separation of the source pair, which can be measured directly 
from the compound-image produced using self-calibration algorithms to solve for the common antenna based phase and phase derivative residual errors. 
HDM was demonstrated for a close pair of sources, 33\arcsec{} apart, using the point-by-point sum of the reference and target source visibilities, at 8.4 GHz, and the results agreed with those from in-beam PR. %, both using AIPS. %; for details on the analysis path and scope of application see the papers.
A more recent demonstration, performed in the low frequency regime, is 
Multi-Source Self-Calibration (MSSC) \citep{radcliffe_16} that combines the signals from multiple weak targets that lie within the FoV, to increase the SNR and allow self-calibration. 
%In this case any individual source position will then be relative to that of the ensemble of sources.
In this case the relative astrometry is preserved between the ensemble of sources, as long as direction-dependent effects are not significant.

%{\bf concluding remark and linking to next stage MV:}\\
\subsection{Advanced PR calibration strategies} %) - Dedicated solutions}
Atmospheric effects are one of the most poorly understood components in the a\,priori theoretical model of VLBI observables, and are the dominant source of astrometric errors. 
This section is concerned with `advanced' strategies to improve the tropospheric and ionospheric modelling (i.e., reduce typical a\,priori model errors $\Delta \ell_z$  and $\Delta I$ in Eq.~\eqref{eq:a07_terms}).
Improved model values in PR result in equivalent benefits to using a smaller $\Delta\theta_{AB}$, that is achieving higher astrometric accuracy,
but with widely separated sources. %, not only the ones that have a very nearby reference source. \\
We dub this approach advanced PR astrometric techniques.
%Here we review the established approaches
%for Advanced Tropospheric Calibrations and Advanced Ionospheric Calibrations.
%We dub them advanced PR astrometric techniques to distinguish them from PR, which relies solely in the angular separation as the dilution factor for model errors.
Because the nature of tropospheric and ionospheric effects are very different, the advanced strategies are specific for application in the high and low frequency regimes. 
We refer to these as Advanced Tropospheric and Ionospheric Calibration strategies (ATC and AIC, respectively).
These comprise using either dedicated blocks inserted during the VLBI observations, the program data itself or external measurements from an independent technique.

% Atmospheric fluctuations are responsible for the dominant calibration errors; these are increasingly significant 
% at lower frequencies where they are due to the ionosphere and direction dependent and are weather dependent.  % the angular scales are smaller
% A precise astrometric strategy requires high quality spatial and temporal phase calibration solutions. Here we review the established approaches
% for Advanced Tropospheric Calibrations and Advanced Ionospheric Calibrations. 

% %
% These advanced strategies improve the tropospheric and ionospheric modelling (e.g., reduce typical model errors $\Delta \ell_z$  and $\Delta I$, see Eq.~\eqref{eq:a07_terms}),
% either by using dedicated blocks inserted during the VLBI observations, the program data itself or external measurements from an independent technique.
% %
% Improved model values result in equivalent benefits to using a smaller $\Delta\theta_{AB}$ in PR, that is, achieving higher astrometric accuracy,
% but with widely separated sources. %, not only the ones that have a very nearby reference source. \\

Huge strides have been taken in the last two decades, particularly at the higher frequencies. ATC strategies can reduce typical tropospheric errors for $\Delta \ell_z$ from several cm to the level of $\sim$1\,cm. 
These tropospheric strategies fail at lower frequencies ($\le$8\,GHz) where the dispersive ionosphere is the dominant contribution. 
The ionospheric disturbances have a high degree of spatial structure, which is responsible for the direction-dependent calibration errors and can not be accurately modelled with a single global value at the Zenith of the telescope.  % removed scaled by the Zenith angle,
GPS-based TEC global models have errors of about $\sim$10-20\% \citep[e.g.,][]{hernandez_09}, or typical nominal residual errors of $\Delta I \sim$6\,TECU \citep{vlba_23}.
Such $\Delta I$ correspond to an excess path error of 1\,m at 1.5\,GHz, which is a hundred-fold larger than the best tropospheric residual errors, and scales as $\nu^{-2}$. 
The AIC strategies improve the $\Delta I$ estimate for the line of sight.
%This level of systematic error is a fundamental barrier towards precision astrometry at low frequencies. 
Here we discuss advanced strategies which can reduce $\Delta I$ to $\sim$0.1\,TECU. % are presented the following sections

\subsubsection{Advanced Tropospheric Calibration methods} % with external measurements}
\label{sec:ATC}

\paragraph{\bf Dedicated tropospheric blocks:}
The ``GeoBlocks'' method \citep{brunthaler_05,reid_09_I} uses dedicated blocks of observations 
($\sim$30 minutes long, scheduled every 3-4 hours)
interleaved with the program observations to improve the a-priori value of the tropospheric model. 
See Table~\ref{tab:sched} for an indicative observing schedule.
The GeoBlocks comprise observations of multiple ICRF sources following a geodetic-like schedule and analysis. 
That is, strong sources are selected which cover a wide range of Zenith angles, and are observed in short scans in rapid succession with wide bandwidths. The derived group delays are used to solve for the Zenith path length error plus clock errors, for each station. %These station-based corrections are applied to all the VLBI data, to remove any quasi-static residual non-dispersive Zenith path errors. 
These estimates, mapped to the corresponding elevations, are used to remove the tropospheric static component from the program observations and the differential observables.

Such an approach can reduce the value of tropospheric zenith delay residual errors to $\Delta\ell_z \sim$\,1cm, which leads to 10\uas/deg relative astrometric errors
at moderate antenna elevations (i.e., $Z \sim$45\degr{} and above). 
Nevertheless, the differential residual errors at low elevations, amplified by the $\sec{Z} \tan{Z}$ term in Eq.~\eqref{eq:a07_terms}, can become very large.
For example, at a zenith angle $Z\sim$60\degr\ the systematic errors are eight times larger than those at $Z\sim$45\degr, which leads to 80\uas/deg relative astrometric errors. %, for the same source pair angular separation. 

Due to the relative simplicity of the implementation of the GeoBlock strategy, and its versatility and universal application, this approach has led to a significant expansion and has had a major impact in the field of precise astrometry at frequencies around $\sim$22\,GHz, spreading the application of astrometry to many diverse fields \citep[see the references in][]{reid_micro}.
The GeoBlocks superseded earlier approaches by the same team \citep{reid_99} % (Reid+1999,Brunthaler 2005?) 
that fitted the residual differential phases of the program observations to 
%, as a function of the elevation and a constant residual 
correct for a tropospheric zenith delay error per station; this method required strong program sources and the estimates, using a single pair of sources, were less precise.
Other approaches comparable to GeoBlocks involve using an image optimisation algorithm, simultaneous tropospheric measurements from 
co-located GPS receivers \citep[see the comparative study in][]{honma_08_trop} or Water Vapour Radiometer (WVR) systems \citep{roy_06}. 

\citet{xu_18} showed the benefits from the combined use of the methods above, compared to outcomes from each individual method, when applied to challenging low elevation observations.
We highlight the GeoBlocks and image optimisation calibration strategies for their widespread usage on many targets over a wide scope of research, 
and as drivers of astrometric surveys, i.e., the VLBA BeSSeL and VERA Key Science Program (KSP) projects, respectively.

%\subsubsection{Advanced Tropospheric Calibration with the observations}
\paragraph{\bf Residual Minimisation of Program Observations:}
The differential phase delay astrometry method uses a least squares fitting of the program observations to estimate the astrometric source coordinate parameters, that can also include fitting of the residual tropospheric zenith path length per telescope, among other parameters.

%The University of Valencia Precision Astrometry Package (
The University of Valencia Precision Astrometry Package (UVPAP) software is an upgraded version of the VLBI3 program \citep{robertson_75}, which
incorporates new capabilities such as a more robust automatic algorithm to solve for phase ambiguities, using closure conservation conditions for triangles of telescopes, and increased processing capability for a joint fit of the observations of multiple source pairs simultaneously \citep{marti-vidal_08}. %{\bf (references?)}. 
The former is fundamental to determine the unambiguous observables.
The latter can reduce the residual zenith excess path length to the order of $\Delta\ell_z \sim$1\,cm from the added constraints, 
%and improved modelling of the tropospheric effects (and reduce the residual zenith excess path length) to the order of 1\,cm, from the added constraints )
%{\bf OR} The later results in an improved modelling of the tropospheric effects to the order/level of $\Delta\ell_z$ \sim$1\,cm \citep{marti-vidal_16}, 
particularly if the program observations span a significant range of elevations in a rapid temporal succession \citep[see discussions in][]{marti-vidal_16}. 
This is comparable to the outcome from GeoBlocks. 

\subsubsection{Advanced Ionospheric Calibration methods}
\label{sec:AIC}

\paragraph{\bf S/X group delay correction:}
The extremely well established geodetic VLBI technique of obtaining an `ionosphere-free' visibility data, by combining group delays from S-band (2.4GHz) and X-band (8.4GHz) from simultaneous dual-frequency observations, is one of the oldest AIC methods.
% measuring group delays at S-band (2.4\,GHz) and at X-band (8.4\,GHz) simultaneously, and then computing $\Delta I$ from the linear combination and removing this to obtain an `ionosphere free' visibility data product,
%subtracting the scaled difference of the two values to obtain an `ionosphere free' visibility data product, 
%is one of the oldest AIC methods in astrometry.
This was also the most common method for relative astrometry, using S/X observations of the pair of sources, particularly before the availability of GPS derived corrections.  
Using this approach one can reduce the residual ionospheric errors to $\Delta$I$\sim$1.5\,TECU \citep{ros_99}. We note that these results were achieved without applying a-priori GPS-based corrections, so the initial errors would have been up to $\Delta$I$\sim$40TECU.  Ionospheric corrections derived from S/X observations of source pairs have been shown to be significantly better than those from GPS-corrected analysis \citep{vlba_23}. 
{ This method, however, includes the implicit assumption that the source positions have a linear dependence on wavelength \citep[see][for discussions of this]{porcas_09}.}
VLBI observations corrected by direct dispersive delay measurements have not been used much recently, as only the Geodetic (i.e., S/X) single polarisation receivers supported such simultaneous dual-frequency  observations. 
However, with the adoption of wideband cm-wave receivers (see Sect.~\ref{sec:brand}) on the next-generation of telescopes, this situation is sure to be reversed and this and related approaches will see a regeneration. 

%\subsubsection{Advanced Ionospheric Calibration on the phases} % (Brisken) - Image and visibility based}
\paragraph{\bf Phase-Fitting:}
%An alternative to the S/X advanced group-delay ionospheric method is to  use phase-fitting to measure $\Delta I$ for each antenna, using the dispersive $\nu^{-1}$ signature in the differential residual phases over the observed frequency range.
An alternative to the S/X group-delay ionospheric method to measure $\Delta I$ for each antenna, is to fit the differential residual phases for the dispersive $\nu^{-1}$ signature, over the observed frequency range.
The improved estimate for $\Delta I$ is used to eliminate the ionospheric contribution from the target data.  
Special considerations are required to avoid phase-fitting problems related to the inherent ambiguity of the phase observable. 
This method has been demonstrated for pulsar parallax measurements, with VLBA observations of multiple  pairs of sources separated a few degrees apart, at L-band.
%This was demonstrated in the pulsar parallax survey (?) of 
\citep{brisken_00,brisken_phd,brisken_02}.  % ref correct?
The approach consisted of making initial phase-referenced snapshot images of the target sources at each sub-band of the frequency span (1.4 -- 1.7\,GHz) using conventional phase referencing imaging.
The astrometric errors in the phase referenced images are expected to be at $\sim$mas level and changing between sub-bands. %, 
By fitting the (sub-band) antenna based phase residuals across the frequency span to a function comprising a point-source model with a non-dispersive (i.e., scaling with $\nu$) and a dispersive (i.e., scaling with $\nu^{-1}$) term, one can determine the $\Delta I$ ionospheric errors per antenna
as a function of time \citep{brisken_02}. These are used to correct the differential phases and improve the accuracy of the astrometric measurement.
Note that the initial phase referenced image is vital to ensure that there were no ambiguity issues in the fitting of phases across the band. 
%{\bf OR} 
Alternatively the residual ionospheric errors can be determined 
%in the image domain, where one can combine the inputs from across the VLBI array to improve the signal strength.
from the characteristic frequency dependent (i.e., dispersive) position-shift signature measured in the image domain. That is, 
%where one can combine the inputs from across the VLBI array to improve the signal strength, 
by measuring the position changes in the snapshot images of the program source across a range of frequencies. 
%By doing this, where one can combine the inputs from across the VLBI array to improve the signal strength,
The image-based approach improves the signal strength by combining the inputs from across the VLBI array; in this case 
average ionospheric wedges over the array  (or more often a subset of the array)  as a function of time
could be derived \citep{brisken_00}.
%Alternatively the average wedge over the array (or more often a subset of the array) could be derived from the snapshot images \citep{brisken_00}.

The residual ionospheric error $\Delta I$ after these fitting strategies could be as low as 0.1 TECU \citep[estimated from][Fig. 4]{brisken_00}, a fifty-fold improvement over the nominal residual values after GPS-based corrections. %($\sim 2 cm at 1.4 GHz$
With this level of systematic errors one could imagine achieving 10\uas/deg astrometry precision at L-band, provided the signal was sufficiently strong. Nevertheless, the complexities of the analysis and the thermal limits of the snap-shot phase-referencing imaging of the pulsars 
%\sout{ every minute or so,} 
has hither-to prevented wide adoption of this method. 
{ Furthermore, this method includes the implicit assumption that the source positions have no dependence on wavelength, which is true for pulsars but not generally true for AGNs.}
However, with the sensitivity of the next-generation instruments, both due to increased collecting area and wide bandwidths, it is possible that these approaches may yet be useful again.

%{\bf this is closer to SX AIC}
\paragraph{\bf Dedicated ionospheric blocks:}
Section~\ref{sec:mfpr} on multi-frequency phase referencing (MFPR) describes a related multi-frequency approach for reducing $\Delta I$ to a fraction of a TECU using dedicated blocks interleaved with the program observations, on the target source; 
%that are used in a similar fashion to GeoBlocks, 
we dubbed this strategy ``ICE-blocks'' { (Ionospheric CorrEction, to resonate with the GeoBlock method for the troposphere)}. % These also may enhance the applicability of this approach.
Such strategies should be widely applicable (as for GeoBlocks) thus improving the performance and applicability at low frequencies.

\subsubsection{Concluding remark on advanced astrometric calibration methods}
%{{\bf concluding remark (of 4.6): ADAPT BETTER}\\}
A precise astrometric strategy requires high quality spatial and temporal phase calibration solutions. The ATC and AIC strategies presented above
have had a large impact in the field of astrometry with PR, particularly in the high frequency regime. 
Still, the scope of PR-related applications remains unchanged, with an upper frequency limit of 43 GHz. 
Sections \ref{sec:methods_sfpr} and \ref{sec:mfpr} discuss methods that allow for precise astrometry above this limit. 
At low frequencies, the AICs can provide a path to exquisite improvements in the ionospheric calibration. Nevertheless, its implementation is limited by the multi-frequency capabilities of current instruments and as a result precise astrometry at the low frequency regime is driven by in-beam PR, which has a more limited applicability compared to ATC.
% above using multiple sources allows the static model errors to be reduced sufficiently for PR calibrators to be $\sim$1\degr{} from the target. As the ionosphere is less well modelled by a single global value at the Zenith of the telescope, the advanced ionospheric calibration strategies improve the $\Delta I$ solution for the line of sight. % removed scaled by the Zenith angle,
Both AIC and ATC are able, in principle, to deliver astrometric precision at 10\uas/deg level, although in practise this precision has been largely limited to the tropospheric corrections and the high frequency regime, to date.
Sect.~\ref{sec:methods_mv} presents a new technique (i.e., MultiView) for ultra precise astrometry with next-generation instruments, which works for both tropospheric and ionospheric corrections for the high and low frequency regimes and is not dependent on a strong target source.

% Advanced PR astrometric techniques combine these calibration strategies with PR to give improved results. 
% These have had a large impact in the field of astrometry, particularly with ATCs in the high frequency regime.
% %
% In the low frequencies, the AICs can provide a path to significant improvements in the ionospheric calibration. Nevertheless, its implementation is limited 
% by the multi-frequency capabilities of current instruments and as a result precise astrometry at low frequency regime is driven by in-beam PR, which has a more limited applicability compared to ATC.

\section{The next-generation of radio instruments and the opportunities for astrometry} % + Pathfinders + Science enabled}
\label{sec:next}
\typeout{TECHNOLOGY}

% {\it  Increased sensitivity is highly desirable. Arrival of new instruments with extremely large sensitivity have revitalised the fields at the very high and very low radio frequencies, with the prospect to expand into new accessible areas of research (science enabled). Enables precision required 
% to detect/measure and discern between different theories... }

We present a short summary of the next-generation instruments that are coming online, which will drive forward the field of ultra-precise astrometry with VLBI observations. { Table~\ref{tab:next-gen} lists crucial parameters for some of these and current arrays.}
This review is primarily interested in the considerations for astrometric surveys, therefore we have focused on the relevant issues for those. 
{ We emphasise our views on the impact that these new instruments can have, in conjunction with established and innovative astrometric methods and technologies.}
% provided the requirements for optimum astrometric calibration methods and usability are not designed out.

%{
\typeout{Not finalised}
\begin{table}[ht]
    \caption{Summary table of some of the current VLBI arrays and next-generation elements with VLBI compatibility (Col. 1) %
    mentioned in this review. Col. 2 and 3 list the frequency range and collecting area. %The array frequency range and collecting area and
    Col. 4 shows the capability for multi-beams, indicating if this is available via formation of simultaneous multiple-pixel beams or by alternating between sources (Yes or Switching, respectively). % or not available (No).
    Col. 5 shows the capability for multi-frequencies, indicating if this is available via observing multiple frequencies simultaneously, by fast frequency switching or not available (Yes, Switching or No, respectively).
    %switching, simultaneously or not available. 
    %The capability for forming multiple beams (or of fast source switching) and observing multiple frequencies (or of fast frequency switching) is indicated in Col. 4 and 5, respectively. FIX
    The VLBA is a full time dedicated VLBI array, formed of 10 25-m antennas, of which 8 can observe at the highest frequency of 86\,GHz.
    EVN is an ad-hoc array, with some large antennas (3$>$60m) and many (9 for these calculations) smaller ones. 
    x-KVN stands for KaVA (KVN and VERA array) plus Yebes (40m).
    VERA is the Japanese astrometric array.
    %Not all antennas can observe at all frequencies, so Col. 3 is the upper limit.
    For the next-generation VLBI elements, the collecting areas of SKA-Low and SKA-Mid Phase-1 are given; Phase-2 values are given in brackets. 
    % mention of FAST and ngVLA??
    ngVLA is proposed to also be a standalone VLBI array, by absorbing the VLBA sites.
    FAST is the 500m diameter Chinese single dish and SKA pathfinder.
    $\dagger$: not continuous for current arrays; 
    $\star$: frequency dependent, as not all stations support all frequencies. $\diamondsuit$ KVN only.}
    \label{tab:next-gen}
    \centering
    \begin{tabular}{|c|cccc|}
    \hline
    Name     &  Frequency Range$^\dagger$ & Collecting Area$^\star$ & Multi-Beams & Multi-frequency \\
        \hline
    \multicolumn{5}{|c|}{Current VLBI Arrays}\\
        \hline
    VLBA & 0.3 to 86 GHz&  6,000\,m$^2$ & Switching & Switching\\
    EVN  & 0.3 to 43 GHz& 28,000\,m$^2$ & Switching & No\\
    x-KVN  & 22, 43, 86, (130$^\diamondsuit$) GHz & 4,000\,m$^2$ & Switching & Yes\\
    VERA   & 22, 43 & 2,000\,m$^2$ & Yes & Yes\\
    \hline
    \multicolumn{5}{|c|}{Next-Generation VLBI elements}\\
    \hline
    SKA-Low  & 50 to 350MHz& 0.4 (1) km$^2$ & Yes & Yes\\
    SKA-Mid  & 0.35 to 15GHz& 0.04 (1) km$^2$ & Yes & Switching\\
    ngVLA    & 1.2 to 120GHz& 0.1 km$^2$ & Yes & Maybe Yes \\
    FAST     & 0.07 to 3GHz& 0.1 km$^2$ & Yes & No\\
    %GAIA     & & - \\
    \hline
    \end{tabular}
\end{table}
%}

\subsection{m and cm radio astrometry in the SKA era}

The Square Kilometre Array (SKA) is an international collaboration to build an array with a collecting area that will be eventually a { hundred} times greater than current facilities. It is being built in two stages, an initial roll-out (Phase-1) and then the final full array (Phase-2).  
It comprises of two arrays and frequency ranges, SKA-Low covering 50 to 350MHz, sited in Western Australia,
and SKA-Mid covering 0.35 to 15.3\,GHz (and potentially higher), sited in South Africa and eventually eight other African countries. 
SKA-Low Phase 1 will consist of 512 stations each of 256 log-periodic dipoles (equivalent to $\sim$40\% of the full SKA-Low), spread over a diameter of 65\,km.
SKA-Mid Phase 1 will consist of 197 $\sim$15m offset-Gregorian parabolic dishes (equivalent to $\sim$3.5\% of the full SKA-Mid), spread over a diameter of 150\,km.
The full SKA will have baselines of thousands of km, and can be considered a VLBI instrument in its own right.

The arrival of SKA has revitalised the research at m to cm wavelengths in many fields \citep{ska_aas} and this includes a long baseline component.
{ To exploit the full potential of the first phase, both SKA-Low and SKA-Mid will perform joint observations with other telescopes, hereafter SKA-VLBI.}
These can potentially perform high precision astrometry (providing the mechanisms are in place). 
Given the collecting area of the SKA telescope, which will be an { order of magnitude (Phase-1) to two orders of magnitude} (Phase-2) times greater than that available currently, the sensitivity of VLBI baselines from existing infrastructure to the SKA is expected to increase by an order of magnitude or more with respect to current levels; this offers a great opportunity for VLBI studies.
% \sout{We discuss  the requirements to ensure  
% that the telescopes are equipped with all the technological capabilities required to deliver ultra-precise astrometric precision.}

Observations with SKA-Low %(50--350MHz) 
will be dominated by ionospheric systematic propagation effects, with spatial variations at scales much smaller than the wide FoV of the 
individual dipole elements. Each SKA station has a diameter of 38m (and a FoV from 15 to 1.3\degr), and can form multiple station beams towards any point visible to the individual elements. Multiple tied-array beams can be formed within these station beams, from the sum of the station outputs; the current specifications are for 4 full bandwidth coherent VLBI beams. The backend processing hardware can support a maximum of 16 full bandwidth VLBI beams, but this would require an increase in the number of planned VLBI servers.

The frequency range for SKA-Mid observations %(350--15,400MHz initially)
comprises regimes with dominance of either the ionospheric or tropospheric systematic propagation effects, and a regime where both are significant. 
The SKA-Mid dishes have a significantly smaller FoV on the sky than for SKA-Low, but will still see nearly 4\degr\ at the lowest frequency. 
Similar backend hardware is planned for SKA-Mid as for SKA-Low, but with the greater instantaneous bandwidth for SKA-Mid  a smaller number of tied-array beams can be formed. Currently 4 full bandwidth (2GHz) coherent VLBI beams  would be possible; the number of beams can be increased  by  compromising on the bandwidth \citep[for details see][]{jj_wp10}. 

With the phased-up SKA Phase-2 acting as an extremely sensitive station in a VLBI array the sensitivity for the SKA baselines will be an order of magnitude better than those for the best 100m-class radio telescopes available today
(see Table~\ref{tab:next-gen}).
Furthermore the  capability to tie together all the outputs of the stations to form multiple tied-array beams for VLBI (and also pulsar studies) 
%will allow surveys to be performed but moreover 
will be essential for the science cases which require  astrometry with SKA-VLBI, as a number of simultaneous tied-array beams are vital to obtain highly precise outcomes; moreover this will allow VLBI surveys to be performed.  
Whilst the VLBI baselines will still  be Earth bound we
may reasonably assume that the achievable sensitivity will be ten
times better than current limits, achieving dynamic ranges (DR) of $\sim$1000:1. 
We note that nearly such levels are being achieved currently for a few objects \citep{yang_16,miller-jones_20}.
A thousandth of the beamsize, if the observations are thermally limited, would potentially allow for 8\uas\ astrometric precision at L-band ($\sim$1.4GHz) and 1\uas\ precision at X-band
($\sim$8.4\,GHz), with global baselines of 6,000\,km; Sect.~\ref{sec:space} discusses Space-VLBI possibilities that allow for much longer baselines.
These limits would open up a huge parameter space for SKA-VLBI to explore, in a multitude of scientific fields \citep[see][]{ska_vlbi}. However this is predicated on the actual measurements matching the potential astrometric accuracy.
\MV{} is a calibration method that can provide the matching accuracy as discussed in Sect.~\ref{sec:methods_mv}.
Sect.~\ref{sec:future} describes some of the science cases for SKA-VLBI where ultra-precise astrometry is key. 

%{{\bf Pathfinders:}}

The SKA precursors and pathfinders play an essential role 
%
%Given that it is essential 
in developing and demonstrating the SKA technologies and techniques now, before the instrument is finalised.
A number of them have performed a limited amount of VLBI. 
The LOFAR has made long baseline observations \citep[e.g.,][]{lofar_lbs} between the international stations.
The Murchison Widefield Array (MWA) has performed VLBI in conjunction with the Indian Giant Metrewave Radio Telescope (Kirsten, per. comms). MeerKAT has formed fringes between their stations and the European VLBI Network (EVN). ASKAP has plans for a VLBI beam-former.
The massive boost in low frequency astrometry that will be provided by the
FAST telescope (China, 500m diameter), the world's most sensitive radio telescope \citep{fast}, should be noted.
FAST is a SKA pathfinder with a collecting area similar to SKA Phase-1, which makes it highly desirable for joint VLBI astrometric observations; see Table~\ref{tab:next-gen}.
The narrow FoV and limited pointing capabilities put a constraint, in principle, on VLBI astrometric measurements.
Nevertheless it is currently equipped with a 19-beam receiver, which could enable precise astrometry using \MV. %, a technique discussed in Sect.~\ref{sec:methods_mv}.
The multiple-pixel effectively extends the narrow FoV to that of a 30m diameter single pixel telescope. There are plans to install a Phased Array Feed (PAF), which will result in an even better  performance with denser coverage of the FoV and wider bandwidths. Detailed discussions on the impact of technological developments, such as multiple-pixel capability, on astrometry are to be found in Sect.~\ref{sec:tech}.

\subsection{cm and mm radio astrometry in the ngVLA era}
%\subsection{Astrometry in the ngVLA era}
\label{sec:ngvla}
%\subsection{Estimates for Multi-frequency performance on the ngVLA}

The next-generation VLA (ngVLA) project is an American proposal for a cm to mm antenna array \citep{ngvla}. It includes long baselines,
spanning at least $\sim$500\,km, and covers the frequency range from 1.2 to 120\,GHz \citep{ngvla_sci}, with a planned collecting area larger than SKA-Mid  Phase-1. 
The ngVLA will be built out from the current VLA site in New Mexico,
which is a high and dry plane with a good view of most of the
sky. 
It consists of a dense core of 160 close-packed 18m diameter parabolic antennas, almost completely covering the inner kilometre, then transitioning to a log spiral pattern out to 20km in radius. Longer baselines, with a further 50 or so antennas, then extend across Southern USA, and even into Mexico \citep[][]{ngvla_tec}. 
Furthermore, the VLBA sites are envisaged as being added to this, with two to three dishes at each site, to increase the sensitivity of these longest baselines; see Table~\ref{tab:next-gen}.
The receiver systems sit at the off-axis Gregorian focus, and the proposed design has six feeds for six bands with a very fast translator, which allows switching between frequencies on the timescale of seconds. 
Such an instrument will be an exciting partner for SKA and the Atacama Large Millimeter/submillimeter Array (ALMA) .

ngVLA, with baselines of at least 500 and potentially 8,000km, is
essentially a VLBI instrument with spatial resolution at the milliarcsecond scale.
ngVLA has larger dishes than SKA-Mid and is not envisaged as a survey instrument. %\sout{Observations  will be PI-driven, and} 
As such it is well matched in both capabilities and in organisation to be an outstanding VLBI instrument \citep{ngvla_lbs}.
%
%For the instrument to perform differential astrometry the data needs to be phase referenced. 
With the enhanced sensitivity from a phased-up core and wide recorder bandwidths, 
multiple calibrator sources within the FoV of an antenna could be expected to be found for most directions up to frequencies of $\sim$8GHz, 
% 8.0 GHz N (>7.00e-05 Jy) = 2.64e+00 (per 18.0 m (0.12 deg) beam)
where the FoV would be 0.1\degr (see Table~\ref{tab:inbeam_mv} { for the predicted number of sources to be found in a primary beam}). % 0.5/(18./25)*(1.4/8) degree
ngVLA will be able to output multiple tied-array voltage streams for simultaneous observations of  multiple sources within the FoV \citep{ngvla_tec}. %, which will enable the latest astrometrical techniques to be used at the lower frequencies.
In general at higher frequencies ($>$8\,GHz) astrometry would be carried out with source switching, and the telescopes are being designed with the capability to switch rapidly between different lines of sight (4\degr{} in 10\,sec). 
With these capabilities ngVLA is highly suitable for observations using the latest astrometrical techniques, such as \MV{} as discussed in Sect.~\ref{sec:methods_mv}.

Astrometry at high frequencies using conventional techniques is extremely challenging, due to the fast fluctuations of water vapour in the troposphere. 
Water Vapour Radiometer (WVR) systems, which have proved invaluable on the ALMA site, will monitor the precipitable water vapour content in
the direction of the antenna pointings to improve the estimates of the tropospheric model contributions. 
Nevertheless there are significant risks in depending entirely on this approach. 
The WVR for ngVLA will be using the less sensitive 22GHz line, rather than that at 183GHz used for ALMA, and this tends to saturate and has a spotted history \citep{clark_15_memo10}. 

Innovative methods to correct for the tropospheric effects that use multi-frequency observations, i.e. Source Frequency Phase Referencing (SFPR),  are described in detail in Sect.~\ref{sec:methods_sfpr}. 
The ngVLA feed translator enables very fast switched-frequency observations (significantly faster than currently possible with the VLBA). 
However, this is a sub-optimal solution; the best solution is to have simultaneous observations at multiple frequencies. These provide enhanced coherence at the highest frequencies and more robust and straight-forward operations.
In the design of a Gregorian feed the focal length is very short; this in
turn makes it very difficult to introduce complex optical paths for simultaneous frequency
systems such as used on the KVN. %\citep{sslee_14,han_08,han_13,han_17} 
Nevertheless, technical solutions have
been proposed; also we have carried out simulation studies to
investigate the performance costs of using fast frequency switching in
comparison to simultaneous observations at multiple frequency
bands. The outcome is described in detail in  Sect.~\ref{sec:study_ngvla}.

%\subsubsection{the Korean VLBI Network and Friends}
Some of the high-frequency considerations for ngVLA are being tested on the KVN, which is the first dedicated full-time mm-wavelength VLBI (mm-VLBI)
array and was commissioned in 2009. 
In this  respect, we consider the KVN as a pathfinder for ngVLA.
Currently KVN consists of three 21-m diameter antennas across 
South Korea, hosted at the campus of the Universities of
Yonsei, Ulsan and Tamna \citep{sslee_14}, with a maximum baseline
length of $\sim$500\,km. Each antenna is equipped with an
innovative quasi optical design of mirrors and low pass frequency
filters that enables simultaneous observations at four frequencies:
%multi-band receiver able to simultaneously observe at 
22, 43, 86 and 130\,GHz \citep{han_13}. 
This design is a technological solution for effective tropospheric calibration and enables astrometry at frequencies three times higher than that achievable previously, using measurements at a lower frequency band. A detailed description of the method is presented in Sect.~\ref{sec:methods_sfpr}.

\subsection{mm and sub-mm radio astrometry in the EHT era}\label{sec:new_alma}
\typeout{This section will cover the parameters of (sub)mm-VLBI, the science goals, and the challenges there-in.}

The latest of the new global radio astronomy facilities to be opened is the Atacama Large Millimeter/submillimeter Array
(ALMA).\footnote{\url{https://www.almaobservatory.org}} This consists 
{ of extremely high precision antennas, with 54 having diameters of 12\,m and another 12 of 7\,m,} in the Atacama desert (Chile), at about 5000\,m above sea-level. The frequency coverage is from 84 up to 950\,GHz, with plans for frequencies as low as 35\,GHz.
The antennas can be phased up to provide the equivalent sensitivity to that of a 85-m diameter single dish, from which the tied-array voltage signal can be recorded for VLBI observation.
The ultra-sensitive phased-up ALMA is providing the corner-stone for VLBI at the highest frequencies ever performed,  at hundreds of GHz. At these frequencies, where the atmospheric coherence times are
extremely short, ALMA has allowed for a huge leap in performance for the Event Horizon Telescope (EHT) \citep{eht}. 
The chief design goal of the EHT is the imaging of the shadow of a black hole, for which both frequencies above 200GHz and spatial resolutions greater than 40\uas\ are required. The former is to overcome the intrinsic scattering around these sources and the latter is to provide the resolution to separate the black hole shadow from the bright disk of in-falling matter \citep{eht_bh}.
So far EHT has exclusively focused on making self-calibrated images, but when the small number of sufficiently strong sources are fully explored there will be a demand for more sensitive observations. 
Greater sensitivity  can be achieved by longer coherent observations, such as provided by PR-type techniques regularly used at lower frequencies. 

A conventional PR experiment at hundreds of GHz is probably impossible, as the atmospheric coherence times at these wavelengths are extremely short. 
Nevertheless an alternative innovative method has been demonstrated up to 130GHz, the highest frequency with KVN, and we foresee no issues in application to higher frequencies, such as those relevant to EHT; ALMA has used a similar technique
(``band-to-band'') on their internal baselines \citep{Asaki_2020}. The approach of SFPR is discussed in Sect.~\ref{sec:methods_sfpr}.
%As long as resonance lines are avoided. See A20
Early discussion about next-generation EHT (ngEHT) is considering such a plan. 

\subsection{Space missions: astrometry from above the atmosphere and extraterrestrial baselines}\label{sec:space}

Millimetron is a cm to sub-mm Russian space mission which envisages a radio telescope located at the L2 Lagrangian point (and later moving to a highly elliptical orbit), observing at frequency ranges from 15\,GHz to { 15THz} \citep{millimetron_18}. 
The space platform would host a single dish of 10m diameter, and will alternate operations between single dish and VLBI observations (at frequencies between 30 and 690\,GHz), jointly with ground telescopes including the EHT antennas.
The long baselines to the satellite telescope will provide a huge leap in the achieved angular resolution of the images and potentially astrometry, compared to ground arrays.
Calibration of this instrument for VLBI will be a huge challenge, and therefore the capability to observe at multiple frequencies simultaneously is part of the design; this is also relevant to enable astrometry.
For terrestrial VLBI the antenna positions are usually known to better than a cm, but this is not the case for the orbit error in Space-VLBI and this becomes the dominant source of astrometric error. VSOP had orbit errors of about 5\,m \citep{porcas_00} and those for RadioAstron were significantly worse. 
It has been very hard to meet the requirements for astrometry with the satellite antennas \citep{vsop-2} \citep[with the exception of two particular cases:][]{porcas_00,guirado_01} thus new approaches for astrometry are required. 
Details on the benefits of multi-frequency observations for astrometric calibration of mm-VLBI using SFPR, such as the removal of orbit errors, are described in Sect.~\ref{sec:methods_sfpr} and in \citet{rioja_11b}.

Another high-frequency VLBI mission not bound to Earth is the proposed Japanese balloon-VLBI mission \citep{doi_19-balloon}, in which the telescope is mounted on a balloon-borne platform. 
Here the purpose is to observe from the stratosphere, avoiding the tropospheric contributions. The baselines are not significantly longer than those of the global Earth-bound stations, and will have significant positional uncertainties. 
Nevertheless, if such a system was fitted with a multi-frequency receiver it would be suitable for astrometric observations, in the same manner as Millimetron.

Cosmic Microscope  is a low frequency dual element space-VLBI concept by the Shanghai
Astronomical Observatory of the Chinese Academy of Sciences,
for state-of-the-art high resolutions studies at low frequencies.
The Chinese space-VLBI mission is still in an exploratory phase, but has ambitious goals including astrometric capabilities.
%It consists of two 30-m-diameter radio telescopes with operational frequency bands at 30, 74, 330 and 1670 MHz,
This Cosmic Microscope mission consists of two space-based 30m
radio-telescopes in highly elliptical orbits with apogee heights of
90,000 and 60,000\,km respectively,
 %The perigee is 2,000\,km 
with operational frequency bands at 30, 74, 330 and 1,670 MHz \citep{an_svlbi_18}. 
The modes of operation will comprise: space-ground VLBI  with the most
sensitive ground infrastructure, such as the SKA, FAST and Arecibo telescopes, resulting in a powerful combination of high sensitivity and extremely long
baselines for astrometric studies;
space-space single baseline VLBI, resulting in observations free of 
% the dominant source of astrometric errors in ground observations arising from 
atmospheric propagation medium effects (but not from orbit positional errors); and single dish mode, to monitor the sky
for transient events that will be followed up with VLBI. 
The astrometric measurements will be carried out using the recently developed \MV\ \citep{rioja_17} technique, with multiple in-beam calibrators (Sect.~\ref{sec:methods_mv}); this technique has important
benefits for space-VLBI since, by correcting for the orbit errors \citep{dodson_13} as well as  atmospheric contributions, it enables astrometry.

\subsection{The VLBI Global Observing System for Geodesy and Astrometry}
\label{sec:vgos}

Accurate geodetic data is vital for astrometry, through the better modelling of source coordinates with the International Celestial Reference Frame (ICRF), the Earth Orientation Parameter (EOP) and telescope coordinates with the International Terrestrial Reference Frame, so astrometry and geodesy improve hand in hand.
The next-generation instrument for geodesy was proposed by the International VLBI Service for Geodesy and Astrometry and is known as the VLBI Global Observing System (VGOS). 
VGOS is designed to deliver an order of magnitude improvement in the 
precision of the delay observable, which propagates into a similar improvement in the geodetic outcomes, amongst these the
quality of the ICRF. 
This comprises a dedicated VLBI network, rather than a single station for VLBI. 
In the last decade it has steadily worked forwards from proposal to delivery, and now the concept and important demonstrations are established. 
%https://cddis.nasa.gov/lw21/docs/2018/papers/Session1\_Wakasugi\_paper.pdf \\
%The International VLBI Service for Geodesy and Astrometry (IVS) has been working towards establishing the VLBI Global Observing System (VGOS) to contribute to the Global Geodetic Observing System (GGOS) by providing more accurate, reliable and continuous geodetic VLBI data products. The operation for geodetic VLBI has undergone an end-to-end redesign, 
%which will lead to an order of magnitude increase in the achieved accuracy of astrometric and geodetic parameters.
To provide an order of magnitude improvement the requirements are: observations of many sources with-in the coherence time; rapid sky coverage to separate systematic terms; and sufficient sensitivity to be able to detect sources with flux densities greater than 250\,mJy, in that time. 
Thus the system is based around fast slewing ($\sim$10\degr/sec), 12-m class antennas (System Equivalent Flux Density (SEFD) of $\sim$2,500Jy), capable of recording broad bandwidths from a 2--14\,GHz continuous frequency band, and data processing technology at a high data rate of up to 16 Gbps (SNR of 20 in 10\,sec).
A number of new telescopes conforming to the VGOS specifications have been built or are planned globally. %The expansion of network and 
These, with improvements in analysis procedures and the optimisation of scheduling, should provide residual pathlength errors of the order of 1\,mm, or a delay precision of four pico-seconds.

The unambiguous `group delay', or delay across the observed bandwidth, has traditionally been the prime observable for geodesy and absolute astrometry since the 1970's \citep[see discussions and references in][]{counselman_76}.
%Counselman 1973, see references in Counselman 76 review; 
%Significant reduction in the uncertainty of the measured group delay over time has come from engineering improvements, from 0.1 nsec to 10 psec.
%The group delay has traditionally been the prime observable for geodesy and absolute astrometry/ 
For VGOS these will come from observations of four 1-GHz bands spread across the frequency coverage.
The demonstrations have shown that the group delay uncertainties, with the improved sampling and bandwidth, are an order of magnitude less than those of the previous products and that the system calibration and stability is sufficient to meet the requirements \citep{neill_18_vgos}. 
%These group delays are used to determine positions to accuracies of about a cm and tenths of mas for the station and celestial coordinates respectively; another order of magnitude improvement is expected with VGOS.
%The continuous long-term campaign observations have resulted in a quasi inertial celestial reference frame (currently ICRF3 \citep{charlot_icrf3}) defined by the precise positions of hundreds of extragalactic sources, mainly quasars, spread across the sky, and is discussed in Sect.~\ref{sec:applications_icrf}.  
%VLBI absolute astrometry relates to measuring positions of extragalactic radio sources with respect to each other in a global sense, i.e., over the entire celestial sphere.
%Those positions constitute a grid of fiducial points on the sky which define a reference frame against which ultra-precise relative astrometric measurements may be accomplished based on differential techniques such as those reported in this review paper.

%It is notable that geodesy is tending towards astrometry, in the proposed %increased use of `phase delay' analysis, and astrometry is tending towards %geodesy, in the use of wide bandwidths and wide angular separations. 
%The known technical challenges for VGOS, of designing wide bandwidth instruments and sensitive schedules, are undoubtedly achievable. The unknown challenges, such as including (potentially wavelength dependent) structural effects, will determine the ultimate geodetic accuracies. It is in these aspects, we believe, where astrometry and geodesy will become symbiotic and mutually supportive. 

\subsection{Cross-over between radio and optical astrometry in the GAIA era}

In the era of \gaia{} some aspects of radio astrometry will change considerably, therefore we quickly review \gaia{} and its impacts. 
The ESO satellite mission \gaia{} was launched in 2013, and is surveying the visible sky from the Earth-Sun Lagrangian point. It is expected to produce a catalogue of parallax and proper motions for one billion stars and AGNs with magnitudes brighter than 20. The final goal for astrometric precision is 7\uas{} for $\sim$10$^{\rm th}$ magnitude stars, 20\uas{} at 15$^{\rm th}$ magnitude, and 200\uas{} at 20$^{\rm th}$ magnitude
\citep{lindegren_18}.
Additionally, \gaia{} performs photometry and radial-velocity measurements on the brighter targets. 
The observations are relative between two telescopes pointing 106.5\degr{} apart. All objects should be observed about seventy times in the lifetime of the mission, which is currently about half way through.
The point spread function is about 300mas, and the astrometric position is derived from a precise fit to the optical centroid.

%We will address the synergies. 
The billion optical sources that \gaia\ \citep{gaia_dr2} will survey will include approximately 550,000 quasars  \citep{gaia_dr2d}, which is better than 10 sources per sq. deg., so will be able to form a high precision optical inertial reference frame that can be compared to the radio-based ICRF. % \citep{charlot_icrf3}. 
%{\bf skipped ma_icrf_98, Ma and jacobs 18 icrf3_18.} 
Given the large number of sources that \gaia\ will measure, the derived reference frame will be of an extremely high quality. %
% One of the greatest weaknesses in the ICRF is that it is based on a small number of targets, particularly in the Southern Hemisphere \citep{titov_07}. %MENTION VGOS.... what is ICRF_SHEM???
% The \gaia\ observations will resolve this issue { providing a link between the Radio-measured Reference Frame and that of \gaia\, based on quasar positions. \textbf{MORE EXPAND HERE} \citep{lindegren_20}.
%}
One of the greatest weaknesses in the ICRF is that it is based on a small number of targets, particularly in the Southern Hemisphere \citep{titov_07}. 
The \gaia\ observations will resolve this issue { providing a link between the Radio-measured Reference Frame and that of \gaia\, based on quasar positions.
\citet{lindegren_20} discusses the important distinction between the `faint' \gaia{} reference frame, which would include the AGN's that form the reference frame and the `bright' reference frame, made up on the whole with stars from the Galactic population.
There are instrumental difficulties directly connecting these two frames, and relative VLBI-astrometry of radio stars in the latter, referenced to the ICRF, will provide a crucial direct link between the two \gaia{} data products.
}
Additionally, given the limited life time of any space mission, the long term radio-derived ICRF provides a vital cross-reference. The ICRF is based on observations from the 1980's onwards, albeit with continuous improvements, and thus is robust against long term systematic drifts such as the rotation of the frame. 
Currently, on going improvements of the VLBI geodetic and astrometric observations within VGOS are expected to result in an order of magnitude improvement, as discussed in Sect.~\ref{sec:vgos}. 

The \gaia\ Data Release 2 included mapping of the Galactic plane
stellar kinematics \citep{gaia_dr2e} with 3.5 million measurements of
proper motion and parallax of stars down to 12$^{\rm th}$ magnitude
from the local Galactic quadrant. I.e., 5 to 13kpc from the Galactic Centre, out to $\pm$2kpc above and below the plane \citep[see Fig. 9][]{gaia_dr2e}.
This allows for a phenomenally rich and detailed reconstruction of the Galactic dynamics and history of the optical stellar population. However, the global parameters for the disk rotation were still taken from \citet{reid_14}, underlining the importance of independent radio astrometric observations.
%
% \sout{VLBI astrometric observations of \wat\ masers in High Mass Star Forming (HMSF) regions have provided a huge leap in our understanding of our Galaxy cite{reid 19,honma 15}.
% The weakness in this analysis lies in that the targets are limited to one subset of Galactic dynamical tracers. The number of classes of stellar sources that \gaia\ will probe provides the solution for these limitations.}
% The radio measurements span a greater extent of the Galactic disk, and are immune to the effects of dust extinction that tends to obscure some of the most interesting regions.
% \gaia, with the later data releases, will be able to improve its reach and should be able to independently derive the Galactic rotation parameters.
%
Hitherto the research areas of geodesy and astrometry have been  dominated by radio VLBI. Clearly the role of the radio investigations will need to adapt to the new landscape forged by \gaia, but given the respective strengths and weaknesses their complementary nature will only benefit both.

\section{Next-generation astrometric calibration methods}\label{sec:methods_new}

There have been a number of new methods developed over the last decade that have had a significant impact on the field of precise astrometry,
as they have provided a breakthrough in its application across the radio spectrum compared to previous methods. 
They are coming to fruition now with wide spread usage, being applied to KSPs of existing instruments, and driving the planning of the next-generation instruments.

In the low frequency regime, \MV{} effectively reduces the impact of the large residual direction-dependent ionospheric effects between the target and the reference sources, and has a wide applicability. 
Section~\ref{sec:methods_mv} describes the basis for \MV{} and the astrometric outcomes.
%delivers state-of-the-art precise astrometry with larger angular separations, and has a wide usability.
% { \MV{} astrometry will provide the thermal noise limit in observations with current instruments and out-perform state-of-the-art precise astrometry, using calibrators at large angular separations. 
% %With next generation instruments, the prospects for ultra precise astrometry,  with an order-of-magnitude improvement, are expected/feasible with next-generation instruments and MultiView.
% %Will discuss the feasibility of 
% The prospects are promising for ultra precise astrometry,  that is an order-of-magnitude reduction in the errors, from observations with next-generation instruments and with the \MV{} analysis technique, thus overcoming the barriers for precise astrometry at low frequencies.}

In the high frequency regime, new approaches such as Source Frequency Phase Referencing (SFPR), and variations of it, breakthrough the upper frequency threshold imposed by the very fast tropospheric fluctuations and enables astrometry at much higher frequencies. 
%performs calibration at a more tractable lower reference frequency and applies these to the higher target frequency. 
Section~\ref{sec:methods_sfpr} describes the basis for SFPR and the astrometric outcomes.
% { These have been demonstrated to provide precise astrometry at frequencies a factor of three times higher than that previously possible, which is the highest frequency of the KVN. 
% The prospects are promising for the method to continue to work at even higher frequencies.}

%In both cases they provide an alternative route for precise astrometry in space-VLBI at all frequency regimes.
In both cases they open the prospects for precise astrometry at all frequency regimes, including with space-VLBI.

\subsection{MultiView astrometric method} % -  best for low frequencies, all spectra; also high freqs: @ low elev and space-VLBI }
\label{sec:methods_mv}

%{{\bf Description and bib reference:} \\}
The \MV\ calibration method \citep{rioja_17} offers the potential to achieve `universal' high precision astrometry, 
%at all frequency regimes, 
including in the low frequency regime dominated by ionospheric disturbances, where other PR methods result in large errors. % (where there are strong restrictions  on the source separation in PR),  and 
%Additionally it is widely applicable.
%
These arise from the spatial structure in the distribution of electrons in the ionosphere above each antenna, which introduce direction-dependent phase changes resulting in quasi-stationary spatial phase gradients;
errors with similar signatures arise from residual tropospheric and geometric errors, albeit with smaller impact at the lower frequencies. 

The strength of the \MV{} method is in that it uses an appropriate
combination of the observations towards multiple reference calibrator sources, at least three,
surrounding the target, to measure those phase gradients and its temporal variation, and thus calibrate the target observations accurately. 
See Table~\ref{tab:sched} for an indicative observing schedule.
All sources are observed at the same frequency. \MV\ calibration is ideally carried out in the visibility domain
and uses  2D linear interpolations of the residual phases sampled along the calibrator directions. 
Effectively, this is equivalent to PR with a source pair whose relative separation is $\Delta\theta_{AB} \sim$0 and therefore minimises the calibration errors. 
Instead, the \MV{} calibration errors arise from the deviations of the ionospheric disturbances from a planar fit, in its most basic implementation with three calibrators. %; if required, higher order fits  could be fitted.

%The observations of all sources are carried out at the same frequency.
%This strategy results in excellent error compensation, %of the order of in-beam PR or better, 
%even using distant calibrator sources that therefore place less stringent constraints for the angular separation and increases the applicability.
Our demonstration of \MV{} astrometry using VLBA observations at L-band \citep{rioja_17} 
%resulted in an optimum compensation of systematic ionospheric errors, 
using three calibrators with angular separations between 2 and 6 degrees away from the target, reached the thermal noise error regime of $\sim$100\uas.
For comparison, this accuracy is equivalent to that achieved with in-beam PR 
%the equivalent conventional position error would correspond to PR 
observations with a calibrator separated by 10\arcmin{}. 
The \MV\ (MV) strategy results in excellent error compensation, %of the order of in-beam PR or better, 
even using distant calibrator sources that therefore place less stringent constraints for the angular separation and thus increases the applicability.

%{{\bf Differential phase equation; astrometric outcome; ambiguity issue} \\}

Following Eq.~\eqref{eq:terms_diff} for the differential phase observables, with $\nu_R = \nu_T = \nu$, a MV-calibrated dataset is formed 
with a linear combination of observations of N calibrators ($C_i$) that surround the target source (A):
%
%A MV-calibrated target dataset is formed with a linear combination of N calibrators ($C_i$) that are around the target source (A), following Eq.~\eqref{eq:terms_diff}:
%
\begin{equation} \label{eq:diff_mv} \begin{split}
   \phi_{T} - \mathcal{R}*\phi_R&= \phi_A (t_o, \nu) - \sum_{i=1}^{N} \alpha_i *\phi_{C_i} (t_i, \nu) \\
   & = (\phi_{\rm A,pos} -  \phi_{\rm C_v,pos}) \pm \sigma\phi^{MV} \pm \sigma\phi_{\rm thermal} +  2\pi (n_A - \Sigma_i \alpha_i n_i),\\
      & \quad {\rm\ with\ \alpha_i\ real}
%  = (\phi_{A, \rm str} +) \phi_{A, \rm pos} - \sum_{i=1}^{N} \alpha_i* \phi_{C_i} + \sigma\phi^{MV} + \sigma\phi_{\rm thermal} +  2\pi \Sigma_i \alpha_i n_i^\prime
%\sigma\phi^{MV} &= \sum_{i=1}^{N} \alpha_i *\sigma_i \phi
  \end{split}
\end{equation}
where $\alpha_i$ are the weights used for the 2D linear spatial interpolation of the residual calibrator phases $\phi_{C_i} (t_i,\nu) + 2\pi n_i$ from the scans on the N calibrators 
%the residual phases $\phi_{C_i}$ measured in the direction of the calibrators $C_i$
to the direction and scan time of observations of the target source A.  $\alpha_i$ are real and their values depend on the distribution of the sources in the sky.
{ The values for $\alpha_i$ used in \citet{dodson_17}, for example, are simply the linear fractional weights between the on-sky source positions; more complicated functions could be fitted, but are unlikely to be significantly better except where angular separations or the phase surface curvature are extremely large.}
The sum of $\alpha_i$  will be one if the calibrators surround the target, so the errors are not inflated.
{ Beyond the preference for the target to lie within the locations of the calibrators on the sky, which ensures the corrections are interpolated rather than extrapolated, there are no particular limitations on their arrangement, other than to note that degenerate configurations (e.g., all in a line) will not allow accurate estimates for the phases as a function of the orthogonal direction.}
If all sources are observed simultaneously (dubbed in-beam \MV) only the spatial interpolation is required. 
%, if the solution to the target is interpolated (but will be greater than one if extrapolated).
\MV\ uses the FT-inversion approach to resolve the ambiguity issue, hence
it is important to ``phase connect'' the sequence of N calibrator phases prior to the interpolation, whilst there is no such requirement for the target.

%where $\alpha_i$ are the weights used for the 2D linear interpolation of the residual calibrator phases $\phi_{C_i}$
%to calibrate the measured target phases $\phi_A$, 
%in the direction of the target source A, 
%as a function of time.
%$\sigma_i \phi$ are the individual $i$ error terms from Eq.~\eqref{eq:a07_terms}, for each of the N calibrators.
%
%Note, it is important that there are no untracked ambiguities in the reference source phases. That is, it is important that $n^\prime_i$ is zero. 
%The values of $\alpha_i$ depend on the distribution of the sources in the sky and the sum of $\alpha_i$  will be one, if the solution to the target is interpolated (but will be greater than one if extrapolated).

The MV-calibrated target dataset is Fourier inverted and deconvolved to yield a synthesis image of the target source, which conveys a measure of its position relative 
%which is astrometrically referenced 
to a virtual point in the sky (i.e., not the position of a particular object) determined by the ensemble of calibrators; 
this is indicated by the term $(\phi_{\rm A,pos} -  \phi_{\rm C_v,pos})$ in Eq.~\eqref{eq:diff_mv}.
%
%The astrometric outcome is retained in the term $(\phi_{\rm A,pos} -  \phi_{\rm C_v,pos})$ in \ref{eq:diff_mv}, where $C_v$ corresponds to a virtual po
%which conveys a measure of the target position relative to a virtual point in the sky, which does not correspond to the position of a particular object.
We note that, for comparison, the measured target positions using PR methods are tied to the assumed position of the corresponding (single) calibrator. 
Nevertheless, as long as the calibrator sources provide good fiducial points (i.e., are stationary), this virtual point is also stationary and any changes between the astrometric measurements 
in multi-epoch observations trace the motion of the target in both \MV\ and PR. 
$\sigma\phi^{MV}$ are the phase calibration errors using \MV\ and are quantified below, with all other terms as before.
%measurement noise related to sensitivity. 
%It is crucial (to ensure) that the multi-calibrator phase fitting strategy is not affected by/is free of phase ambiguity issues. {\bf OR} 
%Special considerations are required to avoid inherent phase ambiguity issues in the multi-calibrator phase fitting strategy.
%, as these are included with non-integer weights.
%
As mentioned above, special considerations are required to avoid inherent phase ambiguity issues in the multi-calibrator phase fitting strategy.
\MV, as implemented in \citet{rioja_17}, includes an automatic
ambiguity check mechanism, based on the approach used in UVPAP \citep{sergio}, to avoid these issues. % in the multi-calibrator phase interpolation. 
Those are expected to be increasingly relevant at lower frequencies ($<$5\,GHz) or in cases when the calibrator source positions are poorly known.
%Moreover, in general, at lower frequencies ($<$5\,GHz) one is likely to need to account for phase ambiguities. 
%In \citet{sergio} we introduced aspects the analysis methods of UVPAP, in that we also searched over multiple phase ambiguities. 
In the demonstration paper of \MV\ \citep{rioja_17}, at 1.7\,GHz, we found that, typically, a few antennas over the VLBA array had one turn of phase to be corrected for, which is consistent with 6TECU residuals.

$\sigma\phi^{MV}$ is given by a linear combination of the calibration errors for the N target-calibrator pairs $(\sigma_i \phi)$, each following Eq.~\eqref{eq:a07_terms}; i.e., $\sigma\phi^{MV} = \sum_{i=1}^{N} \alpha_i *\sigma_i \phi$,
using $T_{\rm swt,tro}=T_{\rm swt,ion}=T_{\rm swt}$, where $T_{\rm swt}$ is the duty cycle of the observations, and 
$\Delta \theta_{\rm tro}=\Delta \theta_{\rm ion} = \sum_{i=1}^{N} \alpha_i\hat{s}_i -\hat{s}_{\rm true} \sim 0$, for the propagation effects. The latter applies under the assumption 
of  the phase errors above a telescope having a planar spatial structure, in this case the static residual errors are fully compensated both for tropospheric and ionospheric components, including  the case of 
low elevation observations that result in significant PR errors. %, even at the higher frequencies. 
This assumption is supported by empirical measurements of the small contribution from fine scale structure or ``roughness'' of the atmospheric disturbances after the subtraction of the low-order fit; these studies are discussed in Sect.~\ref{sec:study_mv}. 
In the case of simultaneous observations of the target and calibrator sources, so called in-beam \MV\,  dynamical terms are largely cancelled as well. Thus we find the predicted phase errors for \MV{}, using the terms from Eq.~\eqref{eq:a07_terms}, to be:
%({ As before, this is predicated on that the cycle time  ($t-t'$) is smaller than the atmospheric variation time.)}
%That is, combining Eqs.~\eqref{eq:a07_terms} and \ref{eq:diff_mv}:
%
\begin{equation*}
  \begin{split} \label{eq:terms_mv}
\sigma\phi^{MV}_{\rm sta, tro} &= \Sigma\alpha_i \sigma_i \phi_{\rm sta, tro} \approx0\\[6pt]
\sigma\phi^{MV}_{\rm sta, ion} &= \Sigma\alpha_i \sigma_i \phi_{\rm sta, ion} \approx0\\[6pt]
\sigma\phi^{MV}_{\rm dyn, tro} &= \Sigma\alpha_i \sigma_i \phi_{\rm  dyn, tro} \approx0\\[6pt] %F(\nu,T_{\rm swt}^{5/6})  \\ 
%7.1  C_w  (\nu/8{\rm GHz}) (\sec(Z_g)/\sec(45\degr)^{1/2} \times [T_{swt,tro}/60s]^{5/6}\\
\sigma\phi^{MV}_{\rm dyn, ion} &= \Sigma\alpha_i \sigma_i \phi_{\rm  dyn, ion} \approx0 %F(\nu^{-1},T_{\rm swt}^{5/6}) \\ 
%\Sigma\lambda \sigma\phi_{\rm dyn, ion} &=  0.95 (\nu/8{\rm GHz})^{-1} (\sec(Z_i)/\sec(45\degr)^{1/2} \times [T_{swt,ion}/60s)]^{5/6}\\
  \end{split}
\end{equation*}
%An additional small term arising from the fine scale structure of the residual phase surface or inherent 'roughness', after the subtraction of the low-order fit to the spatial atmospheric disturbances, remains. 
%Empirical studies to characterise the atmospheric spatial structure are presented in Sect.~\ref{sec:study_mv}.
%{\bf OR} This is discussed further in Section 
Moreover \MV\ also calibrates the residual geometric errors,
such as antenna coordinate errors, which result in spatial phase gradients above the antennas as well, that is: $\sigma\phi^{MV}_{\rm geo} \sim$0. { This would include `tip-tilt' errors in the alignment of a multi-beam feed such as that on FAST.}
%
% { This is relevant for space-VLBI, a regime where inaccuracies in the orbit determination are a dominant source of errors which have prevented astrometric measurements, see Sect.~\ref{sec:space}.} %(with the exception of two pairs of very nearby sources, using in-beam PR references...Porcas.. Guirado..).
Therefore, \MV\ provides a very effective mitigation of the dominant sources of calibration errors that remain with other methods, arising from propagation medium effects and geometry errors, 
%comparable to the thermal noise error for very sensitive instruments, 
leading to astrometry accuracies in the thermal noise regime  across the spectrum. 

Table~\ref{tab:inbeam_mv} lists the estimated systematic astrometric errors $\sigma\Delta\theta^{MV}$ (Col. 3), from the propagation of \MV\ residual atmospheric errors, taking into account the empirical studies (Sect.~\ref{sec:study_mv}), at a range of frequencies.
For comparison, these are about one order of magnitude smaller than those for in-beam PR with a calibrator 10\arcmin{} away (PR$_{10^\prime}$) (see Sect.~\ref{sec:future} and Fig.~\ref{fig:astro_history}, for the direct comparison between the outcomes of both methods). % between the techniques).
As a result, \MV\ astrometric accuracy easily reaches the thermal limit 
%($\sim\theta_{\rm beam}$/DR) 
with current instruments,
%(i.e., $\sigma\Delta\theta^{MV} \ll \sigma\Delta\theta_{\rm thermal}$), 
%with $\sigma^{MV} \ll \sigma_{thermal}$; 
which corresponds to $\sim$100\uas{} at 1.6\,GHz, with DR of 100:1; that is $\sigma\Delta\theta^{MV} \ll \sigma\Delta\theta_{\rm thermal}$.
%\sout{; 
In comparison in-beam PR$_{10'}$ results in a similar astrometric performance, but where 
%$\sigma^{PR} \sim \sigma_{thermal}$. \\} with current instruments.}
$\sigma\Delta\theta^{PR_{10'}} \sim \sigma\Delta\theta_{\rm thermal}$.  % (with increasing uncertainties towards lower frequencies). %% Beacuse true of both \\
\MV\ systematic calibration errors are comparable to the thermal noise levels from the next-generation instruments (i.e., DR of 1000:1),
%in this case, the accuracy of \MV\ astrometric measurements are those listed in Table~\ref{tab:inbeam_mv} column 3, 
for example 6\uas{} at 1.6\,GHz (Table~\ref{tab:inbeam_mv} Col. 3).
The prospect is for \MV\ to achieve ultra-precise \uas-level astrometry at frequencies where today's measurements are carried out, with SKA and ngVLA, and enable precise astrometry at the much lower frequencies with SKA. 
% { Note that at these reduced error levels, contributions that are neglected in the astrometric analysis with current instruments will become relevant. In this respect, issues arising from the definition and stability of reference points (Sect.~\ref{sec:ref_pts}) have to be considered, in multi-epoch studies.}
%
%{\it  Frequency Scope of application:}
% { The biggest impact of \MV\ calibration is at low frequencies from removing the large systematic ionospheric phase gradients across the sky in the differential observables. Nevertheless \MV\ is  also relevant at higher frequencies, particularly %in observations where the differential residuals are large; for example 
% at low elevations, for which the $\Delta \theta \sec{Z} \tan{Z}$ dilution factor for residual zenith tropospheric path length, relevant for PR methods, can increase by an order of magnitude.}
%
%{\it   Next gen instruments:}
 Ultra precise astrometry is one of the most innovative outcomes of VLBI with the next-generation instruments and some example projects are presented in Sect.~\ref{sec:future}.
 %, and its feasibility derived from empirical ionospheric measurements with SKA-Low pathfinders discussed in Sect.~\ref{sec:study_mv}. 
  %The prospects for/feasibility of ultra precise astrometry using \MV\ are presented in Sect.~\ref{sec:future} using an empirical dataset of ionospheric measurements with SKA-Low pathfinders, discussed in Sect.~\ref{sec:study_mv}.
  
%A related issue is for the requirements for multiple-pixel that enable simultaneous observations such as on the number of 
Recently suitable technologies for forming simultaneous multiple beams have been developed. We discuss 
multiple tied-array beams for antenna arrays in Sect.~\ref{sec:tab} and multiple beams for large telescopes in Sect.~\ref{sec:mbeams} and their implementation on the next-generation instruments in Sect.~\ref{sec:next}. % ($T_{\rm swt}=0)$.
Observations of at least three calibrators surrounding the target is the minimum requirement,
%{for 2-D linear spatial interpolation to the direction of the target}, 
which requires $\ge$4 tied-array beams.
%;  when the positions of two calibrators and the target are aligned in the sky 1-D spatial interpolation is possible. %\citep{fomalont_03} %(Fomalont 2003). 
With more than the minimum number of calibrators one could over-fit the plane (or fit higher order surfaces) to reduce the measurement errors. 
%Including a larger number of calibrators allows the over-fitting of the plane (or fitting a surface with a higher order polynomial) to reduce the measurement errors. 
However, the main advantage, in general, would be in the identification and removal of unsTable~reference points (Sect.~\ref{sec:ref_pts}). 
It is such systematic errors that we believe will become a new dominant source of errors and determine the ultimate astrometric limit with the next-generation instruments.
 
MultiView is also relevant for enabling highest precision astrometry from the extra-terrestrial baselines with space-VLBI, where the large orbit errors have traditionally prevented  astrometry using PR methods (see Sect.~\ref{sec:space}). %(with the exception of two cases, \citet{porcas_00,guirado_01}). % Porcas\&Rioja, Guirado. 
Instead, \MV\ results in an effective calibration of geometric errors, such as the position errors of the orbiting antenna, along with the propagation medium effects.
\citet{dodson_13} carried out \MV{} simulation studies with realistic atmospheres and orbit errors, to demonstrate the feasibility of \MV\ for Space-VLBI astrometry and found 3$\pm$1\uas{} systematic errors in the best cases, at 1.6\,GHz.  
%the outcomees for Space-VLBI astrometric outcomes are estimated in \citet{dodson_13}.

Other related implementations are the recent \citet{reid_17} demonstration of \MV{} spatial interpolation in the image domain.
%with observations that alternate between  target and reference pairs, cycling through these pairs in blocks.
Earlier approaches were: `Cluster-Cluster' VLBI, proposed in the 1990's \citep{rioja_97, rioja_02}, similar slewing strategies \citep{fomalont_02,atmca},  and the bi-gradient method \citep{doi_06}. None of the latter have had great usage, perhaps because they lacked methods for correcting for ambiguities.
Future implementations of MSSC plan to fit higher order functions to the  multiple calibrators used in that method  (Sect.~\ref{sec:hdm}), which would be similar.

%{\bf END OF MV HERE?}\\
%Other similar approaches to \MV{} have been proposed in the past: `Cluster-Cluster' VLBI was proposed in the 1990's \citep{rioja_97, rioja_02}, \citet{fomalont_02} discussed a simplification with just two calibrators aligned with the target, \citet{atmca} provided the means to interpolate the phase from three or even two references to a target (developed for the experiment to measure the speed of gravity \citep{fomalont_03}), as does the bi-gradient method of \citet{doi_06}, and finally \citet{reid_17} is a demonstration of \MV{} spatial interpolation in the image domain with observations that alternate between  target and reference pairs, cycling through these pairs in blocks. 
%
%None of these have had great usage because they were not widely applicable. The fact that they lacked methods for correcting for ambiguities (except for the latter, which requires a particular observing schedule) may have contributed to this.
%Future implementations of MSSC are planning to fit higher order functions to the  multiple calibrators used in that method  (Sect.~\ref{sec:hdm}), which would be equivalent. 

\subsection{Source/Frequency Phase Referencing astrometric method} % (highest frequency regime)} 
\label{sec:methods_sfpr}

%{\it  (best with multi-freq simultaneous technology; 
%highlight KVN; pathfinder for ngVLA; for space VLBI too \\}

%The Source Frequency Phase Referencing (\SFPR) calibration method \citep{rioja_11a} offers the potential to achieve high precision astrometry in the high frequency regime 
%where the performance of PR methods is poor and limited. % ($\nu \ge 43$GHz).
%Thus it 
The Source Frequency Phase Referencing (\SFPR) calibration method \citep{rioja_11a}  provides a breakthrough for mm-VLBI astrometry beyond the scope of PR methods. %the upper frequency threshold of PR methods.
SFPR achieves an effective calibration of the residual fast tropospheric error fluctuations and the geometric errors, both with a non-dispersive nature, along with the dispersive ionospheric errors.
%errors are large. % and beyond the upper frequency threshold. 
%
%These errors have a non-dispersive nature, as for the residual geometric errors, which are also eliminated with this calibration method.
%{\bf introduce KVN here}\\
\SFPR\ is widely applicable with no upper frequency limit, if the instrument has the required capability.

%The upper frequency threshold/limit for VLBI astrometry has traditionally been limited to 43 GHz, from the fast tropospheric fluctuations.
%Recent developments 
The fast tropospheric fluctuations limit the coherence time and are the main challenge in VLBI observations at high frequencies, 
which combined with intrinsically lower source fluxes and higher instrumental noise limits the observations to stronger sources.
Astrometry has even more stringent considerations than imaging, therefore PR astrometric measurements have traditionally been limited to frequencies up to 43 GHz.

Multi-frequency calibration is an approach that has only recently begun to deliver on its promise for mm-VLBI, based on the longstanding recognition of the non-dispersive nature of the tropospheric propagation effects.
Strategies involving dual frequency observations, where the high frequency (i.e., target frequency $\nu^{\rm high}$) observations are calibrated using the scaled solutions from a lower (i.e., reference frequency $\nu^{\rm low}$)
and more amenable frequency  have been proposed in the past \citep{asaki_fpt_96,carilli_99,middelberg_05}. We refer to these as ``frequency phase transfer'' strategies  (FPT).
But it was not until the development of calibration methods that precisely accounted for all, non-dispersive and dispersive, error contributions that successful mm-VLBI high precision astrometry was achieved. %, using \SFPR\ and \MFPR.

In \SFPR\ the small and slow changing differential residual ionospheric dispersive error contributions,
%dispersive errors 
are mitigated with interleaving multi-frequency observations of the target (A) and a second source (B, reference source or calibrator); here, the angular separation between the two sources ($\Delta \theta_{AB}$) can be up to many degrees 
 and the duty cycle ($T_{\rm swt}$) up to many minutes. 
This is unlike the case for PR methods which require fast switching to match the fast tropospheric fluctuations.
We note that the SFPR mitigation applies to other dispersive contributions as well, such as instrumental terms.
See Table~\ref{tab:sched} for an indicative observing schedule.

A detailed description of the basis of SFPR and the first demonstration of the astrometric capability 
%using fast frequency switching in VLBA observations at 43/86 GHz in 2007 
was presented in \citet{rioja_11a}, using fast frequency switching VLBA observations at 43/86 GHz, of a pair of sources 10\degr{} apart. %, carried out in 2007.
%The VLBA Memo \#31 \citep{vlba_31} contains full details for the data reduction.
Superior performance comes from the technical solution adopted by the KVN, where the multi-frequency receivers \citep{han_08,han_13}  enable simultaneous observations at
four frequency bands (22, 43, 86, 130 GHz). This results in much improved error mitigation and
therefore allows application to much higher frequencies. \citet{rioja_15} describes a SFPR demonstration at the highest KVN band, at 130 GHz (see Fig. \ref{fig:sfpr130}).

% {{\bf jumping ahead?}
% Multi-Frequency Phase Referencing (\MFPR) is an alternative method that explicitly measures the residual dispersive terms on the target source itself, using dedicated blocks of observations spanning a range between 1.3 and 22-GHz. Hence this method does not require observations of a second source. It is fully described in \citet{dodson_17}, where a 21\uas{} core-shift between 22 and 43-GHz was measured, and briefly covered in the following Section. % \ref{sec:methods_new} in this paper.
% }

Following Eq.~\eqref{eq:terms_diff} for the differential phase observables, with $\nu_R = \nu^{\rm low}$, $\nu_T = \nu^{\rm high}$ and $R={\nu^{\rm high} \over \nu^{\rm low}}$, a SFPR-calibrated target dataset is formed with 
a linear combination of observations at the 
reference and the target frequencies ($\nu^{\rm low}$ and $\nu^{\rm high}$, respectively), and of the target and reference (or calibrator) sources (A and B, respectively) as:

%%{{\bf Differential phase equation: (plus describe astrometric outcome and ambiguity issue}\\}
%%{\bf what about using/extending this nomenclature for the MV and SFPR methods? }
%The SFPR-calibrated target dataset is formed with 
%%The differential residual phase Eq.~\eqref{eq:sfpr_resid} for SFPR 
%a linear combination of the phases from observations at the 
%reference and the target frequencies ($\nu^{\rm low}$ and $\nu^{\rm high}$, respectively), and of the target and reference (or calibrator) sources (A and B, respectively). See Eq.~\eqref{eq:sfpr_resid}.
%
\begin{equation}
  \begin{split}
\label{eq:sfpr_resid}
%$
\phi_{T} - \mathcal{R}*\phi_R &= \left(\phi_A (t, \nu^{\rm high}) - \mathcal{R}*\phi_A (t, \nu^{\rm low})\right)\\
&\quad - \left(\phi_B (t', \nu^{\rm high}\right)- \mathcal{R}*\phi_B \left(t', \nu^{\rm low})\right)\\ 
  &= \left(\phi^{\rm high}_{A_{\rm pos}} - \mathcal{R}*\phi^{\rm low}_{A_{\rm pos}}\right) - \left(\phi^{\rm high}_{B_{\rm pos}} - \mathcal{R}*\phi^{\rm low}_{B_{\rm pos}}\right) \pm \sigma\phi^{\rm SFPR}\\
  &\quad \pm \sigma\phi_{\rm thermal} + 2\pi \left(\mathcal{R} n^{\rm low}-n^{\rm high}\right)
  %,\, {\sc R real, n inte} \\ %, n$^\prime$ integer} % \iff \mathcal{R} \in integer
  %\delta\theta_A^{high-low} &={{(\phi_{A_{\rm pos}}^{\rm high} - %\mathcal{R}\phi_{A_{\rm pos}}^{\rm low})}\over{360^o}}\theta^{\rm high}_{\rm beam}
  \end{split}
\end{equation}
%
%where $\mathcal{R}$ stands for the frequency ratio ${\nu^{\rm high} \over \nu^low}$, 
where the superscripts `high' and `low' are for observations at $\nu^{\rm high}$ and $\nu^{\rm low}$, respectively.
$\sigma\phi^{\rm SFPR}$ stands for the SFPR calibration errors, and all other terms as for those in Eq.~\eqref{eq:terms_diff}.
The SFPR-calibrated dataset is Fourier inverted and deconvolved to yield a synthesis image of the target source at the target frequency ($\nu^{\rm high}$); the SFPR-map. 
The position offset of the target source relative to the SFPR-map center is a bona-fide astrometric measure of the combined spectral position-shifts (i.e., $\delta \theta^{\rm high-low}$) for sources A and B, between the observed frequencies (e.g., core-shifts in AGNs, emission from different molecular species or transitions, etc).
This astrometric outcome is retained in the terms $(\phi^{\rm high}_{A_{\rm pos}} - \mathcal{R}*\phi^{\rm low}_{A_{\rm pos}})$ and
$(\phi^{\rm high}_{B_{\rm pos}} - \mathcal{R}*\phi^{\rm low}_{B_{\rm pos}})$ in Eq.~\eqref{eq:sfpr_resid}.
In general, using multiple combinations of source pairs, it is possible to disentangle the individual contribution from each source as for PR measurements; for an example see the solution for the five sources scheduled in the observations for \citet{rioja_15}.
%relative ``core-shifts'' in the calibrator and target sources between the two observed frequencies. 
%This astrometric outcome is retained in the term "$(\phi^{\rm high}_{A_{pos}} - \mathcal{R}*\phi^{\rm low}_{A_{pos}})$ - 
%$(\phi^{\rm high}_{B_{pos}} - \mathcal{R}*\phi^{\rm low}_{B_{pos}})$" in Eq.~\eqref{eq:sfpr_resid}.
%which convey a measure of the relative "core-shifts" in the calibrator and target sources between the two observed frequencies. 

%The astrometric shift of source A between the high and low frequencies, $\delta\theta_A^{high-low}$, as measured from the map corresponds to:

%\begin{equation*}
%\delta\theta_A^{high-low} ={{(\phi_{A_{pos}}^{\rm high} - \mathcal{R}\phi_{A_{pos}}^{\rm low})}\over{360^o}}\theta^{\rm high}_{\rm beam}
%\end{equation*}

%The differential observable retains the signature of the combined spectral position-shift for the two sources, between both frequencies. 
Simultaneous multi-frequency observations greatly facilitate the phase connection, as the typical phase rate ($\simeq 10^{-13}$s/s) at these high frequencies ($\sim$100\,GHz) requires the frequency switching duty cycle to be significantly less than a minute.
If $\mathcal{R}$  is an integer, then $\mathcal{R}n$ is also an integer and the $2\pi$ phase ambiguity issue is implicitly dealt with in the Fourier mapping approach. For this reason it is highly recommendable to select integer frequency ratios. % whenever it is possible.
The special considerations for observations with non integer $\mathcal{R}$  values (particularly for unrelated maser species) are described in \citet{dodson_14}. 
In essence, 
%it is vital to solve for all ambiguities, so that $n$ is zero; 
this requires ensuring good a-priori model values (so that $n^{\rm low}$ is zero).
%{\bf case of non integer frequency ratio requires ... (Dodson+2014?}.

The direct outcomes of \SFPR\ are: i) inter-band astrometry (or $\lambda$-astrometry), for precise registration of images at different bands, 
ii) increased sensitivity, from the extension of the coherence times, which can be extended up to 
several hours at 130\,GHz \citep{rioja_15}, and, moreover 
%in (in the high frequency regime) a regime where (in addition) the sources are intrinsically weaker an the instrumental noise is higher.
iii) astrometry of the target source at $\nu^{\rm high}$ with respect to an external reference source in the sky, by also including PR observations between the pair of sources A and B 
at $\nu^{\rm low}$ \citep[as for the example in][]{dodson_14}.

%{{\bf Calibration errors:} \\}

The propagation of errors for SFPR is discussed in detail in \citet{rioja_11a}.  
Briefly, the magnitude of $\sigma\phi^{\rm SFPR}$, at the target frequency $\nu^{\rm high}$, can be estimated using the Eq.~\eqref{eq:a07_terms}, taking into account: i) the scaling by $\mathcal{R}$ and  $\mathcal{R}$-1/$\mathcal{R}$ of the tropospheric and ionospheric terms at $\nu^{\rm low}$, respectively, along with ii) $\Delta \theta_{\rm tro} = \delta \theta^{\rm high-low} \sim 0$ and $T_{\rm swt,trp} = 0$, iii)  $\Delta \theta_{\rm ion}=\Delta \theta_{AB}$, $T_{\rm swt,ion}=T_{\rm swt}$. That is:

\begin{equation*}
  \begin{split}
%\begin{tabular}{llllr} 
\label{eq:terms_sfpr}
\sigma\phi^{\rm SFPR}_{{\rm sta, tro}} (\nu^{\rm high}) &= \mathcal{R}\ \sigma\phi_{{\rm sta, tro}} (\nu^{\rm low}) =0\\[6pt]
\sigma\phi^{\rm SFPR}_{{\rm sta, ion}} (\nu^{\rm high}) &= (\mathcal{R}-1/\mathcal{R}) \ \sigma\phi_{{\rm sta, ion}} (\nu^{\rm low}) = O(\Delta \theta_{AB},\mathcal{R}),  \\[6pt]
\sigma\phi^{\rm SFPR}_{{\rm dyn, tro}} (\nu^{\rm high}) &= \mathcal{R}\ \sigma\phi_{{\rm dyn,tro}} (\nu^{\rm low}) \approx0\\[6pt]
\sigma\phi^{\rm SFPR}_{{\rm dyn, ion}} (\nu^{\rm high}) &= (\mathcal{R}-1/\mathcal{R}) \ \sigma\phi_{{\rm dyn, ion}} (\nu^{\rm low}) = O(\Delta \theta_{AB},T_{\rm swt},\mathcal{R}) 
  \end{split}
\end{equation*}

The residual ionospheric terms in $O()$ represent small contributions.
Moreover, SFPR also calibrates the residual geometric errors, being non-dispersive, and $\sigma\phi^{\rm SFPR}_{\rm geo} \sim 0$; this is particularly relevant for space-VLBI, as discussed above.
Note that both components of the residual tropospheric errors are completely calibrated, $\sigma\phi_{\rm tro}^{\rm SFPR}\approx 0$, 
thanks to the same-line-of-sight calibration and simultaneous observations at the two frequencies;
the same applies to the contributions arising from residual geometric errors (i.e., errors in reference source and antenna position coordinates),
% in \SFPR\, which are also zero, 
%$\sigma\phi^{\rm SFPR}_{\rm geo}$=0, 
and in general any other non-dispersive residual contributions.
%Of note is that: $\sigma\phi_{\rm trop}^{\rm SFPR}\sim 0$, as the expected positional changes with frequency are very small and, if the frequencies are simultaneously observed, there is no temporal interpolation; 
Therefore, $\sigma\phi^{\rm SFPR}_{\rm sta,ion}$ is likely to be the dominant source of errors, particularly in cases where the angular separation between the two sources is several degrees and $R$ is large.
%these errors are 
%scaled up by the frequency ratio). 
%The best strategy to minimise these is to ensure that the residual ionospheric phase error contributions are naturally small.
Its magnitude can be minimised 
%This can be achieved 
by using values for $\mathcal{R}$ and $\nu^{\rm low}$ as low and as high as possible, respectively.
For example, the value for $\sigma\phi^{\rm SFPR}_{\rm sta,ion}$ 
at $\nu^{\rm high} = 132$\,GHz 
%at $\nu^{\rm high} = 88$\,GHz 
%using SFPR between 
with $\nu^{\rm low}=22$\, GHz,  
%and $\nu^{\rm high}=132$\,GHz, 
where the scale ratio  
$\mathcal {R}-1/\mathcal{R}$ is 5.8
%3.75 %% for 86GHz
and the residual excess ionospheric path length at 22 GHz is $\sim$0.5cm, is an order of magnitude larger than using 
$\nu^{\rm low}=44$GHz, for which $\mathcal {R}-1/\mathcal{R}$=2.7
%1.5 %% for 86GHz $\Delta l_I \over \nu^{\rm low}^2
and the residual path length is $\sim$0.1cm. 
%ionospheric path length error of 
% { 
% We can estimate likely limits for astrometric precision in SFPR by equating $\sigma\Delta\theta_{\rm thermal}$ and $\sigma\Delta\theta^{\rm SFPR}$ at 44\,GHz; from the nominal values in Eq.~\eqref{eq:a07_terms} this predicts a systematic error limit that matches a dynamic range of $\sim$94, which would give astrometric errors of $\pm$3\uas.}
%
In this case the systematic limits to the SFPR astrometric accuracy $\sigma\Delta\theta^{\rm SFPR}$ at 88\,GHz would be
0.9\uas/deg and at 132\,GHz it would be 1.6\uas/deg, using $\nu^{\rm low}=44$\,GHz and nominal parameters for the atmosphere; the corresponding
thermal astrometric limits $\sigma\Delta\theta_{\rm thermal}$ would be 1.2 and 0.8\uas{}, at 88 and 132\, GHz, respectively, for a DR 100:1.
%Systematic limits to the astrometric precision for SFPR between 22 and 88\,GHz would be
%10\uas/deg, where as for 44 to 88\,GHz would be 1\uas/deg; the equivelent dynamic ranges are 39 and 280:1, achievable with current instruments.

%{{\bf Instrumental features with benefits:} \\}
With an increasing number of telescopes equipped with simultaneous multi-frequency capabilities the perspectives for mm-VLBI are most promising. The KVN is leading the effort towards exploring the advantages of
this technological solution at the highest frequencies, together with the East Asian VLBI Network for longer baselines such as those to the VERA antennas \citep{zhao_19}.
In addition, in Europe, Yebes-40m (Spain) has a multi-frequency system installed and carries out regular
observations with KVN and KaVA \citep{bws_evn_18}; Medicina-32m, Noto-32m and SRT-64m telescopes in Italy are being equipped with compatible multi-frequency receivers, and several other telescopes are expected to be simultaneous multi-frequency ready in the near future, both in Europe and across the globe. 

%{{\bf Impact of calibration method in the context of Next Gen instruments:} \\}
We expect that the current promising results will further improve with the
ongoing developments and observations with the next-generation instruments.
The
specifications of ngVLA include the (near) simultaneous multi-frequency capability, and this is further
discussed in Sect.~\ref{sec:study_ngvla}. 
%(PLOT FROM SIMULATIONS FOR NGVLA FOR COMPARATIVE BETWEEN SWITCHING AND SIM OBS?)
SFPR is also relevant for high frequency space-VLBI missions, such as Millimetron, a regime where astrometry is challenging (see Sect.~\ref{sec:space}).
The large a-priori orbit errors of the satellite antenna(s) are an unsurmountable issue for PR methods; 
instead these are readily compensated for, with the dual frequency calibration.
%because orbit errors are non-dispersive and therefore $\sigma\phi^{\rm SFPR}_{\rm geo} \approx$0. % and because frequency switching/simultaneous operations are easier. \\

\subsection{Multi-Frequency Phase Referencing astrometric method} 
\label{sec:mfpr}

%{{\bf Description method and reference:} \\}
Multi-Frequency Phase Referencing (MFPR) is a multi-frequency technique that builds on \SFPR\ and is equally of relevance for high precision astrometric measurements in the highest frequency regime. It also relies on near simultaneous multi-frequency observations for tropospheric calibration, 
but unlike \SFPR\, it does not require a second source to remove the remaining dispersive residual terms.
It is unique in enabling precise phase astrometry with observations of only the target source. 
Instead \MFPR\ explicitly measures the dispersive terms on the target source itself, with dedicated blocks of observations at multiple bands spanning a wide frequency range
between 1.3 and 22\,GHz {($\nu_i, \nu_j$ and $\nu_k$ in the observing schedule in Table~\ref{tab:sched})}. The basis of the method and the first demonstration are described in detail in \citet{dodson_17}, along with special considerations for instrumental calibration and other effects during the analysis. 
These ionospheric calibration blocks (ICE-blocks) are interleaved with the program multi-frequency (near) simultaneous observations in the highest frequency regime, and do not require 
the same simultaneous, or near-simultaneous, frequency coverage. Therefore standard frequency switching capabilities between the frequency scans in the ICE-blocks  provides a sufficiently accurate measurement of the dispersive residual ionospheric contribution (i.e., $\Delta I \lesssim 0.1\,{\rm TECU}$) which is calibrated out from the program observations. 
The experiment measured a 21\uas{} core-shift between 22 and 43-GHz for BL-Lac (for which no suitable high frequency calibrator could be found), which deviated from the expectation based on the \citeauthor{bk_79} standard model.
An instrument with great frequency agility is required to carry out \MFPR\ observations and
currently only the VLBA, with its fast frequency switching capability and wide frequency coverage, is suitable. 

The MFPR-calibrated target dataset is formed with  a linear combination of the phases from observations of the target source A at the reference and the target frequencies ($\nu_R=\nu^{\rm low}$ and $\nu_T=\nu^{\rm high}$, respectively) and $\mathcal{R} = {\nu^{\rm high} \over \nu^{\rm low}}$, following Eq.~\eqref{eq:terms_diff}:
\begin{equation}
  \begin{split}
\label{eq:mpfr_resid}
%$
  \phi_{T} - \mathcal{R}*\phi_R &= \left(\phi_A (t,\nu^{\rm high}) - \mathcal{R}*\phi_A(t, \nu^{\rm low})\right) \\
  &=  \left(\phi^{\rm high}_{A, \rm pos} - \mathcal{R}*\phi^{\rm low}_{A, \rm pos}\right) \pm \sigma\phi^{\rm MFPR} \pm \sigma\phi_{\rm thermal} + 2\pi (\mathcal{R} n-n^\prime) %,\, {\sc R real, n$^\prime$ integer} % \iff \mathcal{R} \in integer
  \end{split}
\end{equation}
where  $\sigma\phi^{\rm MFPR}$ stands for the MFPR calibration errors and the other terms are as for Eq.~\eqref{eq:sfpr_resid}. The same considerations as for SFPR apply if $\mathcal{R}$
%={\nu^{\rm high} \over \nu^low}$ 
is a non-integer ratio.

%The SFPR-calibrated target dataset is Fourier inverted and deconvolved to yield a synthesis image of the target source at the target frequency ($\nu^{\rm high}$). We call this the SFPR-map. 
Following the same steps as for SFPR, the position offset of the target source relative to the MFPR-map center is a bona-fide astrometric measure of the ``core-shift'', or more generally the spectral position shift, in the target source between the two observed frequencies, $\delta\theta_A^{\rm high-low}$.
The  calibration errors are obtained following Eq.~\eqref{eq:a07_terms}, with $\Delta I \sim 0,\, \Delta \theta_{\rm tro}=0$, \, $\Delta \theta_{\rm ion}=0$, \, $T_{\rm swt,trp}=0$ and where $T_{\rm swt,ion}$ is the interval between the ICE-blocks. 
%
%The measurement $\delta\theta^{\rm high-low}$ corresponds to the astrometric shift of target source, between the two frequencies.
However, any core-shift in the prime calibrator would introduce a fixed phase offset between frequency bands. Care is needed to avoid or correct for this \citep[see][for an example]{dodson_17}. The propagation of errors for MFPR is discussed in detail in \citet{dodson_17}, and is similar to that for SFPR. That is:

\begin{equation*}
  \begin{split}
%\begin{tabular}{llllr} 
\label{eq:terms_mfpr}
\sigma\phi^{\rm MFPR}_{\rm sta, tro} (\nu^{\rm high})   &= \mathcal{R}\ \sigma\phi_{{\rm sta, tro}} (\nu^{\rm low})   =0\\[6pt]
\sigma\phi^{\rm MFPR}_{{\rm sta, ion}} (\nu^{\rm high}) &= (\mathcal{R}-1/\mathcal{R}) \ \sigma\phi_{{\rm sta, ion}} (\nu^{\rm low})  =0\\[6pt]
\sigma\phi^{\rm MFPR}_{{\rm dyn, tro}} (\nu^{\rm high}) &= \mathcal{R}\ \sigma\phi_{{\rm dyn,tro}} (\nu^{\rm low})  =0\\[6pt]
\sigma\phi^{\rm MFPR}_{{\rm dyn, ion}} (\nu^{\rm high}) &= (\mathcal{R}-1/\mathcal{R}) \ \sigma\phi_{{\rm dyn. ion}} (\nu^{\rm low})  = O(T_{\rm swt,ion},\mathcal{R})
  \end{split}
\end{equation*}

%All contributions are zero except for the term $\sigma\phi_{{\rm dyn,ion}}$, which is a function of $T_{swt,ion}$ and $\mathcal{R})$, and is small.
%
In MFPR, like SFPR, both tropospheric residual contributions are perfectly compensated for ($\sigma\phi^{\rm MFPR}_{\rm tro}=0$), as they are derived from simultaneous multi-frequency observations of the target source. 
Likewise, the ionospheric residual errors are solved for in the direction of the target, hence $\sigma\phi^{\rm MFPR}_{\rm sta,ion}=0$.
The residual dynamic ionospheric errors $\sigma\phi^{\rm MFPR}_{\rm dyn,ion}$, which are a function of $T_{\rm swt,ion}$ and $\mathcal{R}$, will be small as the multi-frequency ICE-blocks are closely interleaved with the program observations. 
Therefore the systematic astrometric precision for MFPR will easily be able to exceed that of the thermal limits from a DR of 1000:1, i.e. $\sigma\Delta\theta^{\rm MFPR}<\sigma\Delta\theta_{\rm thermal}$.

%{{\bf Instrumental capabilities with benefits:} \\}
MFPR requires an instrument with great frequency agility, and enables precise $\lambda-$astrometry, between high-frequencies, using only observations of a single source. 
This is relevant because the availability of calibrator sources is an increasingly significant issue at higher frequencies.
 %{\bf what do you mean with 'assumed'?????? Maybe it is highly desirable...} 
 We note that, however, the instrument stability is vital, as that can not be separated from the measurement of the ionospheric dispersive terms. 

%, the target /the observations at a lower calibrator frequencies, of the same source. 
Wide band receivers that allow for simultaneous observations at multiple bands spanning
a large frequency range are ideal for the ionospheric calibration using the ICE-blocks 
(e.g., the BRAND and VGOS technological developments, see Section~\ref{sec:brand}). 
Simultaneous multi-band high frequency receivers are the optimum solution for the tropospheric calibration, as for SFPR. 
We believe that ngVLA could fulfil all of these requirements.

\subsubsection{Beyond Frequency Phase Transfer: FPT-squared} 
\label{sec:fpt2}

FPT-squared is an alternative calibration method to \MFPR\ that uses
%Alternatively one can extract the same information using 
two pairs of simultaneous dual-frequency observations at mm-wavelengths, of the target source only. 
{ It builds on the Frequency Phase Transfer (FPT) method described above but applies the scaled-correction twice, to allow for the cancellation of ionospheric contributions, using the multiple frequency pairs from the KVN.}
\citet{zhao_17} describe the basis of this approach and its suitability for high-frequency all-sky surveys including very weak sources, such as in iMOGABA (Sect.~\ref{sec:imogaba}) providing coherence times up to several hours at 86\,GHz. The demonstration using KVN observations at 22/43 GHz and 22/86 GHz confirmed the validity of the FPT-squared calibration across large angular separations and temporal gaps and confirmed the feasibility to calibrate sources distributed across the sky with a few calibrators.
This method is, as with MFPR, very sensitive to the instrumental stability. 
Additionally the astrometrical signature, which is a mixture from both frequency pairs, can be complex to disentangle. 
On the other hand it provides an approach which does not require low frequency measurements such as ICE-blocks; therefore this is of interest for arrays which do not have that capability, such as the KVN. 
Given the rapid expansion of the SFPR network this technique may find useful astrometric applications. % {\color{red}{\bf what does this last sentence means?}.}

\section{Historical development and advances}
\label{sec:past}

\begin{figure}
    \centering
    \includegraphics[width=\textwidth]{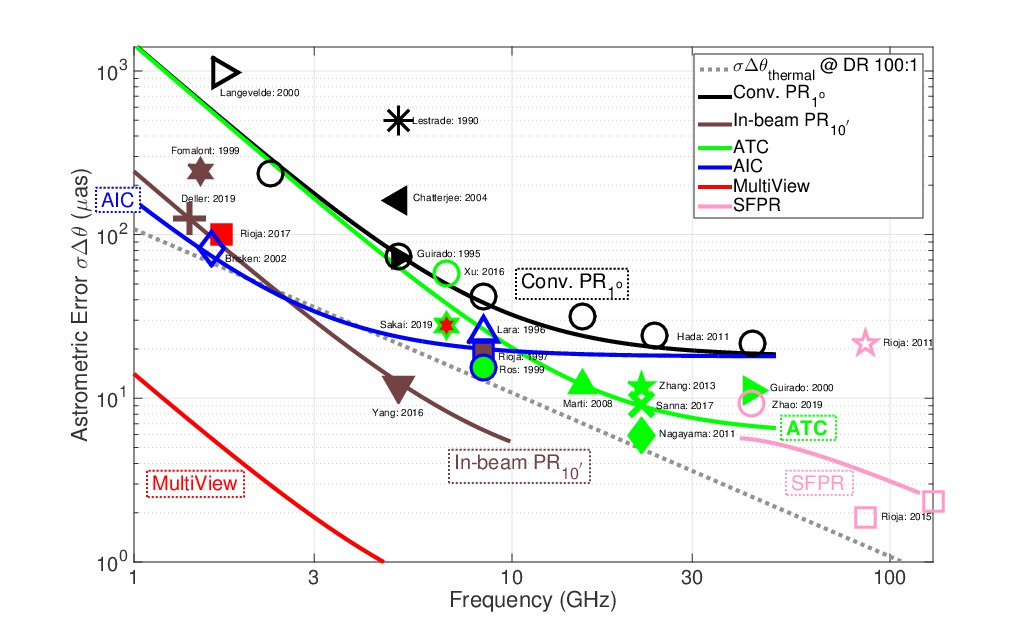}
    \caption{
    Plotted with lines are the estimates of astrometric performance, over the range of frequencies with published results, arising from thermal noise with a DR of 100:1 (grey dotted line; $\sigma\Delta\theta_{\rm thermal}$) and systematic calibration errors $\sigma\Delta\theta^{\rm cal}$ from the techniques described in Sections \ref{sec:methods_old} and \ref{sec:methods_new}, with a baseline of 6,000\,km.
        %Plotted are the limits of astrometric accuracy that can be expected given the resolution ({ for a baseline of} 6,000km) as a function of frequency, assuming a DR of 100:1 (grey dotted line), and from the 
    For conventional PR with an angular separation of 1\degr{} (black solid line; $\sigma\Delta\theta^{\rm PR}_{1\degr}$), typical atmospheric residuals are $\Delta{\rm I}$=6\,TECU for the ionosphere and $\Delta\ell_z$=3\,cm for the troposphere; these are the same for in-beam PR, except for the angular separation being 10\amin (brown solid line; $\sigma\Delta\theta^{\rm PR}_{10^\prime}$).
       % given formulae for conventional PR with an angular separation of 1\degr\ and atmospheric residuals of $\Delta{\rm I}$=6\,TECU from the ionosphere and $\Delta\ell_z$=3\,cm from the troposphere (black solid line), and in parallel, for in-beam PR with an angular separation of 10$^\prime$ (purple line). 
    The former line extends up to 50\,GHz and the later extends up to 10\,GHz, to indicate the typical limits of those techniques.
    Also plotted are the expected performance for PR with Advanced Tropospheric Calibration (ATC), with the same parameters as PR except for $\Delta\ell_z$=1\,cm (green solid line), and for PR with Advanced Ionospheric Calibration (AIC), with the same parameters as PR except for $\Delta{\rm I}$=0.6\,TECU (blue solid line).
    { For comparison the astrometric errors from \MV{} (red solid line; $\sigma\Delta\theta^{\rm MV}$), with parameters from Sect.~\ref{sec:study_mv}, are significantly smaller than the thermal limits and would extend up to 50\,GHz. 
    SFPR enables astrometry at the highest frequencies, beyond the thresholds of PR.
    The astrometric errors from SFPR, with a 5\degr{} angular separation (pink solid line; $\sigma\Delta\theta^{\rm SFPR}$), are shown between 40 and 130\,GHz, using $\nu^{\rm low}$=22\,GHz observed simultaneously. 
    The final astrometric error is the combination of the thermal noise and the systematic contributions, for each technique.}
        % C PR
    Over plotted with symbols are some empirical results for a single epoch from the literature. %, normalised to 1\degr{} angular separation, except for the in-beam results.
    They illustrate that the theoretical formulae match the empirical trends, and also indicate the improvements over time to arrays and a-priori models.
    The results correspond to a range of different targets (pulsars, masers, AGNs), source flux densities, schedules, weather conditions, analysis pipelines and instruments; thus significant differences are to be expected.
    For a meaningful direct comparison they are normalised to correspond to 1\degr{} angular separation (except for in-beam PR, where there was no normalisation, and SFPR where they are normalised to 5\degr{}), single epoch accuracies and to approximately 6,000km baselines. 
    The colours match the techniques as described above and the symbols reproduce the trend of the systematic limits, except where other limitations apply as discussed fully in Sect.~\ref{sec:past}.
        \label{fig:astro_history}}
\end{figure}

In this section we review the historical progress on astrometric precision and the scope of application, since the foundational work. 
Here we are focusing on the developments that enabled the improvements in astrometric performance. %; for a review on the impact on the scientific outcomes provided by astrometry see \citet{reid_micro}. 

Figure~\ref{fig:astro_history} illustrates the comparative performance between astrometric techniques as a function of the observing frequency, assuming a baseline of 6,000\,km. 
The grey dotted line is for the astrometric limits imposed by the thermal noise (or sensitivity) of the instrument, in the absence of systematic errors, assuming a DR of 100:1. 
The other lines show the single epoch systematic astrometric limits $\sigma\Delta\theta^{\rm cal}_{\rm sta}$, estimated for nominal values of $\Delta\ell_z$ and  $\Delta I$ that model the dominant propagation medium effects, as presented in Sections \ref{sec:methods_old} and \ref{sec:methods_new}, for the different methods. % (see Sections 6.1 to 6.5).
Between these two competing contributions, sensitivity versus calibration errors, whichever is the largest determines the expected measurement precision. 
%Note that the black line is indicative of thermal noise limits, for a DR of 100:1, for weak sources the thermal noise limits can be larger.
Overplotted symbols correspond to observational results from the literature, to show how these conform to the expectations.
% To allow for comparison of the multitude of different observations and reported information the results are scaled:
% to 1\degr{} angular separation between  the target and calibrator for conventional PR, except for in-beam PR;
% to 5\degr{} angular source separation for SFPR;
% to 6,000\,km baselines, for VERA, KVN and KaVA observations;
% to the equivalent of a single epoch, i.e., parallax errors are are scaled by the square root of the number of observations. 
To allow for comparison of the multitude of different observations and reported information the results are scaled
to 1\degr{} angular separation between the target and calibrator for conventional PR and
to 5\degr{} for SFPR;
no scaling is applied for in-beam PR.
VERA, KVN and KaVA observations are scaled to 6,000\,km baselines.
All are converted to the equivalent of a single epoch, i.e. parallax errors are scaled by the square root of the number of observations. 
This ignores significant effects from: source declination, weather conditions, error analysis, source flux density and scheduling strategy.  
These normalised values provide guidance to how astrometric methods have improved over time, and therefore how we may improve further, as required with the next-generation instruments (Sections \ref{sec:next} and \ref{sec:tech}).

\subsection{Conventional PR}\label{sec:hist_conv}

        % Representing conventional astrometry, in black, are the observations of M87 by \citeauthor{hada_11} (2011, open circles) at multiple frequencies.
        % \citeauthor{lestrade_90} (1990) (star) published PR measurements of the weak radio-star ALGOL, without GPS Ionospheric corrections (i.e., $\Delta{\rm I} \gg 6$TECU),
        % \citeauthor{langevelde_00} (2000) (open triangle-right) published observations of OH-masers at 1.7GHz that also did not have GPS corrections, and \citeauthor{chatterjee_04} (2004) (triangle-left) published observations of pulsars at 5GHz. 
        % \citeauthor{guirado_95}, (1995, black triangle-right) was PR without ATC. 

The solid black line in Fig.~\ref{fig:astro_history} corresponds to the theoretical expectations for astrometric errors in PR, arising from residual tropospheric and ionospheric errors characterised with $\Delta\ell_z = 3$cm and $\Delta{\rm I} = 6$TECU, for a source pair separation of 1\degr.
One can clearly appreciate how the astrometric errors are expected to rapidly degrade towards low frequencies. 
As PR is impracticable above 43-GHz, the line truncates at 50-GHz.
Symbols in black are historical astrometric measurement errors selected from the literature to represent the performance of conventional PR methods across the spectrum,
%; for a meaningful comparison the errors quoted in the  publications have been 
scaled to a common angular separation of 1\degr.
Open circles are from the M87 observations by \citet{hada_11}, which mapped out the frequency dependent core-shift effect with exquisite accuracy. 
Matching precision was found by \citeauthor{guirado_95}, (1995, black triangle-right) at 5\,GHz, using phase-delay differential astrometric analysis, for a single pair of sources.
This can be compared to the accuracy from only 5 years earlier for \citeauthor{lestrade_90} (1990, star). As GPS corrections were not available at that time $\Delta{\rm I}$ could have been as high as 40.
The early observations of OH-maser parallaxes at 1.6\,GHz by \citeauthor{langevelde_00} (2000, open triangle-right) were for many years the most accurate low frequency astrometric maser result, but also limited by %the DR of $\sim$10. 
the accuracy of the ionospheric model. 
\citeauthor{chatterjee_04} (2004, triangle-left) showed that better pulsar astrometry precision could be obtained with conventional PR by observing at (much) higher frequencies, 5\,GHz in this case, even though the targets were much weaker. 
The sensitivity is not given for the latter observations, but we estimate that the DR would have been $\sim$30. 
%In both of these cases the astrometric accuracy is limited by their thermal errors.

%
\subsection{In-beam PR}
        % %in-beam
        % In-beam astrometry is represented by the purple star for \citeauthor{fomalont_99} (1999), plus for \citeauthor{deller_18} (2019) at L-band, \citeauthor{yang_16} (2016, triangle-down) at 5\,GHz and \citeauthor{rioja_inbeam} (1997, square) at 8.4\,GHz. Rioja also included S/X ionospheric corrections, so is outlined in blue.

The in-beam PR strategy to mitigate the astrometric challenges relies on using very nearby reference sources; its main application is at the low frequencies. 
The brown line in Fig.~\ref{fig:astro_history} gives the astrometric performance expected, for the same nominal conditions as in Sect.~\ref{sec:hist_conv}, but scaled to an angular separation of $10^\prime$. The dilution factor is six times greater and therefore the errors are six times smaller, compared to the black solid line for conventional PR.
It becomes harder to find in-beam calibrators for a target within the smaller antenna FoV at higher frequencies (see predictions in Table~\ref{tab:inbeam_mv}), therefore the usability of in-beam PR is reduced above L-band. Thus the line truncates at 10-GHz.
%corresponds to the theoretical expectations arising from residual tropospheric and ionospheric errors characterised with $\Delta\ell_z$ = 3$cm and $\Delta I = 6$TECU, respectively, using PR methods.
The brown symbols in Fig. \ref{fig:astro_history} are demonstrations of its effectiveness; 
the pulsar astrometry of \citeauthor{fomalont_99} (1999, star) at 1.5\,GHz was the first demonstration using a close-but-weak reference source at 12\arcmin.
%Fomalont 99 is about 240 uas and will be ~5 times above the line This is why it was not included
The pulsar astrometry from \citeauthor{deller_18} (2019, plus) at 1.4\,GHz are for a target-calibrator angular separation of 10\arcmin. 
The difference between these two represents the improvement in the systems (instrument, and a\,priori models) over the two decades separating them.
\citet{rioja_phd} and \citep{rioja_inbeam} observed the 1038+528 source pair at 8.4\,GHz, which is the closest pair of strong sources, being separated by only 33\arcsec{} (it is worth noting that the latter analysis also included S/X AIC corrections, so it is marked as a brown square outlined in blue).
Their analysis is dominated by the thermal noise, with a DR of $\sim$30. % 0.6mas beam 36uas error = 16! or 20uas =30
\citeauthor{yang_16} (2016, triangle-down) performed a very deep EVN observation of a Swift-detected tidal burst with an in-beam calibrator only 2\arcmin{}  away and with a DR of up to 537 that had extremely high precision, leading to a very strict limit on the speed of the outflow.

\subsection{Advanced Tropospheric Calibration} % (ATC) or Advanced PR for high frequency regime}
        % % ATC
        % One of the great leaps forward was the introduction of GeoBlocks, which is represented by the results of \citeauthor{zhang_13} (2013, star), \citeauthor{sanna_17} (2017, cross) at 22GHz and 
        % \citeauthor{xu_18} (2018, open circle) at 6.7GHz in green.
        % VERA uses image optimisation to achieve similar results, and is represented by \citeauthor{nagayama_11} (2011, green diamond).
        % %The UVPAP results from Lara (1996, triangle) in green and blue have both S/X ionospheric and Zenith path tropospheric corrections.
        % % UVPAP
        % \citeauthor{guirado_95} (1995, 2000), \citeauthor{lara_96} (1996), \citeauthor{ros_99} (1999)  and \citeauthor{marti-vidal_08} (2008) observed continuum sources using UVPAP or its precursors, at 5-, 43-, 8.4-, 8.4- and 15-GHz, respectively. 

ATC methods to reduce the value of $\Delta\ell_z$ to $\sim 1$ cm are described in Sect.~\ref{sec:ATC}. Amongst these are GeoBlocks and simultaneous source-pair fitting, with phase reference imaging and differential phase delay astrometry, % (UVPAP), 
respectively. %; we refer to these strategies as Advanced PR. 
Its main impact is at the high frequencies regime, whilst at low frequencies the limitations would be the same as for conventional PR.
%($\ge 6-8 GHz??$) {\bf or maybe not need to include here, we should have described before our definition of high/low freq regimes}. 
The green line in Fig.~\ref{fig:astro_history} corresponds to the theoretical expectations for residual tropospheric and ionospheric errors characterised with $\Delta\ell_z$=1\,cm and $\Delta I$=6\,TECU, respectively, for a target-calibrator angular separation equal to 1\degr{} and terminating at 50-GHz, as for conventional PR.
Examples of published measurements are shown with symbols in green, scaled to 1\degr{}. \citet{guirado_00} and \citet{marti-vidal_08} carried out at 43 (triangle-right) and 15\,GHz (triangle-up) respectively and used UVPAP. 
%{\bf note this is also uvpap}
A similar analysis in \citeauthor{ros_99} (1999, green filled circle with blue outline), at 8.4\,GHz (X-band) using dual band S/X observations, where both advanced tropospheric and ionospheric corrections were made, lies slightly below the nominal ATC line.
We include only two of the many BeSSeL results at 22\,GHz that use GeoBlocks, \citeauthor{sanna_17} (2017, cross), which measured a record distance to a star-forming region on the far side of the Galaxy of $20.4_{+2.2}^{-2.8}$kpc and \citeauthor{zhang_13} (2013, star), which has the record parallax precision of $\pm$3\uas.
\citeauthor{nagayama_11} (2011, diamond) reports the best VERA parallax, with a precision of $\pm$7\uas. This VERA result is scaled down by a factor of 3, to allow direct comparison with the VLBA results.
%We note the result of \citet{ros_99}, at 8.4GHz using S/X observations, where both advanced tropospheric and ionospheric corrections were made, that lies slightly below the nominal ATC line.
The BeSSeL observations of \citeauthor{xu_16} (2016, open circle)  at 6.7\,GHz sits in the frequency regime where ATC corrections have little effect and it is appreciably above the other ATC results.

\subsection{Advanced Ionospheric Calibration} % (AIC) or Advanced PR for low frequency regime}
        % %The UVPAP results from Lara (1996, triangle) in green and blue have both S/X ionospheric and Zenith path tropospheric corrections.
        % % AIC
        % Ionospheric corrections from \citeauthor{brisken_02} (2002, blue open diamond) allowed significantly improved low frequency astrometry.

AIC methods to reduce the value of $\Delta I$ are described in Sect.~\ref{sec:AIC} and their main impact is at the low frequency regime, whilst the high frequency regimen remain as for conventional PR.
The blue line in Fig.~\ref{fig:astro_history} corresponds to the theoretical expectations for systematic astrometric errors with residual tropospheric and ionospheric errors characterised with $\Delta\ell_z = 3$cm and $\Delta I = 0.6$TECU, respectively, for a target-calibrator angular separation equal to 1\degr{}. 
Note that the value of $\Delta I$ is reduced by a factor of ten, compared to conventional PR, achieving similar precision as in-beam PR.
Examples of measurement errors per epoch from the literature are scaled to 1\degr{} and shown as blue symbols: 
%The astrometric results with AIC of 
\citeauthor{brisken_02} (2002, open diamond), for pulsar parallaxes measurements at 1.6\,GHz with fits to the residual phases to reduce $\Delta$I; and \citeauthor{lara_96} (1996, triangle), for AGN astrometry at 8.4\,GHz, using dual frequency S/X observations where analysis of the delays provide `ionosphere-free' data products.

\subsection{MultiView} % (\MV\ )}
        % % MV & SFPR
        % MultiView results from \citeauthor{rioja_17} (2017, square) and \citeauthor{sakai_19} (2019, star) are in red and red with green, respectively. 

\begin{figure}
    \centering
 \includegraphics[width=0.9\textwidth]{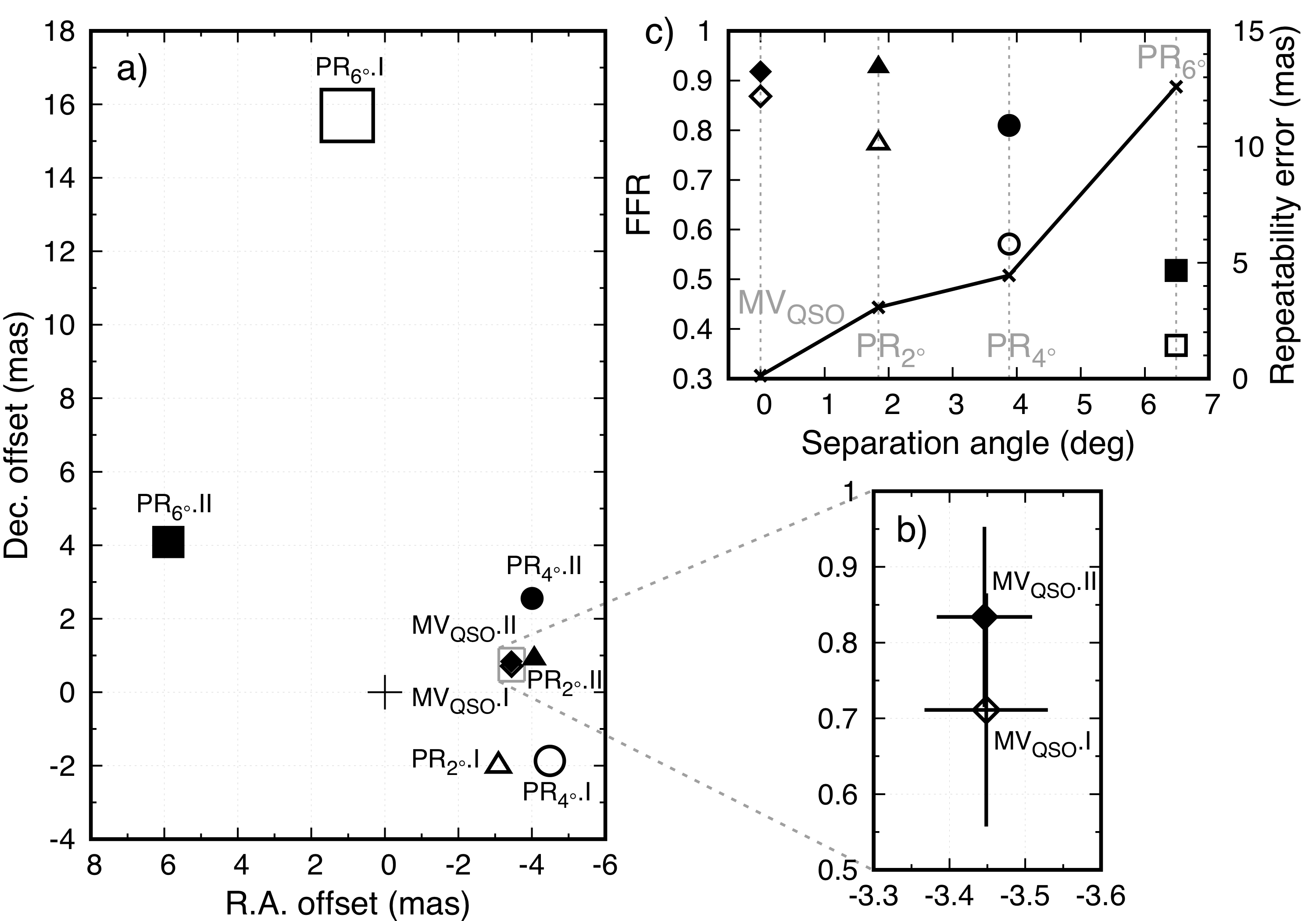}
\caption{ {\it a)} Astrometric measurements with \MV\ and  PR at 2\degr,  4\degr{} and 6\degr{} \citep[see][for description]{rioja_17},
  from the observations of quasars at two
  epochs, with respect to the catalogue position. The size of the plotted symbols
  corresponds to the estimated thermal noise error in each case. The labels describe the calibration methods (MV, PR with the angular separation as a subscript) and epoch of observations (I, II). {\it b)}  Zoom for \MV\ astrometric solutions, which lie within 100\uas of each other. The error bars are for the estimated thermal noise; both epochs agree to within thermal errors.
 {\it c)} Solid line shows the corresponding repeatability astrometric errors versus
 the angular separation between target and calibrator for PR analysis,
 and for an effective 0\degr{} separation for \MV. Filled and empty
 symbols show the Flux Fractional Recovery (FFR) quantity versus angular
 separation for \MV\ (diamond), PR$_{2\degr}$ (triangle), PR$_{4\degr}$ (circle),
 and PR$_{6\degr}$ (square), for epochs I (empty) and II
 (filled). \label{fig:astro_mv}}
\end{figure}

MultiView results in a superior compensation of both $\Delta\ell_z$ and $\Delta I$ by using multiple calibrators as described in Sect.~\ref{sec:methods_mv}.
The red line in the bottom left corner of Fig.~\ref{fig:astro_history} corresponds to the theoretical expectations for systematic astrometric errors with residuals as described in Sect.~\ref{sec:study_mv} and listed in Table~\ref{tab:inbeam_mv} (Col. 3).
As both the tropospheric and ionospheric errors are near-perfectly compensated \MV{} outperforms all other PR-related methods across the frequency spectrum.
It is relevant across the radio spectrum, but particularly in the low frequency regime where the conventional errors are large. % and it enables astrometry beyond the low frequency thereshold of any PR-method.
MultiView calibration is most naturally carried out in the visibility domain. The measurement of \citeauthor{rioja_17} (2017, square) in Fig.~\ref{fig:astro_history} is a \MV{} demonstration on an AGN with calibrators up to several degrees from the target, where the final astrometric errors are dominated by the thermal errors that greatly exceed the systematic errors, which are negligible. 
Fig.~\ref{fig:astro_mv} demonstrates the astrometric repeatability of these observations and compares them to PR.
%In this demonstration the thermal errors exceeded the systematic limits, which were very small.
%T5: MVQSO	0.17	0.14	0.1
%OH: with the repeatability errors being ${\sigma }_{\mathrm{pos},\mathrm{rep}}^{{\mathrm{MV}}_{\mathrm{OH}}}=0.29$ mas for MVOH and ${\sigma }_{\mathrm{pos},\mathrm{rep}}^{{\mathrm{PR}}_{\mathrm{in} \mbox{-} \mathrm{beam}}}=0.64$ mas for PR${}_{\mathrm{in} \mbox{-} \mathrm{beam}}.$ 
%For OH-maser astrometry at 1.6 GHz result in an effective improvement that can be compared to those from \citet{langevelde_00}.
%; note that both measurements use calibrators several degrees away. % and results are scaled to 1\degr.  
\citeauthor{sakai_19} (2019, red star with green outline) used an alternative implementation, in the image domain \citep[so is only effective for the errors coherent over the array during the observations]{reid_17}, applied to BeSSeL measurements of Methanol masers at 6.7\,GHz. \citet{sakai_19} used both \MV{} in the image domain and ATC corrections and can be compared to \citet{xu_16}, which only had the latter, to demonstrate the impact.
% In this case ATC corrections using GeoBlocks were also applied.
% The impact of \MV{} can be seen in the comparison with the result from \citet{xu_16} that just uses ATC.

\subsection{Source Frequency Phase Referencing} % (SFPR)}
        % { SFPR results, marked with open  pink circles, are from \citet{zhao_19}, with simultaneous frequency observations on KaVA and \citet{rioja_11a,rioja_15} with fast-frequency switching on VLBA at 86\,GHz and simultaneous frequencies on KVN at 86 and 130\,GHz respectively. The latter results were referenced to 43\,GHz.
        % All SFPR results are scaled to 6,000km and 5\degr{}. 
        % %All \MV{} and SFPR results are thermally limited.
        % }

\begin{figure}
    \centering
    \includegraphics[width=\textwidth]{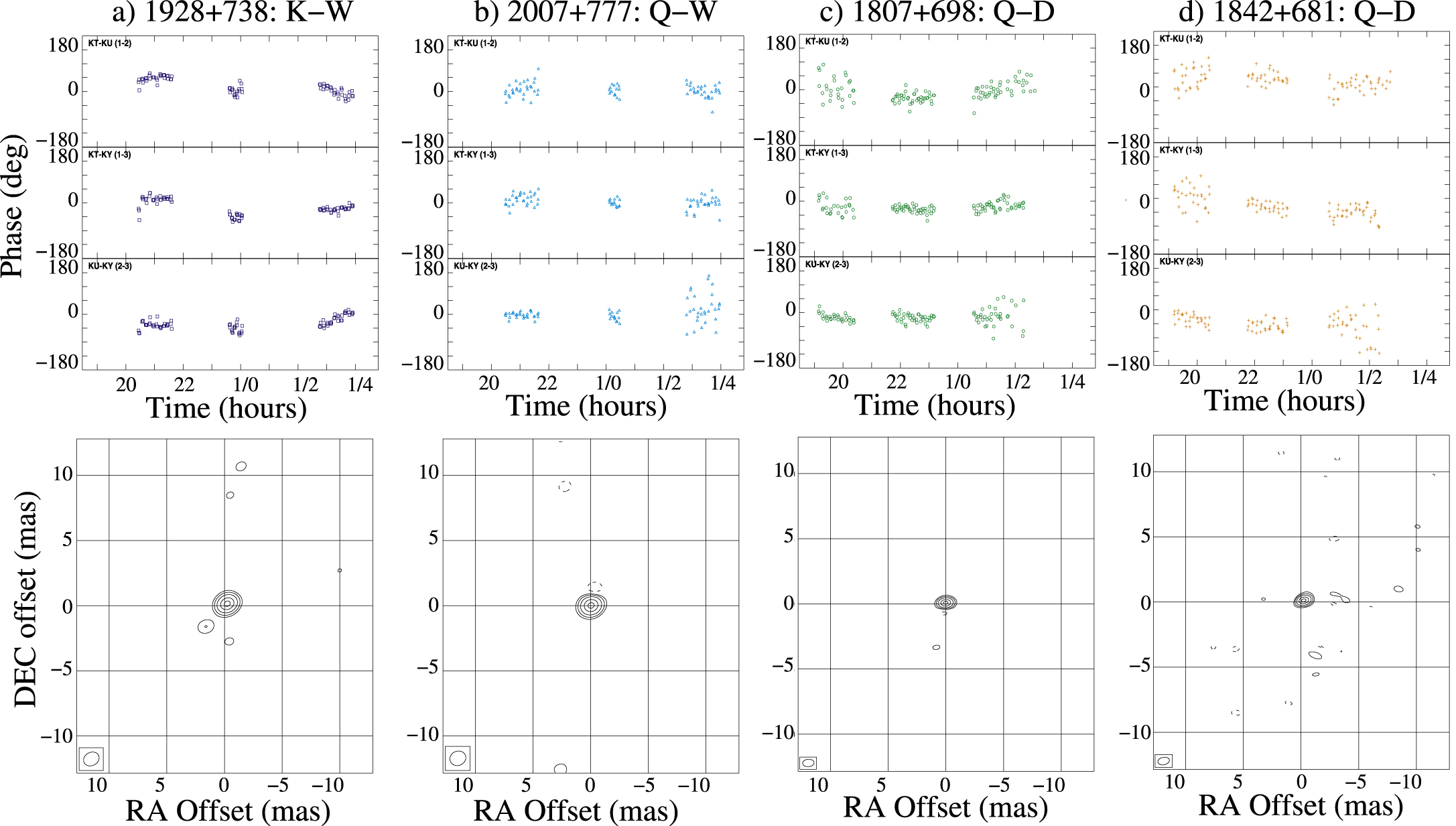}
    \caption{%{\bf shortened} 
        Outcomes of SFPR astrometric analysis using KVN observations at the four frequency bands K/Q/W/D (22/43/86/130 GHz, respectively),  from \citet{rioja_15}. Upper row: SFPR residual visibility phases for source and frequency pairs. The target source and target frequency band (i.e., ${\nu }^{\rm high}$), along with the angular separation from reference source, the reference frequency band (i.e., ${\nu }^{\rm low}$), and the frequency ratio $\mathcal{R}$, in parentheses, are specified next  for each plot. From left to right: 1928+738 at W band (6\degr\ separation from reference, K band, $\mathcal{R}$ = 4); 2007+777 at W band (6\degr\ separation, Q band, $\mathcal{R}$ = 2); 1807+698 at D band (8\degr\ separation, Q band, $\mathcal{R}$ = 3); 1842+681 at D band (10\degr\ separation, Q band, $\mathcal{R}$ = 3). Lower row: SFPR astrometric maps from left to right: 1928+738 at 87\,GHz (W band), 2007+777 at 87\,GHz; 1807+698 and 1842+681 at 130\,GHz (D band). Peak fluxes are 2 Jy beam$^{-1}$, 266 mJy beam$^{-1}$, 415 mJy beam$^{-1}$, and 216 mJy beam$^{-1}$, respectively. The contour levels in the maps start from 0.75\% of the corresponding peak fluxes, respectively, and double thereafter in all cases. Each map includes a negative contour level at the same percent level of the peak flux as the first positive one. The beam size is indicated at the bottom left of the image.}
    \label{fig:sfpr130}
\end{figure}

SFPR results in a superior compensation of the fast tropospheric fluctuations in $\Delta\ell_z$ that sets the upper frequency threshold of any PR-method, and opens a new window for VLBI-astrometry. %Therefore its impact is  particularly relevant in the highest frequency regime,
%beyond the usability of any other existing methods; 
The pink line in Fig.~\ref{fig:astro_history}, starting at 40\,GHz and continuing to 130\,GHz, is for the expected SFPR astrometric errors $\sigma\Delta\theta^{\rm SFPR}_{\rm sta}$ with a reference frequency ($\nu^{\rm low}$) of 22\,GHz and a calibrator 5\degr{} away.
See Sect.~\ref{sec:methods_sfpr} for details.
Examples of published measurement errors are from KaVA \citep[open circle]{zhao_19}, at 43\,GHz. Higher frequencies plotted are from \citet[]{rioja_11a} using fast switching observations with VLBA at 86\,GHz (open star) and \citet[]{rioja_15} using simultaneous multi-frequency observations with KVN at 86\,GHz and 130\,GHz (open square), referenced to 43\,GHz. 
All of the KVN and KaVA SFPR results are scaled to 6,000\,km and 5\degr{} and are dominated by the thermal errors.
Figure~\ref{fig:sfpr130} shows the astrometric measurements at the highest KVN band, 130\,GHz, using pairs of sources several degrees apart; full details are in \citet{rioja_15}.
The SFPR-calibrated visibility phases are shown along with the SFPR images at 86- and 130-GHz, after correcting with simultaneous observations at 22- or 43-GHz. 
%All target sources were selected to be compact and strong, with sufficient flux to allow a careful analysis of the residuals and to demonstrate the power of the method. 
Small core-shifts for these sources could be detected, although this analysis was limited by the resolution from the relatively short KVN baselines. 
Recent and on-going developments towards a global simultaneous multi-frequency ready network will enable high precision astrometry measurements at frequencies never achieved before.
SFPR underpins the successful KVN AGB KSP, which is discussed in Sect.~\ref{sec:applications_sfpr}. % ({\bf say something about matching resolution and science case}). 

\section{Technological developments}% \sout{ / Hardware advances} }
\label{sec:tech}

Since its beginnings astrometry has benefited from the relentless pursuit of technological developments in radio astronomy, that have allowed leaps in performance.
Historical examples of this would be the ever increasing capability to record wider bandwidths (that improves the sensitivity, therefore reduces the thermal noise errors and enables faster source switching for more precise phase and delay measurements)  and the dual band S/X receiver system (that allowed for the formation of `ionosphere-free' data products).
%The advantages come from the increase of sensitivity in the interferometric imaging, reduces the thermal noise errors, enables faster source switching, and the ability to account precisely for the dominant atmospheric propagation residual errors in astrometric measurements.
%
%Of the external inputs for the data-reduction, improved station coordinates and atmospheric models (from GPS data for example) have reduced the intrinsic residuals to manageable levels.
In parallel improvements in the a\,priori models have reduced the intrinsic residuals to manageable levels. These include improved station coordinates and atmospheric models (from GPS data for example), amongst others.
Current ongoing developments are providing the required technological solutions to enable optimum performance of the next-generation calibration astrometric methods described above. 
They address the required capacity for broadband and simultaneous multiple-frequency bands and multiple-calibrator  observing modes, for antenna arrays and single dish telescopes.
%
%The KVN system of multi-frequency receivers is being adapted in response to interest from the community to provide a more compact system. 

\subsection{Capability for multi-frequency observations}

There are clear science benefits from observations of the sky in a large range of radio frequencies. %, and these are relevant for ngVLA and SFPR. 
%Large Frequency coverage and high Frequency agility.  
%Incorporate with fast frequency switching, it is evolving towards having a single 
The previous state-of-the-art has been to use a combination of different radio receivers, each optimised for a specific range of frequencies to provide a wide frequency coverage.
Frequency agility comes from, for example, switching  between these receivers, which can be  done  within  10\,secs for VLBA (USA), 30\,secs at Effelsberg-100m (Germany) and 1\,min at the  { TianMa-65m (China)}. 
%and to switch receivers whenever it was necessary ({\it alter text}), with systems to enable fast frequency switching taking ca. 10 seconds between some of them, like the VLBA.
New developments include the capability for simultaneous observations at multiple frequency bands and broadband systems, with continuous frequency coverage that is up to one order of magnitude increased over the previous generation of receivers.
Broadband, multi-frequency systems have played an essential role in breaking through  the  frequency thresholds for precision astrometric measurements. Likewise the next-generation of instruments will open up both the low and high frequency regimes.
%They have advantages for astrometric projects in the context of next gen instruments. 

\begin{figure}
    \centering
    \includegraphics[width=0.48\textwidth]{fig5a.jpg}
    \includegraphics[width=0.41\textwidth]{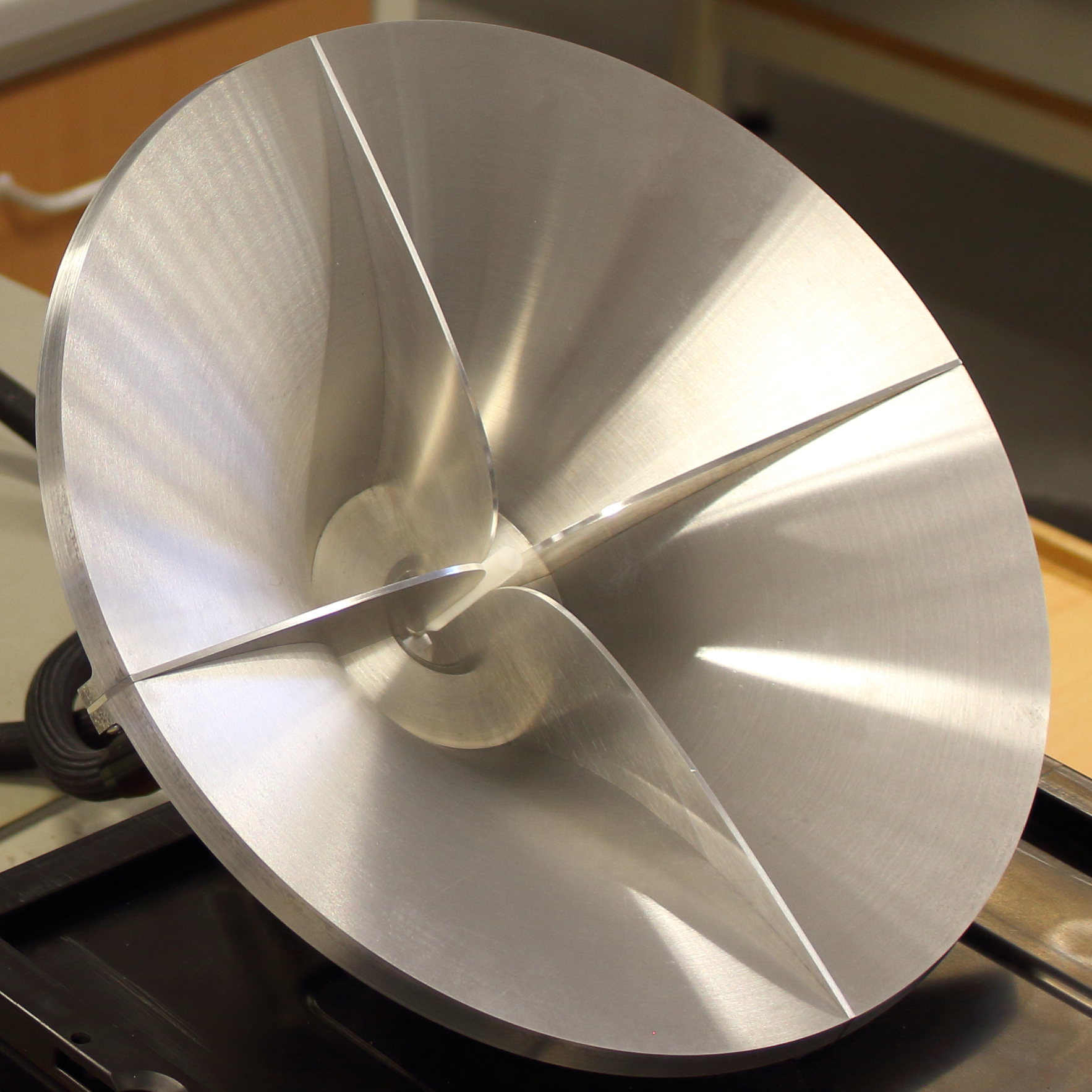}
    \caption{Two broad band technological solutions: {\it left)} The mm-wave KVN optical bench, with separate receivers (with equalised path length) for simultaneous observations at 22, 43, 86 and 130GHz.
    {\it right)}  the RadioNet BRoad bAND (BRAND) QRFH cm-wave receiver, which is able to record a decade of bandwidth from 1.5 to 15.5\,GHz \citep{flygare_19}.}
    \label{fig:mf_sols}
\end{figure}
\subsubsection{Simultaneous multiple frequency systems} % {\bf (not really receivers isn't it?)}}

The KVN innovative engineering solution for mm-VLBI provided a breakthrough in both astrometry and also the detection of weaker sources through extended coherence times, 
using a simultaneous multi-band quasi-optical design for 22/43/86 and 130\,GHz \citep{han_13}, which is shown in Fig. \ref{fig:mf_sols} {\em left}.

%{\bf not here - tech not sci}
%\sout{This enabled astrometry up to 130 GHz, the highest frequency at the KVN, and a factor of 3 higher than previous measurements.  BETTER+>
The impact of this multi-band system on spectral line VLBI is significant, as several different maser types in different frequency bands could be observed simultaneously, to obtain images of the emission with bona-fide astrometric registration for the first time;
\citet{dodson_17_sim} provides a compilation of possible scientific applications and the performance impact is estimated in Sect.~\ref{sec:study_ngvla}.

The Compact Triple-band Receiver (CTR) \citep{han_17}, is a new and much more compact implementation of the KVN multi-frequency receivers, with three frequency bands, K (18-26GHz), Q (35-50GHz) and W (85-115GHz), fitted into a single cryostat.
The reduction in size makes this technical solution very versatile for installation in telescopes where space availability is an issue, as is usually the case.
%places the KVN quasi optical system for 22, 43 and 86\,GHz in a single Dewer, suitable to be installed in a typical antenna reciever mount. The optics need to be tune for the individual antennas, but 
%
A similar configuration is being developed for the Yebes-40m (Spain) \citep{tercero_19}. Their new Q-band (31.5--50.0\,GHz) and W-band (72--116\,GHz) receivers, combined with the existing K-band receiver, provide an alternative solution to the CTR that has slightly wider bandwidths.
Both these systems show that there is great promise for providing mm-wavelength multi-band capability to a wide number of observatories, greatly enhancing the VLBI array suitable for SFPR and the next-generation instruments (as discussed in Sections \ref{sec:methods_sfpr} and \ref{sec:ngvla}, respectively). 

%In the mm-wavelengths  examples of operational VLBI receivers are the ....    (CAY-Yebes, reference) with an outstanding perfomance.... 

\subsubsection{Broadband receivers}
\label{sec:brand}

We note that a wide-band receiver can cover multiple `traditional' frequency bands, hence also providing the simultaneous multi-band observing capability.
In the cm-wavelengths, BRAND-EVN (BRoad bAND European VLBI Network)  \citep{brand, flygare_19} is a RadioNet project to build a prototype  quad-ridge flared horn (QRFH) prime-focus receiver for Effelsberg-100m  (and other telescopes) that spans 1.5 to 15.5\,GHz, which is shown in Fig.~\ref{fig:mf_sols} {\em right}. 
It follows on from earlier designs, such as the decade-bandwidth log-periodic dual-dipole geodetic VGOS system called the `eleven-feed', which covered the ultra-wide bandwidths required for that project (2--13\,GHz) \citep{eleven-feed}.
These systems enable the observations of radio sources over the whole usable band, simultaneously.  

%he different observedsub-bands of the total frequency range can be connected coherently via thisfringe-fit,

Achieving its full potential sensitivity requires the sub-bands to be coherently connected (via fringe-fitting) over the full band, which in turn preserves the chromatic astrometry information and results in astrometrically registered images of the emission from a source at the different observed sub-bands (in a similar fashion as above, in mm-waves). % (see science cases in section XX). 
The RadioNet project RINGS (Radio Interferometry Next Generation Software) is a processing development to address the challenges resulting 
from the fringe-fitting analysis of such wide-band data, among them a new functionality for the simultaneous calibration of both atmospheric contributions to the observables (with their distinct frequency dependencies) as well as the intrinsic chromatic effects from the sources themselves.  

Other recent similar receiver designs are the ultra-wideband low (UWL) receiver for Parkes  \citep{dunning_15,hobbs_19} that covers 700 to 4030 MHz;  two complementary wideband systems (UWMid, UWHigh) that would operate from 4 to 25\,GHz are in the design phase.

\subsubsection{Simulation studies with simultaneous multi-frequency observations for mm-VLBI astrometry with the ngVLA}
\label{sec:study_ngvla}
%Looking forward to the next generation of instruments, 
In this section we consider the impact on the calibration performance from using frequency-switching compared to the simultaneous multi-frequency capability, for the ngVLA (or any other interferometer).
This work was carried out under the ngVLA Community Studies program.
We used the ARIS simulator \citep{a07}, which has very complete tropospheric and ionospheric models, to generate synthetic datasets with realistic atmospheric conditions over a subset of ngVLA stations. 
These comprised of SFPR observations with 16 antennas, with baselines from 27 to 500\,km (spanning the edge of the core and the longest baselines of the proposed array), with a range of duty cycles for the frequency-switching between zero (i.e., simultaneous) and up to 60\,sec. 
The array response to point sources was generated, with zero thermal errors and nominal atmospheric conditions, at the four frequency bands of the KVN. The lowest frequency (22\,GHz) was used as the reference band for the SFPR analysis for all the other bands.

\begin{figure}
    \centering
    \includegraphics[width=0.9\textwidth]{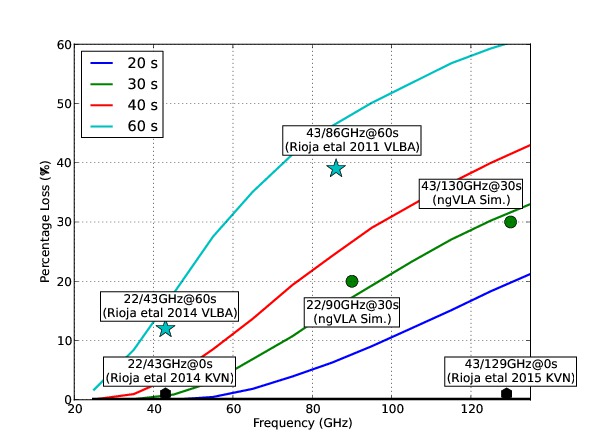}
    \caption{Simulated phase-connection losses in multi-frequency calibration, as a function of frequency and of frequency-switching duty cycle, with lines for 60\,sec (cyan), 40\,s (red), 30\,s (green) and 20\,s (blue). Simultaneous observations have zero losses and are indicated with a black line superimposed on the x-axis. 
    In all cases the simulations were thermal noise free.
    Overplotted as symbols are the fractional flux losses in the images made from the same simulated datasets, using SFPR with 30\,s frequency-switching, in green circles, and empirical results from the literature using VLBA and KVN observations \citep[cyan stars and black hexagons respectively][]{rioja_11a,rioja_14,rioja_15}; 
    labels list the reference and target frequencies and the duty cycle, along with the publication.
    Losses are over 10\% at the highest frequencies ($>$100\,GHz) even for the fastest conceivable duty cycle of 20\,s, whereas simultaneous observations would have zero losses.}
    \label{fig:ngvla}
\end{figure}

Our first metric was the fractional flux loss (FFL; 1-FFR) measured from the SFPR images compared to the input model, when the simulated data were calibrated using interleaved observations with a 30sec frequency duty cycle.
This allows a direct measurement of the phase variance via the FFL.
%=$\Delta \Phi$ ($e^{-\Delta \Phi^2/2}$ \citep{TMSv3}).
%
Additionally we considered the effect of phase-connection errors, which approximates the first metric.
We estimated the loss of phase-connection by comparing the forward predicted phase (using the rate and phase solutions) to the actual phase, for a range of duty cycles. 
We define failure as where the error is greater than a half cycle. 
Longer intervals, obviously, more often fail to correctly predict the actual phase.
Figure~\ref{fig:ngvla} shows, with lines, the fraction of incorrectly predicted solutions, up to a target frequency of 130\,GHz.
Overplotted as symbols are results from the SFPR imaging of the simulated ngVLA data and empirical results from the literature. The two metrics agree well with each other, and are in reasonable agreement with the empirical KVN and VLBA results, for which the weather conditions were excellent rather than nominal.

Figure~\ref{fig:ngvla} shows that, above 100\,GHz, the losses are over 10\%, even for the fastest conceivable duty cycle of 20\,sec, whereas simultaneous observations would have zero losses. 
Note that even at lower frequencies, for weaker sources, one would still introduce significant errors, as the duty cycles would have to be longer to overcome thermal limits; however simultaneous observations would still result in zero interpolation errors.
These studies underline the benefits of simultaneous multi-frequency observations, and support the additional effort required to design systems with this capability, at least for arrays observing at mm-wavelengths ($\ge$22\,GHz). 

\subsection{Multiple-pixel radio telescopes}\label{sec:mbeams}

%{\bf EXPAND: }
The scientific impact of the next-generation instruments can be improved not only by increasing sensitivity, but  also  by  increasing  the  FoV, as both of these increase the survey speed. 
Here we discuss innovative technologies that provide multiple tied-array beams and multi-beam receivers, for connected arrays and single dishes elements of a VLBI array, respectively. 
%There are two approaches, to form multiple beams for connected arrays and to form multiple beams for single-dishes.

\subsubsection{Multiple tied-array beams from antenna arrays}\label{sec:tab}

When using a connected array as a VLBI element, such as the VLA, ATCA or WSRT, it is common to combine the signals from the array's individual antennas  and provide this as a single tied-array output. 
The connected array needs to be self-calibrated, that is `phased-up', before the formation of the VLBI datastream.\footnote{\url{
http://science.nrao.edu/facilities/vla/docs/manuals/obsguide/modes/vlbi}\\
\hspace*{0.55cm}\url{http://www.atnf.csiro.au/vlbi/dokuwiki/doku.php/lbaops/lbascheduling}}
A tied-array beam (TAB) has the same FoV as the connected array synthesised beam; that significantly restricts the FoV, compared to the primary beam of a single antenna.
%The normal solution for this is to/This situation can be alleviated by using only antennas that are close together, which increases the beamsize.
A solution to increase the TAB size consists of using  only the antennas that are close together.
A more effective alternative would be to provide multiple streams, each corrected for different pointings (or phase centers) within the single antenna FoV.
Such functionality would be essential for in-beam \MV{}, as discussed in Sects.~\ref{sec:next} and \ref{sec:methods_mv}, for ultra-precise astrometry with the next-generation instruments. It would also enable more efficient VLBI surveys with the SKA.
Historically no connected array has offered the possibility of forming such multiple TABs, but this will change in the near future. 
APERTIF, on the Westerbork Synthesis Radio Telescope, has had plans for this functionality (although they are currently on hold); the next correlator upgrade for ATCA (BIGCAT) will provide this, and similarly for ASKAP. %It has not yet been used in 
This would result in the full sensitivity of the connected array, with the same potential FoV as for the individual antennas.

Multiple phase-centres (i.e., tied-array beams towards multiple directions) will be a standard offering for tied-array SKA operations; 
this is essential given its small synthesised beamwidth. 
The tied-array FoV for SKA-Mid would be around or less than an arcsecond, and this would be combined with antennas having FoVs of many arcminutes in VLBI observations.

\subsubsection{Multi-beam receivers}
Receiver packages hosting multi-beam receivers, that is multiple independent receiver horns, have an excellent track record in accelerating survey speeds by enlarging the FoVs. For example the Parkes Multi-Beam, which increased the effective FoV of the  Parkes dish (64m) by a factor of 13, was essential for the rapid HI all-sky survey HIPASS \citep{hipass} and the HITRUN pulsar surveys \citep[e.g.,][]{hitrun}. Note however that these traditional multi-beam receivers do not provide continuous sky coverage as the individual beams do not overlap. 

The capabilities of the multi-beam receiver also allows for simultaneous VLBI datastreams to be collected from those beams pointing at multiple VLBI targets, to be cross correlated with the single datastream from a smaller antenna. 
Such a proof-of-concept was performed with the Parkes Multi-Beam to the  ASKAP test-bed (12m), also located at Parkes, 
as part of the preparatory work for ASKAP \citep[][]{askap_vlbi}. % ASKAP-VLBI TEST
A more interesting demonstration would be to cross correlate the SKA-scale collecting area of FAST (500m), which is equipped with an operational Multi-Beam receiver providing 19 dual polarised beams spread over the same FoV as a 30m antenna,  with other VLBI antennas. As yet no such demonstration has been performed. 

\subsubsection{Phased Array Feeds}

\begin{figure}
    \centering
    \includegraphics[width=0.475\textwidth]{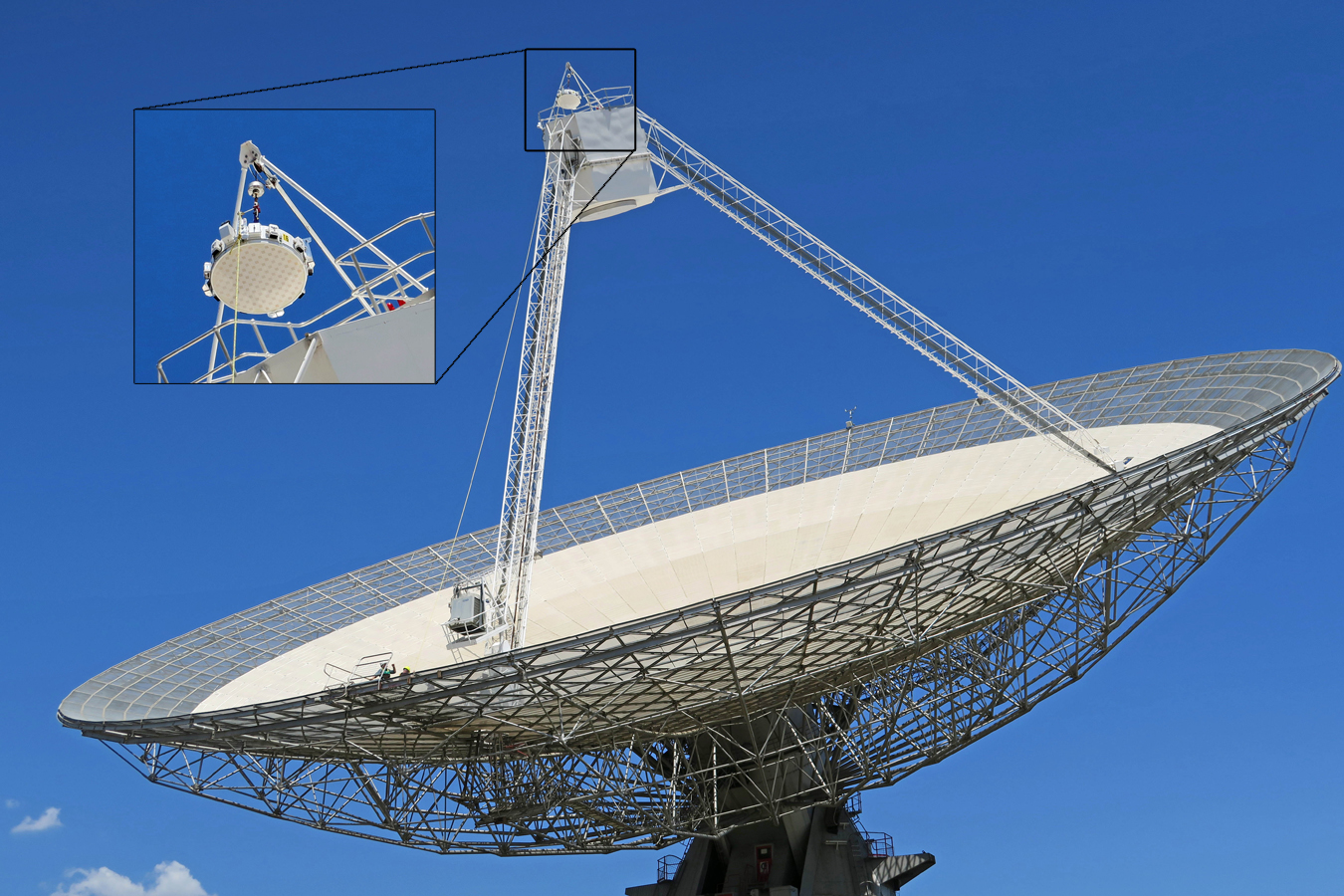}
    \includegraphics[width=0.42\textwidth]{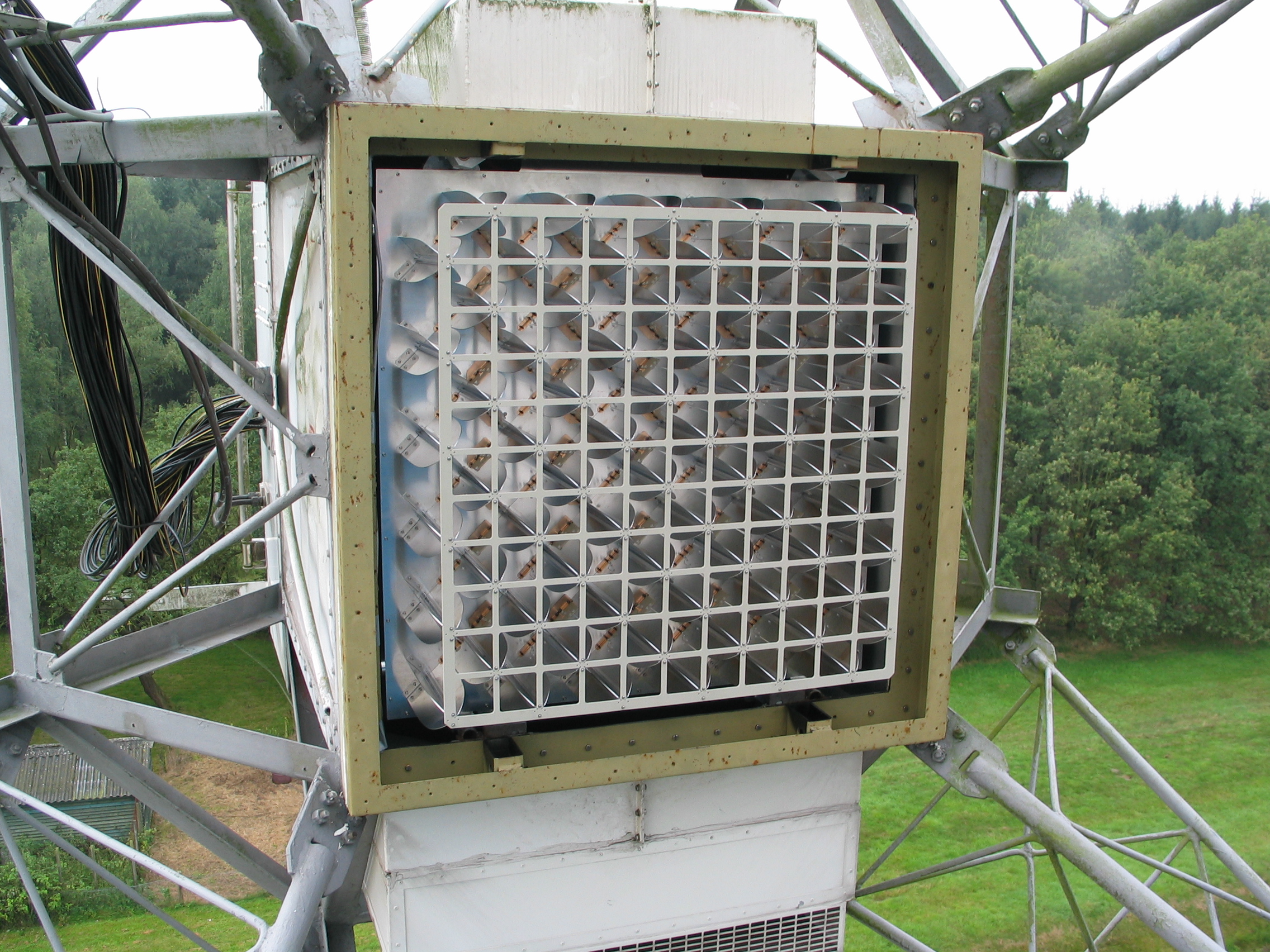}
    \caption{Phase Array Feeds on Parkes-64m (left) and WSRT (right) telescopes, respectively. Shown are the Mk-II CSIRO PAF (originally designed for ASKAP and modified for Effelsberg-100m), where a PCB checker-board element approach is taken, and APERITIF for WSRT, which uses an array of Vivaldi elements. Image credit: CSIRO and ASTRON.}
    \label{fig:paf}
\end{figure}

The most applicable example of increasing the FoV is provided by phased array feeds, which provide a uniform access to any part of the sky within the enlarged FoV, see Fig. \ref{fig:paf}. 
%can allow any part of the  sky within the enlarged FoV to  be  accessed. 
%
Phased Array Feeds (PAF) at the focus of a parabolic radio telescope address one of the great limitations of single telescopes and arrays; that they only see one point in the sky, the equivalent of a single pixel camera. When used as part of a connected array this limitation is offset by the fact that one can form a high resolution image of the radio sky within that single pixel. 
Nevertheless, it was recognised that with the capabilities of modern backends one can do much better than this \citep[see discussion in][]{fisher_00}. 
PAFs on large single dish antennas allow these highly sensitive telescopes to observe a greater portion of the sky.
The technologies behind the APERTIF and ASKAP PAFs apply multi-beam approaches to interferometres, each pixel being in essence a separate interferometric array with a slightly offset pointing centre.
The benefits for precision astrometry is when these PAF-enabled instruments form an element in the VLBI array.
The most effective implementation of this is to match the FoV of the highly sensitive, larger single dishes, to the larger FoV of the less sensitive, smaller single dishes \citep[see discussion in][]{rioja_08}, for calibration using \MV{} (Sect.~\ref{sec:methods_mv}).  
The  antenna arrays currently fitted with PAFs are all comprising smaller dishes, so the benefit for VLBI is less.
In these cases the multiple tied-array beam forming (Sect.~\ref{sec:tab}) is the most applicable capability.
In the future we can expect to perform SKA-VLBI observations with the multiple TABs of SKA, plus a network of large telescopes with PAFs and medium size single pixel telescopes, as illustrated in Fig. \ref{fig:radio_int}.
Currently large telescopes equipped with PAFs, such as Effelsberg (100m), Jodrell Bank (76m) or (planned) FAST (500m), Parkes (64m) and GreenBank (100m), could serve as pathfinders for \MV{} in the SKA-era. Parkes has just been funded to implement a new version of the ASKAP PAF \citep{dunning_16}, which combines the Vivaldi feeds of APERTIF and the checker-board technologies of ASKAP with cryogenic cooling. It is expected that this will reduce the effective system temperature T$_{\rm sys}/\eta_{\rm eff}$ from about 74K for the ASKAP system to 27K for the CryoPAF, delivering a large boost in performance.

%\sout{Cryogenic PAFs .... ``rocket'' paf (prototype)}

\subsubsection{Empirical atmospheric studies for estimating MultiView astrometric limits}\label{sec:study_mv}

\begin{figure}
    \centering
    \includegraphics[width=0.375\textwidth]{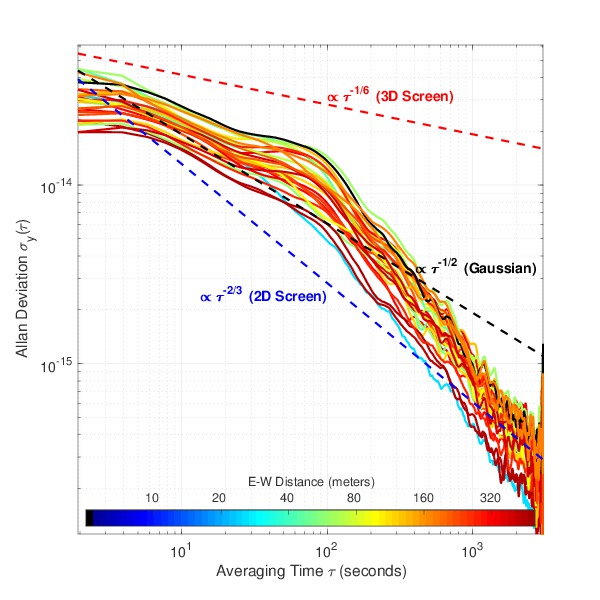}
    \includegraphics[width=0.58\textwidth]{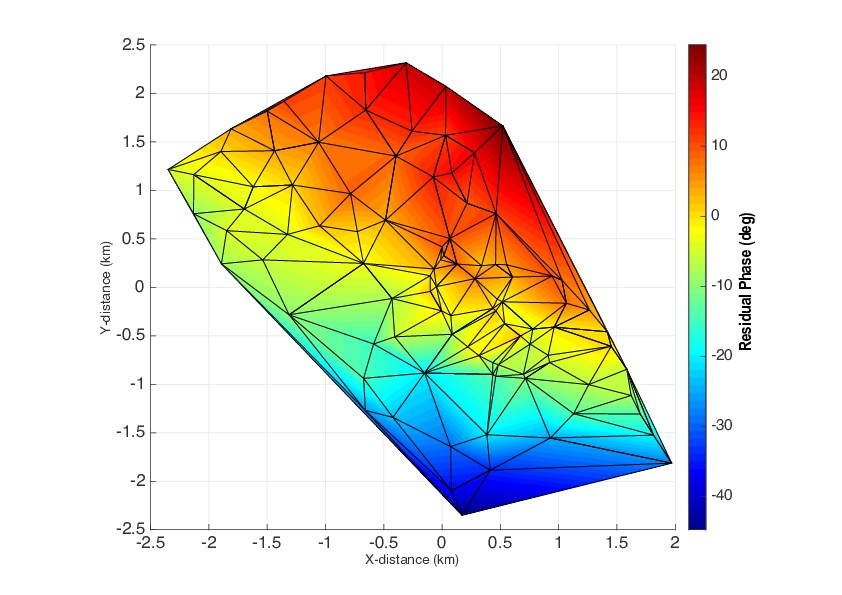}
    \caption{{\it Left}: An example of the tropospheric structure function from tip-tilt corrected observations with the SMA at 230\,GHz from Keating etal. (in prep.). The Allan Standard Deviation for each telescope pair is plotted, with colours indicating the baseline length along EW. The longer-term trend suggests the limiting systematic residual errors are from a 2D thin screen over the array, which extends to $\sim$6\degr{} (assuming the dominant water layer is at 3km).
    % SMA is at 5km itself? 
    {\it Right}: An example LEAP measurement of the ionospheric directional dependent phase surface over MWA Phase-2 at 150\,MHz. The typical phase residual after subtracting a planar fit is equivalent to 1\,milliTECU (Dodson etal., in prep). { The phase screen extends out to $\sim$1\degr{} for MWA Phase-2, with a dense sampling due to the 128 antennas.} % \citep[Rioja etal., in prep][]{rioja_leap2}.
    \label{fig:struct_func_sma}}
\end{figure}

A crucial question for \MV\ astrometry is to characterise to what level the atmospheric residuals can be minimised by the spatial interpolation between the solutions of multiple calibrator sources. 
G. Keating has been testing strategies to correct for tropospheric gradients for Sub-Millimeter Array (SMA) observations at 230GHz using tip-tilt corrections (Keating, in prep.). % \citep{keating_20}. 
Fig. \ref{fig:struct_func_sma} {\em left} shows an example of the Allan Standard Deviation (ASD) formed from the post-correction residual phases. Baselines of 300m are equivalent to about 6\degr\ on the sky, which is the maximum scale of interest for phase referenced VLBI. 
% sqrt(300)*3e-15*3e8=0.016mm -- is this correct? Units? 
% 3e-14 (at 1sec) *3e11=0.01mm 
% (300)^(2/3)*3e-15*3e8=0.04mm  
In the example shown the ASD falls as $\tau^{-2/3}$ at long timescales, which indicates that the residuals act as if they are from a thin 2D screen. 
%For other days\footnote{https://www.cfa.harvard.edu/$\sim$gkeating/SMA/phase\_screen/struc\_func.html}, 
On shorter timescales, when the thermal noise levels dominate, the ASD initially follows a $\tau^{-1/2}$ relationship, before transitioning to $\tau^{-2/3}$ after about 100 seconds. 
The ASD at 300\,seconds is always better than $\sim3\times10^{-15}$ for 100m baselines and is the equivalent to about 0.01mm of uncompensated residual excess path length (i.e., $\Delta\ell^{MV}_z$). 

To quantify the ionospheric residuals we use measurements at m-wavelengths from the SKA-Low precursor, the Murchison Widefield Array (MWA), which are fully discussed in \citet{dodson_evn_18}. 
To characterise the effects of the fine scale spatial structure we have used the by-products of the calibration method LEAP \citep{rioja_18}. This method provides the station-based phase solutions of all calibrators lying within the very large FoV, as shown in Fig. \ref{fig:struct_func_sma} {\em right}. %, in parallel. 
These solutions can be used to characterise the smoothness of the TEC screen, at the sub-degree scale for the calibrator directions. This is the regime most relevant to the SKA. %  \sout{(phase referencing with)}  \sout{(to in-beam sources for the single dish antennas)}. 
We have taken a set of calibration solutions from observations carried out on 18 Dec 2017, which has been classified as of moderate weather.
%, and another day which was classed as unmanageable. 
All of these datasets were from the extended MWA Phase-2 array, with a maximum baseline of about 6km; this corresponds to a maximum angular scale is 1\degr{} for a nominal ionospheric height of 300\,km.
We assumed that the measured TEC surfaces could be approximated by planar surfaces (or other low-order polynomials) and measured the residuals of the fits. 
We limited ourselves to the strongest ($>$25\,Jy) sources within each FoV, because it is vital to obtain results with the smallest possible intrinsic thermal noise.
The LEAP measurements show an RMS of uncompensated residual excess TEC of $\sim$1\,milli-TECU (i.e., $\Delta{\rm I}^{MV}$) after a planar fit has been subtracted for 90\% of cases.
Therefore in the majority of cases there are no significant deviation from a linear gradient
%so these results suggest no angular dependence 
at small ($<$1\degr) angles. These results predict that \MV, with three calibrators up to $\sim$60\arcmin{} away, would result in $\sim$5\,\uas\ astrometric precision at 1.6 GHz for most weather conditions with next-generation instruments. Additional calibrators ($>$3) allow higher order surfaces to be fitted, and increase the range of weather conditions in which high precision astrometry can be achieved. For this reason, among others, the current recommendation for \uas-precision \MV{} with SKA-VLBI is that eight VLBI beams will be required. 
{ We would recommend that these calibrator sources should evenly span the FoV of the smaller antennas in the array, with the larger antenna forming multiple pointed beams towards the sources. These studies suggest that the calibrator and target sources can be separated by up to a degree (for ionospheric corrections) or more (for tropospheric corrections) without introducing non-trackable changes in the phase screen. 
}
It is worth noting that we also used results from the MWA Real Time System \citep{mitchell_08}, to show that the larger residuals from \MV\ with angular separations up to 6\degr{} \citep{dodson_evn_18} should match the sensitivity of current instruments.
This has been confirmed in \citet{rioja_17} with 6\degr{} source separation at 1.6\,GHz and with recent VLBI test observations { using OzScope \citep{hyland_phd}} with sources separated by up to 8\degr{} at 8\,GHz.
These atmospheric studies provide the basis for the \MV{} astrometric error estimates listed in Column 3 of Table~\ref{tab:inbeam_mv} and in Fig.~\ref{fig:astro_history} and Sec.~\ref{sec:future}.
Note that there is no dilution effect for these \MV{} residuals.

\section{Astrophysical applications with current instruments}
\label{sec:applications}
\typeout{Present Scientific applications. Grouped by technique rather than science}
Here we review some of the recent published results from high precision radio astrometric surveys in the last few years, drawing out the important contributions from large and un-biased surveys, and link these to improvements in methods.
For a more general review of the field see the references in \citet[][]{reid_micro}.

\subsection{Pulsar parallax at 1.4\,GHz: PSR$\pi$} 

\begin{figure}
    \centering
    \includegraphics[width=0.47\textwidth]{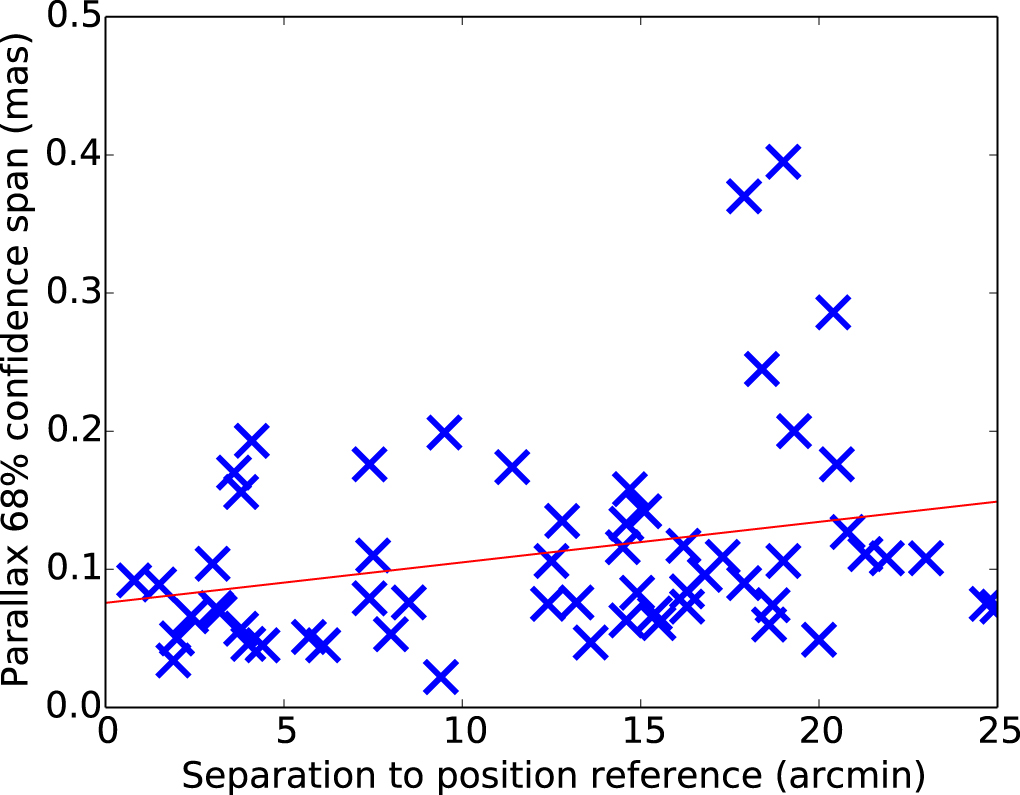}
    \includegraphics[width=0.4\textwidth]{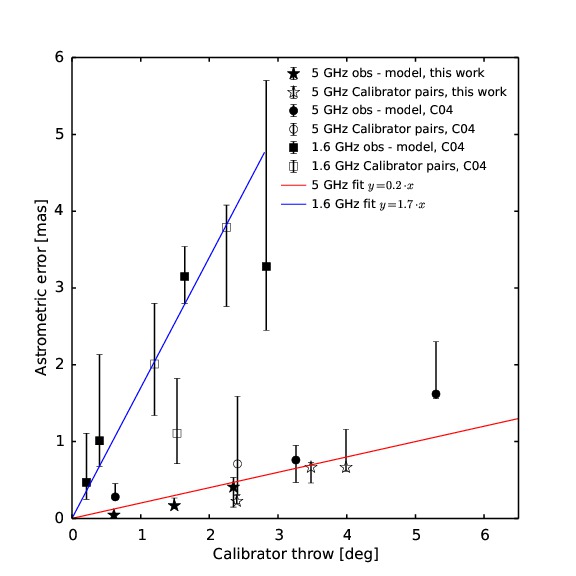}
    \caption{Comparisons of PR-type astrometric errors achieved with existing instruments as a function of angular separation between target and reference sources, from {\it left)} \citet[][Fig. 8]{deller_18} at L-band, for the parallax error and from {\it right)} \citet[][Fig. 1]{kirsten_15} for the per epoch astrometric error that includes data from \citet[][Fig. 3]{chatterjee_04}, at 1.6 and 5\,GHz.
    For the conventional PR ({\it right}), where calibrators will not be thermally limited in general, the dependence on angular separation is clear, which indicates that the errors are systematic. On the other hand for in-beam PR ({\it left}) the dependence on solely angular separation breaks down, due to the low SNR; that is as in-beam calibrators tend to be weaker the thermal and the systematic errors are approximately equal.
    %this results in thermal-error limited astrometry. 
    Thus increased sensitivity with the next-generation of instruments will not result in higher astrometric precision, with these traditional methods.
    This is because the sensitivity will either allow the use of closer but weaker calibrators that match the current DR and result in thermal-error limited astrometry, or calibrators at approximately the same separation as currently, that would then result in systematic-error limited astrometry.
    The right-hand plot also demonstrates the difference in PR astrometric quality at 1.6 and 5\,GHz from real experiments, as indicated by the fits to the 5GHz and 1.6GHz data; the slopes are steeper than the expectations from \citet{a07} (which is 1\,mas/\degr at  1.6\,GHz), but the ratio of the slopes agree, to within errors.
    \label{fig:ang_sep}}
\end{figure}

Pulsar trigonometric distances provide important insights into pulsar emission physics, their equation of state, gravitational waves and SNR associations. 
Surveys allow for the construction of the 3D-structure of the free electrons in our galaxy, by connecting the pulsar dispersion measures to their trigonometric distances \citep{YMW16}. 
Also, the interpretation of kick velocities to understand the properties of the binary progenitors requires accurate distances for the pulsars.
The major 1.4\,GHz VLBA pulsar astrometric program PSR$\pi$ has released the `data-paper' \citep{deller_18} for the project, which increased the number of more distant ($>$2kpc) pulsars with accurate distances by a factor ten. 
%, which provides astrometric measurements for 60 pulsars. 
%The median astrometric error was 40$\mu$as in parallax, with a median pulsar-calibrator separation of 60$\amin$.
PSR$\pi$ is a beautiful demonstration of a successful in-beam PR survey. 
%The 30\amin\ primary beam was searched for phase referencing sources. For 60 pulsars out of 110 initial target list a nearby calibrator could be identified, with one at only 0.8\amin, and a median separation of 14\amin. % The maximum was 60\amin, and the mean parallax error was 70uas
Out of the 110 initial pulsars proposed an in-beam reference source could be detected for 60 targets. The minimum pulsar-reference source separation was only 0.8\amin\ and the median angular separation was 14\amin.
%, with one at only 0.8\amin, and a median separation of 14\amin.
The median pulsar parallax error is 45$\mu$as, from typically nine epochs of observations (i.e., $\sim$135\uas{} astrometric error per epoch). 
%It is worth noting that the errors resulting from the case with smallest angular separation are exactly those for the cases with median separation, because the astrometric errors for that weak source are dominated by the thermal errors. 
It is worth noting that the 
%errors in the case with smallest angular separation are exactly those of the median separations, because the 
astrometric errors for these close-but-weak sources are dominated by the thermal errors. 
This is shown in Fig. \ref{fig:ang_sep}, where the analysis from \citet{deller_18} is contrasted to that in \citet{kirsten_15} (which includes data from \citet{chatterjee_04}).
%{\color{green} We stress that in this situtation, to achieve higher precision astrometry we need to reduce the thermal noise {\em and} the systematics. Closer targets would be expected to be weaker, so we can use the increased sensitivity of the next-generation instruments to reduce the systematics or the thermal noise. {\bf bound to be a repeat }}
%
\citet{deller_18} (Eqs. 1 and 2) reach the same conclusions as described in Sect.~\ref{sec:methods_mv} above; that to achieve systematic errors at the level of 10\,$\mu$as per epoch of in-beam PR observation the angular separation should be $\sim$1\amin, and that the sources should be strong enough so that thermal errors do not dominate, that is the DR should be 1000:1.
This underlines the strength of \MV, for which the effective angular separation can be essentially zero, whilst using strong and widely separated reference sources to provide the matching precision.

\subsection{Galactic astrometry}
\subsubsection{Galactic structure: BeSSeL and VERA}

It is extremely challenging to derive the structure of the Milky Way using observations from the inside, particularly with optical observations that are effectively blocked by dust extinction along the plane. Radio observations of HI and particularly maser astrometry therefore provide a unique insight.
%Our knowledge of the Milky Way is being significantly revised, driven by 
The combined on-going efforts from VLBI astrometric projects carried out mainly with the VLBA and VERA %such as the BeSSeL survey, 
are significantly revising our knowledge of the Galaxy \citep[respectively]{reid_19,vera_cat}.
These observations result in high-precision astrometric measurements of the source coordinates, trigonometric parallax distances, proper motion and radial velocity towards a few hundred high mass star-forming regions \citep[references therein]{reid_19,vera_cat}. %  (Reid+2009c, 2014, Honma 2012 and references therein - from Reid+Honma 2014). 
The targets are mainly massive young stars with 22\,GHz H$_2$O and 6.7\,GHz methanol maser sources, with a few additional methanol masers at 12\,GHz.
These astrometric surveys are revealing the three-dimensional spiral arms structure of the Milky Way with unprecedented accuracy.
A highlight is a record breaking distance measurement, for the \wat\ maser G007.47+00.05, with a parallax of $49\pm6\mu{\rm as}$, equivalent to a distance of $20.4^{+2.8}_{-2.2}\,{\rm kpc}$ \citep{sanna_17}.
In the latest results \citet{reid_19,vera_cat} %Reid 2009, Honma 2012; Reid 2014; Reid 2019 (in prep) 
have used the combined set of astrometric measurements to refine the underlying dynamical model described by the 
fundamental constants $\rm R_0$ and $\theta_0$, the distance to the Galactic Center and the circular rotation speed of the Local Standard of Rest (LSR), respectively. 
The parameter values from the latest modelling of \citet{reid_19} are $\rm R_0=8.15\pm0.15\, {\rm kpc}$ and $\theta_0=236\pm7\, {\rm km\, s}^{-1}$. 
For \citet{vera_cat} the results are $\rm R_0=7.92\pm0.16_{\rm stat}\pm0.3_{\rm sys}\, {\rm kpc}$ and $\theta_0/{\rm R}_0=30.17\pm0.27_{\rm stat}\pm0.3_{\rm sys}\, {\rm km\, s}^{-1}\, {\rm kpc}^{-1}$.
These results include, for the first time, precise astrometric results from observations of 6.7\,GHz methanol masers, e.g., \citet{xu_16,wu_19,sakai_18,sakai_19,zhang_19} and \citet{rygl_evn_18}. 
%% Others wu_19,sakai_18,zhang_19} 
The important advance over the pure ATC corrections from the use of the image-based variation of \MV{} for 6.7\,GHz observations is indicated by the improvement between the first and the later results (see Fig. \ref{fig:astro_history}).
Additionally, the LBA is now delivering its first methanol maser parallax measurements \citep{krishnan_15,krishnan_17} that fill in the fourth quadrant.
{ Observations towards the Galactic centre are extremely important to improve the dynamical models, and these could be provided by observations from the Southern hemisphere, such as with the SKA.
In the meantime this can be attempted with \MV{} astrometry with established arrays in the Northern hemisphere to correct for the larger errors due to the low elevation observations.
}
% \subsubsection{Individual results}

% The astrometric results have allowed detailed descriptions of the individual sources, and their intrinsic properties (such as size, luminosity, mass), to probe the structure and dynamics of jets and bipolar outflows and confirm associations by comparing astrometric results from cluster members.  
% %
% Among these, a highlight is a record breaking distance measurement, for the \wat\ maser G007.47+00.05, with a parallax of 49$\pm$6$\mu$as, or a distance of $20.4^{+2.8}_{-2.2}$kpc \citep{sanna_17}.
% Hitherto distances to the far side of the Milky Way have been impossible to measure accurately, because the parallax is very small, making it challenging for VLBI, and because the interstellar dust blocks optical light from those regions, making it also impossible for \gaia. 

\subsubsection{Spiral arms and density waves}

The BeSSeL results are also revising our knowledge of the structure of the local arm, revealing it as a major spiral structure with star-forming activity similar to that in the Galaxy's major spiral arms \citep{xu_16}. % (Xu+2016)  
\citet{honma_15} presented simulation studies that suggested the feasibility of using astrometric measurements of a few hundred sources at the current level of accuracy to discriminate between different dynamical models proposed to explain the nature and characteristics of the spiral arms.
That is, between the quasi-stationary density wave theory \citep{lin_64} %(Lin \& Shu 1964) 
and non-steady spiral arms, based on numerical simulations \citep{baba_15,baba_13}. %(Baba 2013). 
The former predicts that the dust and stars will be separated within the arms, whilst the latter predicts that they will be co-located. 
Work by \citet{hachisuka_15,sakai_15,griv_17,sakai_18,sakai_19} 
%testing the scenarios for the evolution and dynamics of spiral arms in a disk galaxy, {\bf add some ref from Sakai?} 
compared the empirical results with the predictions. \citet{griv_17} argued that the data supports density wave theory, based on the earlier dataset in \citet{reid_14}. This is refined in the work of Sakai etal. \citeyearpar{sakai_15,sakai_19} on the Perseus arm that suggests the addition of shocks to the density wave theory is required to explain their results. However they note that the full picture is almost certainly more complicated than this, as results for other regions appear less clear or even contradictory.
%from the density wave theory for masers in the Perseus spiral arm... Sakai 2018)

\citet{sakai_20} has recently published a VERA study on the outer rotation curve, in which they add a further four \wat\ maser distance measurements to the literature. 
They use these to determine that the maximum height of the Galactic warp increases with Galactocentric distance and has an oscillatory behaviour.
They suggest that these are possibly induced by the close passage of the Sagittarius dwarf galaxy.

%AuScope antennas is being equipped with wide band C-band receivers to carry out the counterpart of Bessel from the southern hemisphere...

\subsubsection{Early stages of stellar evolution: POETS} %POETS: Proto-stellar Outflows}
%Here some discussion of poetry.

\citet{Mos16} presented the first results of the Proto-stellar Outflows at the EarliesT Stages (POETS)  survey.
The POETS survey has the goal of imaging the outflows for hundreds of Young Stellar Objects (YSOs), at resolutions of a few mas and a few \kms.
As such it has afforded, for the first time, a VLBI survey of the properties of the most widespread interstellar masers based on a large, homogeneous dataset.
POETS uses the BeSSeL measurements of parallax and proper motion of \wat{} masers, combined with additional proper motion studies of 6.7\,GHz methanol masers, and continuum observations between 4 and 26\,GHz carried out at lower-resolution with the JVLA.
\citet{San18} reported on and interpreted the radio continuum data of the whole sample, and \citet{Mos19} completed the combined analysis of the radio continuum and \wat{} maser observations for all the targets, focusing in particular on the  \wat{} maser kinematics.
Recently, \citet{Mos20} has presented the statistics of the \wat{} maser properties.
The main result of the work so far is that the 3D velocity distribution of the \wat{}  masers near the YSO in all the sources of the sample can be interpreted in terms of a single physical scenario: a jet emerging from a disk-wind system. The statistical results indicate that the large majority ($>$80\%) of the \wat{} masers originate near (within 1000\,AU) the YSOs, and that about half of the maser population (with intensities $<$100 mJy/beam) is still undetected.

%POETS uses the BeSSeL observations of parallax and proper motion of \wat{} masers, combined with additional proper motion studies of 6.7~GHz methanol masers, and \wat{}  maser observations for all the targets, focusing in particular on the \wat{}  maser kinematics.

\subsubsection{Simultaneous multi-frequency astrometric survey of evolved stars with KVN}
\label{sec:applications_sfpr}

The Evolved Stars KVN KSP is providing new insights into the late stages in stellar evolution, illuminating the mass loss processes and the development of asymmetry in CircumStellar Envelopes (CSE), using a systematic astrometric study.
The project comprises of simultaneous multi-frequency observations of 22 GHz \wat{} and  43.1/42.8/86.2/129.3 GHz SiO masers towards a sample of 16  evolved stars sources, over multiple epochs with sufficient temporal sampling to trace the variations over multiple stellar pulsations.
It is performed using SFPR measurements and the outcomes are bona-fide astrometrically registered high resolution VLBI images of the spatial and temporal distribution of the emission at all transitions, free of any assumptions.
This has enabled astrometry for asymptotic giant branch (AGB)  stars at frequencies up to a factor of three higher than any previous work.
First results from these observations are now being published (e.g., VX Sagittarii \citep{yoon_18}, R Crateris \citep{kim_18} and V627 Cas
\citep{cho_19}).
Thus, the unbiased measure of overlap or offset between masers at different transitions, ring sizes and temporal evolution provides a
wealth of information to:
i) provide the basis to investigate the spatial structures and dynamics from the SiO to \wat{} maser regions associated with a mass loss process and to trace the development of asymmetry in CSEs;
ii) probe the astrochemistry of the matter thrown off into the ISM, as it is excited by the stellar pulsation of the host star;
and iii) discriminate between  competing proposed maser pumping mechanisms (i.e., radiative or collisional) and to deduce the physical conditions in the CSE.

\subsubsection{Long-period variable star astrometry for the cosmic distance ladder}\label{sec:lpv}
An important step on the cosmological distance ladder comes from period-luminosity relations in variable stellar sources.
There are several classes of objects, and accurate distances to Galactic examples are vital for the calibration of the relationship.
%Other stellar populations can also contribute to the understanding of the Galactic structure. An example would be 
Mira variables and other long-period variables, which are low- to intermediate-mass (1--8\Msun) AGB stars that pulsate with a period range of 100--1000 days, are the classic example for these studies. These sources are detectable to great distances, especially in the era of space infrared telescopes \citep{wood_15}, making them potentially a very powerful distance estimator.
%These sources form part of the cosmic distance ladder, through the correlation of their pulsation period and absolute brightness. %, especially in the era of space infrared telescopes \citep{wood_15}.
Optical astrometric observations of these sources have not been very reliable as AGB stellar diameters are of a similar magnitude to the parallax position shift, and their surfaces are mottled and variable.
VLBI astrometry contributes to this field by providing precise trigonometric parallaxes to the 1.6 GHz OH, 22GHz \wat\ and 43 GHz SiO masers associated with Galactic examples of these stars \citep{nakagawa_18}.
Recently, high-precision astrometric results, using ATC, have been obtained from observation of \wat\ masers \citep{zhang_17,nakagawa_16,sudou_19}. Additionally, a VERA survey using SiO masers is being developed. Alternatively, a OH maser survey using \MV\ astrometry could be useful \citep[see discussions in][]{orosz_17}.

% {
% \subsubsection{Improved data processing techniques}
% \label{sec:bayesian}
%  Not astrometry ....\\
%  Innovative statistical analysis techniques are being used to improve the robustness of astrometric results, by dropping the assumption of Gaussian error distributions. Bayesian analysis has been used by \citet{reid_16} to significantly improve the accuracy and reliability of distance estimates from more traditional observables, leveraged by the accurate parallaxes in the BeSSeL survey to other sources with similar parameters. They fold in the location of the source in the Galactic plane, including the kinematic distance and the known arm structures to provide a much more meaningful probability distribution function. 

% Not enough ... \\
% In a similar context, \citet{quiroga_17} performed a detailed simulation study to establishing the confidence with which the Galactic parameters can be determined from the parallaxes. 
% %(e.g Baade (huib), Reid+ galactic distance)
% The usage of more advanced mathematical and statistical methods in the data processing techniques such as: bootstrapping, Bayesian approaches, and Markov chain Monte Carlo (MCMC) analysis will improve the quality and robustness of astrometric results generally.}

\subsection{Extragalactic astrometry}

\subsubsection{Proper motion surveys of AGNs}
Large surveys with precise measurements of the proper motion of distant objects allow the extraction of global signatures, such as detection of the Gravitational Wave background, the Hubble expansion and the evolution of Large-Scale Structure, when combined with the redshift measurement.
\citet{truebenbach_17} %, \sout{following the approach in \citet{titov_13}}
has published a catalog of 713 VLBI proper motions of AGNs, derived from both archival and new measurements, 
which they have used to measure the Secular Aberration Drift. 
As the authors point out, this demonstrates that large precise catalogs are as useful in radio as they are in optical bands.
The measurements of each individual quasar will be affected by intrinsic core-jitter effects, however averaging over many of these uncorrelated errors using a large survey allows one to extract the global signatures. 
Their catalog is formed from the time-series of absolute astrometric VLBI Geodetic observations from the Goddard archive, and additional relative astrometric observations to improve on that dataset.

\subsubsection{Towards global absolute phase astrometry} %AGN core-shift studies: Wide angle astrometry at 43\,GHz}
%\subsection{Wide angle astrometry at 43,GHz - AGN core-shift studies}
% %34 High-Precision wide-angle astrometry at 43 GHz                              (VLBA, UVPAP) 

\citet{abellan_18} presented a first successful demonstration of precise differential phase-delay astrometry of the angular separations between 13 radio-loud AGNs around the North Celestial Pole, that are up to 15\degr\ apart, using VLBA observations at 43\,GHz. 
This is probably reaching the limit of the applicability of the analysis with UVPAP, given the increasing challenges for higher observing frequencies using PR.
Previous demonstrations from this team were at lower frequencies \citep[and references therein]{marti-vidal_16}.  
%
%âCore-shiftâ values agree with predictions from SSA effects
The precise astrometry measurements at 15 and 43\,GHz were compared, to study the jet physical conditions,  using the core-shift measurements, and to compare astrometric methods, among them SFPR. 
Their results agree with predictions from opacity and synchrotron self-absorption effects following the standard model for AGNs \citep[][]{bk_79}.

% {\bf is this in results?}
% Analysis with UVPAP holds the record for the phase-based (relative) astrometric measurements using the largest pair angular separations, up to 15\degr.
This determination is an important step towards wide field or global astrometry using the very precise phase observable.
The precise determination of the coordinates of extragalactic sources across the sky is relevant to measuring the ICRF.
Traditionally, wide field astrometry has been carried out using group-delay based analysis of geodetic VLBI observations, which traditionally has been about 30--40 fold less precise than the phase observable. 
%the upper frequency limit for application is 43 GHz (Abellan+ 2018f?), similar to phase referenced imaging technique. 
Nevertheless note that geodetic VLBI is also undergoing a regeneration, as discussed in Sect.~\ref{sec:vgos}, and the approaches are converging.

\subsubsection{Gamma-ray bright AGN survey: iMoGaBA}\label{sec:imogaba} %Origins of Gamma-ray Flares in Active Galactic Nuclei}
%{\bf Not sure about this one}
Interferometric Monitoring of Gamma-ray Bright AGNs (iMoGaBA) is a survey carried out with the KVN  to investigate the origins of gamma-ray flares in AGNs \citep{sslee_16}.
It is known that most compact flaring gamma-ray sources are AGNs, and determining the origin of these flares hinges on discovering their location in the jet.
%, allowing a clearer correllation of the Gamma-ray and Radio flares and structural changes to be made.
At the higher frequencies one expects the AGN core and the base of the jet to be exposed, as the otherwise bright synchrotron emission is reduced.
Combining simultaneous high-frequency high-resolution imaging of 34 radio-loud AGNs with gamma-ray emission and radio-flux monitoring, at 22/43/86 and 130\,GHz, has the potential to provide direct observations of where the emission comes from, during a flare.
IMoGaBA is not strictly an astrometric survey, but uses FPT-squared \citep[Sect.~\ref{sec:fpt2} and][]{algaba_15,zhang_17} to detect the weaker sources; also in the extension to global baselines (AiMoGaBA) astrometry with SFPR will be added. Therefore we have included it in this review.
%At the higher frequencies one expects the AGN core and the base of the jet to be exposed, allowing a clearer correllation of the Gamma-ray and Radio flares and structural changes to be made.
%It uses the calibration solutions at the lowest frequencies to extend the phase coherence at the highest frequencies, following FPT-squared method as described in Sect.~\ref{sec:fpt2} \citep{algaba_15,zhang_17}. 
Their observational interpretation is that the radio and gamma-ray locations are sometimes offset or orphaned \citep[for 1633+382 see][]{algaba_18a} and sometimes associated with the radio lobe hot-spot for longer term variations and the core for the short term variations \citep[for 3C\,84 see][]{hodgson_18}.
AiMoGaBA will target the issue of jet base wandering after high-energy flares, first seen in Mrk\,421 \citep{niinuma_15}. This will determine whether the changes are due to alterations in the orientation of the jet or in the environment and magnetic field conditions.

\subsection{Radio astrometry in the \gaia era}
%\section{Radio astrometry in the \gaia-era}

With the publication of \gaia\ Data Release 2 \citep[DR2;][]{gaia_dr2}, the comparison between % parallaxes of stars from VLBI astrometry to  %those in the Gaia DR2 catalog)
the \gaia\ positions, parallaxes, and proper motions, and accurate VLBI astrometric measurements for sources that have 
both optical and radio detections has become a very active field of research. 
Multiple studies focus on different objects, among them, AGN positions \citep{petrov_19_diff}, star-forming regions \citep{kounkel_18} and evolved stars \citep{langevelde_evn_18}.  

One of the interests of these comparisons is to verify these initial \gaia\ results. 
Among these, the parallax value for the Pleiades cluster is particularly important because of its role  in the astronomical distance ladder. Previous comparisons of VLBI and Hipparcos parallaxes resolved the distance controversy by ruling out the Hipparcos value \citep{melis_14}. 
The new \gaia\ results agree with all VLBI Pleiades parallax measurements of eight stars that show compact emission \citep{MelisNew}. Similar studies of a sample of YSO candidates \citep[GOBELINS:][]{ortizleon_18} also show agreement, but hint where further improvements in the measurements of \gaia\ systematics could be made. { Individual examples of disagreement can be found, for example on SV Peg \citep{sudou_19}.}

\citet{xu_19_gaia_comp} used a sample of 108 sources with both VLBI and \gaia{} DR2 astrometric results.
The sample included many different types of stars, including YSOs, AGB stars, pulsars and other radio stars.
They found excellent agreement between VLBI and DR2 parallaxes for all subsamples, except for the AGBs.
These stars are generally large, variable and embedded in dusty environments, as mentioned in Section \ref{sec:lpv}. Furthermore many are in binary systems, for which the orbital motions are not modelled in DR2.
For these cases radio observations of the masers around AGB stars can yield parallax accuracies roughly an order of magnitude better than \gaia.
The compendia of comparative studies dealing with the \gaia{} parallax zero point issue confirm that these are in agreement with expectations at this point in the mission \citep{lindegren_18}, being negative and under $100 \mu$as with an average bias of $-29\mu$as.

For AGNs, 
Mignard et al. \citeyearpar{gaia_dr2e} reported that in general, the VLBI/DR2 position differences are small (close to their uncertainties of a few hundred \uas) with some exceptions. 
Petrov and collaborators \citep{kovalev_17,petrov_19_agn,petrov_19_diff} 
used the  distribution of the \gaia{}/VLBI position angle offsets 
with respect to VLBI jet directions to determine the nature of the outliers; 
they find that the position offsets show a preference to align along the jet. %, and at a smaller extent in the direction opposite to the jet.
%propose that examime the nature of the outliers
They argue that this could be caused by the presence of unresolved optical jets below the resolution of the \gaia\ point spread function, i.e., 100--300 mas, and propose that these VLBI/\gaia{} position offsets could be used for probing properties of the innermost regions (i.e., accretion disks and the relativistic jets) of the AGNs.

Several papers have been published that compare \gaia\ results to the proposed ICRF3 sample \citep{charlot_icrf3}. These suggest that ICRF3 is a better match to \gaia\ than ICRF2 \citep{frouard_18}. 
The work to identify a suitable common set of radio stars that are to be found in both datasets, to bring the two reference frames into agreement has begun \citep{makarov_19,lindegren_20}. Again this work is helping in identifying systematic errors in the \gaia\ results, to be corrected in the future releases.

\section{Future ultra-precise astrometric possibilities with the next-generation instruments}
\label{sec:future}
\begin{figure}
    \centering
    \includegraphics[width=\textwidth]{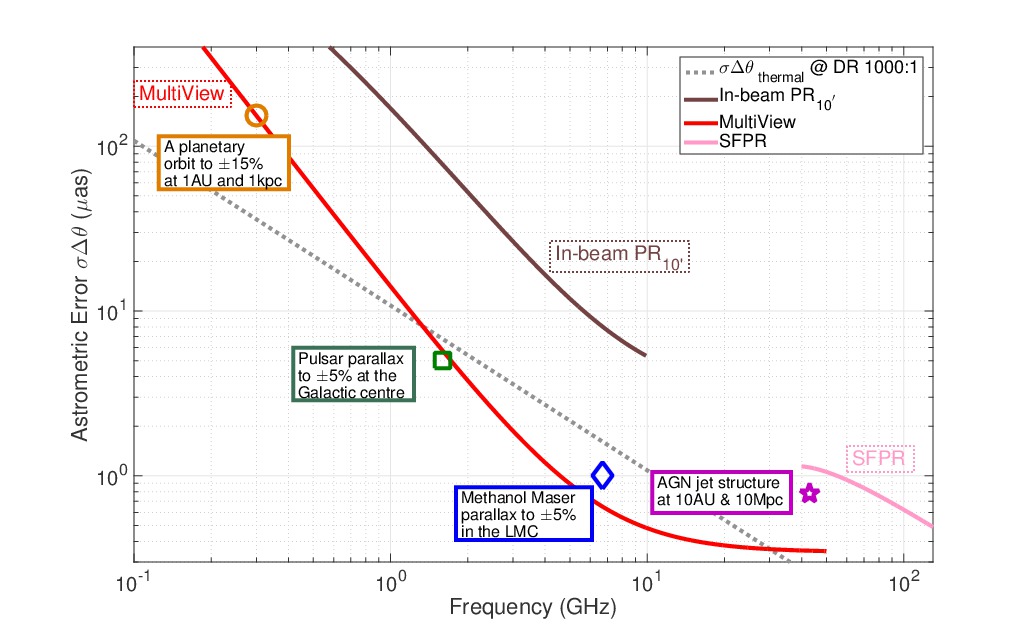}
    \caption{Estimates of single epoch astrometric performance for the next-generation instruments, as a function of frequency between 100\,MHz and 130\,GHz. 
    The grey dotted line is for astrometric errors arising from thermal noise with a DR of 1000:1 and 6,000\,km baselines ($\sigma\Delta\theta_{\rm thermal}$). The solid lines are for systematic calibration errors ($\sigma\Delta\theta^{\rm cal}$) with atmospheric residuals of $\Delta{\rm I}$=6\,TECU and $\Delta\ell_z$=3\,cm, and angular separations of 10\amin{}, $\leq$1\degr{}, 1\degr{} for in-beam PR, \MV{} and SFPR respectively.
    %    from the techniques The solid lines are the systematic calibration errors expected for a 
    Systematic in-beam PR calibration errors (brown solid line; $\sigma\Delta\theta^{\rm PR_{10^\prime}}$), are now far above the thermal limits indicated in grey (c.f. Fig. 2), and would therefore limit the astrometry.
    On the other hand the limits from \MV{} (red solid line; $\sigma\Delta\theta^{\rm MV}$), above 1\,GHz is below or around those from the thermal noise.
    The limit is around a \uas, which is the probable limit of the fiducial reference points (Sec.~\ref{sec:ref_pts}).
    \MV{} promises to provide astrometric accuracies at an order of magnitude lower in frequency, which match those of current best in-beam PR$_{10^\prime}$ astrometry at 2\,GHz.
    Therefore SKA with \MV{} opens up a new window, down to 100\,MHz, for precision astrometric applications and for ultra-precise astrometry at frequencies $>$2\,GHz. 
    %close in-beam calibrator at 10\amin{} with , shown in purple up to 10-GHz. These are far above the thermal limits, and therefore would set the astrometric limitation.
    %The red line is for residuals after MultiView calibration, with residuals of 1\,mTECU and 0.01mm, up to 50-GHz. At frequencies below a few GHz this is the dominant source of error, but mas-level astrometric measurements are possible at, say, 300MHz. Above a few GHz the ionospheric residuals after MultiView are lower than the thermal limits, until the MultiVew tropospheric residuals are reached. 
    %We note the tropospheric residuals are a very conservative limit from { \citet{carilli_99}, which does not include any planar correction and includes the dynamic contributions. Therefore we also include a lower limit, where we assume we can reduce the tropospheric residuals by an order of magnitude as for the ionosphere. }
    %in this example. No direct measurement of the tropospheric residuals would be after MultiView have been made. We assume the tropospheric residuals are improved by 2 orders of magnitude, as the linear approximation of $\sec(Z)\tan(Z)$ above 20\degr\ for a 1\degr \ separation is of that order or better. This is indicated by the dashed vertical line.}
    Finally SFPR (pink solid line; $\sigma\Delta\theta^{\rm SFPR}$), referenced to 22\,GHz and a second source at 1\degr{}, which runs from 40 to 130\,GHz. This would be applicable to provide ultra-precise astrometry up to the highest frequencies of ngVLA with simultaneous multi-frequency observations.
    Superimposed with symbols are indicative experiments: with SKA-Low at 300\,MHz we could track the orbit of a planet at 1AU around a star at 1kpc to 15\% accuracy (per epoch), with SKA-Mid or ngVLA at 1.4GHz we could measure the parallax to 5\% accuracy at the Galactic centre, at 6.7GHz we could measure the parallax of methanol masers in the LMC to 5\% accuracy and with ngVLA at 43GHz we could measure the jet structures and core-shifts on a 10AU scale for a AGN at 10Mpc. 
    \label{fig:ska-limits}}

\end{figure}

We have presented the  best  outcomes at this point in time for astrometric surveys in Sect.~\ref{sec:applications}, and linked these to the improvements in methods. 
We have listed the next-generation of methods in Sect.~\ref{sec:methods_new}, and how those can improve the astrometric outcomes.
The next step is to speculate what is possible with the combination of the next-generation of instruments (Sect.~\ref{sec:next}) with these improved methods, and what science could be targeted. 
These possibilities can be summarised in a similar plot to that with which we explored the current status (Fig.~\ref{fig:astro_history}), but for the future, in Fig.~\ref{fig:ska-limits}.
With the next-generation of instruments the frequency range can be enlarged to cover 0.1 to 130\,GHz and with an increase in the sensitivity limits and smaller astrometric errors. 

VLBI with SKA will be much more sensitive than our current thermal limitations but the baselines will still be global, so we have assumed a DR of 1000:1 and baselines of 6,000\,km to provide a thermal astrometric error limit (grey dotted line) in Fig.~\ref{fig:ska-limits}.
Also plotted are the estimated astrometric limits from in-beam PR systematic calibration errors with a $10^\prime$ pair separation and $\Delta{\rm I}$=6\,TECU and $\Delta\ell_z$=3\,cm residuals (brown line, and as for Fig.~\ref{fig:astro_history}), which matches the thermal potential that is achieved with the current instruments and a DR of 100:1. 
We repeat that to improve the astrometric accuracy significantly with a single in-beam reference source requires that both the thermal limits provide DR 1000:1 and that the reference source is as close as 1\amin. 
The improvement of sensitivity from the SKA can provide the former or could improve the latter, but not both simultaneously. 
However we have shown in Sect.~\ref{sec:study_mv} that \MV{} calibration errors (red line, as for Fig.~\ref{fig:astro_history}) result in systematic ionospheric residuals of the order of a few milli-TECU and insignificant tropospheric residuals (equivalent to less than 1\uas{} above 5\,GHz).
The increased sensitivity will improve the chances of finding multiple in-beam calibrators at frequencies $\le$ 1.4\,GHz for Phase-1 and at frequencies $\le$ 6.7\,GHz for Phase-2 (see Table~\ref{tab:inbeam_mv}, Col. 8). %, but as the potential search area falls as square root of the separation whilst the number of sources is approximately linear with sensitivity. 
Of course at higher frequencies source-switched \MV{} will be possible with SKA and ngVLA, while correcting for static systematic errors in the analysis.
With such limitations we expect the \MV{} astrometric errors to dominate the nominal thermal limit below 1\,GHz (albeit far better than the astrometric limits currently achievable), and be below the nominal thermal limit above 1\,GHz.
 %These should be just the non-linear component of the mapping function (i.e., $\tan(Z)\sec(Z)$) corrected by sources separated by 1\degr, will be less than 13\% above 5 degrees elevation. For most of the sky it $<$1\%. Therefore, for the high frequency regime the main benefit of \MV\ will be when the target source is low in the sky.
{ On the other hand SFPR (pink line) potentially delivers $\sim$1\uas{} astrometry in observations with ngVLA at high frequencies, and would probably only be limited by stability of the fiducial reference points (see Sect.~\ref{sec:ref_pts}).}

Given these limits, science goals for VLBI with SKA-Low, with astrometry at the mas level, could be parallaxes of nearby ($\le$1kpc) sources. 
For example, with SKA-Low, decametric emission from a Jupiter-like planet located at 1AU from its star would be an easily detectable (although complex to untangle) astrometric signal.
Astrometric observations of pulsars at the Galactic Centre at 1.6\,GHz with SKA-Mid or ngVLA would have matching thermal and systematic limits, and we should be able to reach astrometric errors of 10\uas{} per epoch, providing a 5\% error in the distance estimates.
%
%This level is essential to remove  Galactic rotational acceleration contributions and push the limits of General Relativity tests in strong gravitational fields. 
%
Astrometric observations of 6.7\,GHz Methanol masers with SKA-Mid or ngVLA would be limited by the achievable thermal limits, and we should be able to reach errors of a few \uas. %Such sources would also have excellent SNR, which is required to allow DR of 1000:1.
%Science possibilities for SKA-VLBI are to be found, among other places, in \citet{cristina_ska_vlbi}.
Parallax and proper motion measurements of such sources in the Magellanic Clouds would determine whether the clouds are gravitationally bound to our Galaxy, or not. %just passing by.
{ For ngVLA, continuum observations with sub-\uas{} astrometric precision would be possible with \MV{} and SFPR, that would probe down to 10AU for a source at 10Mpc. This is the scale where the AGNs launch the jet, so studies of the jet launching mechanism and the evolution will be very illuminating. 
}
Detailed considerations of many other potential SKA-VLBI projects have been recently published  \citep{ska_vlbi,paragi_evn_18,jj_wp10,vlbi_2030}.

The ngVLA planning for `Astrometry and Long Baseline Science' \citep{ngvla_lbs} includes the goal of 1\uas\ astrometry. Their objectives, among other possibilities, are:
  improved pulsar distances, to remove the Galactic rotational acceleration uncertainty from the measurements of the post-Keplerian parameters \citep{kramer_timing};
  to measure the proper motion of the post-merge ejecta from neutron star-neutron star gravitational-wave sources, as demonstrated for GW-170817 by \citet{mooley_19_gw}, 
    which can be used to improve the measurement of H$_0$ using geometric constraints; % Not 1uas
  tracking \wat\ maser motions in the sub-parsec disks around AGNs; to probe the magnetic fields at these scales through their polarisation properties; and
  the proper motions of AGNs with \wat\ masers in the Local Group of galaxies, following on from the work of \citet{brunthaler_07}. 
%
%the Methanol maser project would also be possible, but in addition the field of mm-VLBI astrometry holds many interesting possibilities.
%A wider survey of the maser transitions around AGB, at mill... improves on KVN? 
All of these are relative ultra-precise astrometry projects. The neutron star projects would be carried out at frequencies below 8\,GHz, therefore would require \MV\ calibration, for which multiple TABs will be essential. 
The other projects are at 22\,GHz and 
%in general if observed at high declination could be better with in-beam PR, if suitable very close ($\sim$1\arcmin) reference sources could be found, which is unlikely; otherwise 
source-switched \MV\ should deliver the required astrometric precision.
Not considered in  \citet{ngvla_lbs} are the possibilities for SFPR projects with frequency phase referenced $\lambda$-astrometry. 
An astrometric study of core-shifts and jet structure with ngVLA sensitivities, for example, will provide exquisite insight into the jet-physics and launching mechanisms from AGN black-holes, which will complement the results from the EHT. 
Additionally, the ngVLA would have the potential to perform astrometry on multiple maser emission lines around their stellar hosts.
A large survey of the masers around the full gamut of AGB stars, with sufficient temporal sampling to track their responses to the pumping mechanism in different astro-chemical environments,
would provide a huge increase in our understanding of the seeding of the ISM.
We are sure that the readers could provide a host of perhaps even more exciting ideas.

\section{Summary}\label{sec:summary}

% Astrometry to improve your science.

% Bona fide Precise Astrometry Adds a new dimension to your research,
% with positions, proper motion, distances, & direct registration of
% multi-epoch and multi-frequency temporal and frequency monitoring

% Fundamental contribution to many research fields in astrophysics:
% widely applicable to many targets   and At a wide range of frequencies
% (m to (sub)mm wavelengths)

% All regimes, Ground vlbi, space vlbi

% Development of new calibration methods and (to optimize the potential
% of) new instruments a leap in the astrometric performance

% (and increased sensitivity by compensation of pm effects)

% The field of astrometry, the precise measurement of the positions,
% distances and motions of astronomical objects, has been revolutionized
% in recent years. As we enter the high-precision era, it will play an
% increasingly important role in all areas of astronomy, astrophysics
% and cosmology. This edited text starts by looking at the opportunities
% and challenges facing astrometry in the twenty-first century, from
% space and ground.
{
We are in an exciting epoch in radio astronomy, with a huge investment in future Great Observatories such as SKA and ngVLA. 
Additionally there are pathfinder instruments such as FAST, ASKAP, MeerKAT, KVN, LOFAR and MWA, among others. 
These will all be used in conjunction with other telescopes across the globe, for high resolution VLBI observations.
Space-VLBI is an active field with proposals for Millimetron and Cosmic Microscope under discussion.
The next-generation of instruments and the opportunities that these provide will allow for unprecedented sensitivity with increased frequency coverage and should lead to improved astrometry, provided systematic effects are compensated. VLBI provides the highest astrometry precision in astronomy currently; the next-generation instruments will maintain this leadership.
Bona-fide precise astrometry adds a vital new dimension to astronomical measurements, providing positions, proper motion, distances and the direct registration of multi-epoch and multi-frequency temporal and frequency monitoring.
As such it makes a fundamental contribution to many research fields in astrophysics.

%#2
This review has therefore focused on the fundamentals of high-precision astrometric VLBI, how it relates to other methods, how it has developed over time and how it could develop into the future. 
We have presented a common framework of the basis for all astrometric VLBI, which allows us to understand how the different approaches are related, where they will perform best and how they could further improve. 

The propagation of the astronomical signal through the atmosphere introduces the dominant source of systematic errors and this then has set the limit to the frequency range, using established calibration methods. 
Advances have come from addressing this limitation, and have required new methods and technologies, which we describe. 
The frequency range for astrometry with established methods has been between 1.4 and 43\,GHz, achieving the level of 10\uas{} per epoch, in the best cases. 
We showed how observations at these frequencies with established methods are matching the thermal noise levels of current instruments. 
Additionally, established astrometric methods are currently limited to ground based arrays with accurate station positions.

To take advantage of the order of magnitude increase in sensitivity from the next-generation instruments for astrometry, systematic error calibration will have to be improved. 
Having made the link between the various astrometric strategies using the common framework allows us to identify the key issues.
These can be addressed with next-generation calibration methods, and therefore indicate what are the technological capacities that must be in place %to achieve the best outcomes from these new methods 
on the next-generation of instruments.
These new astrometric strategies are being demonstrated, operate beyond the range of frequencies possible with established methods and are in active development. They are reducing the systematic limitations from the atmospheric contamination by orders of magnitude, as shown in the studies reported here.
The most promising of the new methods are \MV{} and Source Frequency Phase Referencing, which provide precise astrometry at the low and high frequencies, respectively. 
These in turn would be best implemented using simultaneous multiple beams and multiple frequency observations, respectively. 
Additionally, these methods are suitable for astrometry with Space-VLBI.

We reviewed the technological approaches that are providing simultaneous multiple beams and multiple frequency capabilities. For the former these are multi-beam receivers, Phased Array Feeds and Tied-Array Beams; for the latter these are multi-band and broadband receivers. 
The next-generation methods and technological solutions are essential considerations for the next-generation instruments, to achieve thermal noise limited astrometry. 
The potential, we argue, is to enable precise astrometry over 0.3 to 130\,GHz and above at the level of 1\uas{}, in the best cases.
At this point the stability of the fiducial reference points would probably be the new dominant astrometric error.

%Applications
We reviewed recent astrometrical publications focusing on surveys with current instruments. 
We emphasise the contributions to astrophysics that come from large and un-biased astrometric surveys, which is where the next-generation instruments will have the greatest impact.
These surveys covered a large range of targets, such as pulsars, masers, stars and AGNs, to derive neutron-star physics, dynamic Galactic parameters, astrochemical conditions, gamma-ray origins and cosmology.
We explored potential science cases for the next-generation instruments given the increased frequency range and precision.

The next-generation observatories, and their pathfinders, are working towards ultra-precise ($\sim$1\uas) astrometry,
%improving on established astrometric methods and ensuring the capabilities required for the recently developed methods.
and ensuring that the technological capabilities for optimum performance of next-generation methods are implemented. 
The prospects are therefore excellent for precise to ultra-precise astrometry across the whole radio band and the future for radio VLBI-astrometry is bright.}

\begin{acknowledgement}
This work was supported in part by funding from the European Union's Horizon 2020 
research and innovation program under grant agreement No 730562 [RadioNet] and 730884 [VLBI with the SKA].
% 'R.D. and M.R. acknowledges funding from the [``VLBI with the SKA'' work
% package of the] EU's Horizon 2020 research and innovation programme
% JUMPING JIVE (grant agreement No 730884).'
It was also supported by the ngVLA Community Studies program, coordinated by the ngVLA design and development project of the National Science Foundation and operated under cooperative agreement by Associated Universities, Inc. 
% For a more generic acknowledgement:
%
% The next-generation Very Large Array is a design and development project of the National Science Foundation operated under cooperative agreement by Associated Universities, Inc.
ADS was used heavily to search for relevant publications and provide references.
We have discussed the ideas and issues in this document with many experts and young enthusiastic up coming researchers, 
to whom we are grateful for their time. 
%The authors acknowledge constructive conversations.
In particular Ed Fomalont, Walter Alef, Mark Reid, Patrick Charlot,
Mareki Honma, Sim\'on Al-Tariqa, Richard W. Porcas, Luca Moscadelli and Huib van Langevelde  have been reference points for us, as for many other people.
The comments from the anonymous referees were much appreciated. 
F. Kristen provided the raw data for Fig.~9b, to allow the re-fitting.
We thank A.~Deller, F.~Kristen, J.~Flygare, G.~Keating,  ASTRON and CSIRO for the reproduction of their figures.
% FARCHAPML: farl champ
\end{acknowledgement}

\appendix
\section{Symbols}
\normalsize

\begin{table}[ht]
      \caption{Symbols and abbreviations used in this review.}
    \label{tab:symbol}
    \centering
    \begin{tabular}{l|l}
\hline
\multicolumn{2}{c}{Quantities}\\
\hline
$\nu$&Observing frequency.\\
$\phi_{A,B,C}$& Residual phase observable for sources A, B or C.\\
$\phi_{T,R}$& Residual phase observable for target or reference.\\
$\tau_\phi$& Residual phase Delay, ${{1}\over{2\pi}}{{\phi}\over{\nu}}$. \\ %unambiguous phase over frequency.\\
$\tau_g$& Residual Group Delay, ${1\over{2\pi}}{{\delta\phi}\over{\delta\nu}}$. \\ %phase gradient over the bandwidth.\\
$t$& time.\\
$\overrightarrow{b}$&VLBI baseline vector.\\
$\hat{s}$&unit vector in the direction of the source.\\
$\theta_{\rm beam}$&Synthesised beamwidth of the array.\\
$\delta\theta$&The final astrometric product.\\ %, between directions $\hat{s}_{\rm true}$ and $\hat{s}_{\rm model}$\\
%$N_{ant}$&Number of VLBI antennas.\\
\hline
\multicolumn{2}{c}{Acronyms}\\
\hline
AIC & Advanced Ionospheric Calibration.\\
ATC & Advanced Tropospheric Calibration.\\
DR  & Dynamic Range.\\
EOP & The Earth Orientation Parameter.\\
ICRF& International Celestial Reference Frame.\\
MFPR& Multi-Frequency Phase Referencing.\\
MV  & MultiView.\\
RMS & Root Mean Square.\\
SEFD& System Equivalent Flux Density.\\
SFPR& Source/Frequency Phase Referencing.\\
SNR & Signal to Noise Ratio.\\
TAB & Tied Array Beam.\\
TEC & Total Electron Content { measured in TEC units. 1\,TECU = $10^{16}$ electrons.m$^{-2}$} \\
%ITRF&The International Terrestial Reference Frame\\
\hline
\multicolumn{2}{c}{Residual Error Contributors}\\
\hline
$\phi_{\rm pos,str,geo,ion,tro}$& Residual phase terms from source position and structure, array geometry, \\
&  ionosphere and troposphere propagation respectively.\\
$\Delta\theta_{\rm ion,tro,AB}$&Angular separation relevant for ionospheric calibration, \\
& tropospheric calibration or between sources, respectively.\\
$C_w$& Unitless parameter for the tropospheric dynamic scale: 1,2 or 4 for good, normal or poor.\\
%cale of dynamic troposphere (1,2 or 4 for good, normal and poor), the observing frequency,  the residual tropospheric excess path length and  the residua
$Z_{g,i,F}$&mean Zenith angle between sources from the ground, at ionospheric pierce points for the \\
 & the bottom and the peak of the F-layer, respectively.\\
%zenith angle from the ground, from the ionospheric pierce point and the mean zenith angle for the two sources
$n_{T,R}$&number of phase ambiguities for target or reference observations, respectively.\\
$\Delta\nu_{\rm eff}$&Effective synthesised frequency bandwidth relevant in geodetic observations.\\
$\sum_{i=1}^{N} \alpha_i$&Weighted sum of calibrator corrections relevant for MV. \\% for application to the target.\\
$\mathcal{R}$&General scale factor, for example: the frequency ratio in SFPR or $\sum_{i=1}^{N} \alpha_i$ in MV.\\
$\Delta \ell_z$&Residual tropospheric excess path length at Zenith.\\
$\Delta I$&Residual ionospheric TEC.\\
$\Delta \ell_I$& $\Delta I$ expressed as excess path length at 1GHz.\\
\hline
\multicolumn{2}{c}{Measurement precision}\\
\hline
$\sigma\phi^{\rm cal}$& Combination of all systematic phase errors relevant for a given calibration method (e.g. PR, MV or SFPR).\\
$\sigma\Phi$& Standard Deviation of the phase observable.\\
$\sigma\phi_{\rm thermal}$& Thermal phase noise contribution.\\
$\sigma\Delta\theta_{\rm thermal}$& Thermal astrometric error contribution.\\
$\sigma\Delta\theta^{cal}$& Systematic astrometric error contribution for a given calibration method (e.g. PR, MV or SFPR).\\
$\sigma\Delta\theta$& Total RMS of the astrometric measurement (systematic plus thermal).\\
\hline
    \end{tabular}
\end{table}

\end{document}